\documentclass[twocolumn]{aastex62}

\newcommand\aastex{AAS\TeX}

\submitjournal{ApJ}
\received{25 March 2018}
\revised{21 April 2018}
\shorttitle{\aastex\ Gaia Co-Movers}
\shortauthors{Faherty et al.}

\begin{document}

\title{New and Known Moving Groups And Clusters Identified in a Gaia Co-Moving Catalog}

\correspondingauthor{Jacqueline K Faherty}
\email{jfaherty@amnh.org}

\affil{American Museum of Natural History \\
Department of Astrophysics, 
Central Park West at 79th Street, New York, NY 10034, USA}

\author{Jacqueline K Faherty}
\email{jfaherty@amnh.org}

\affil{American Museum of Natural History \\
Department of Astrophysics, 
Central Park West at 79th Street, New York, NY 10034, USA}

\author{John J. Bochanski}
\affiliation{Rider University\\
2083 Lawrenceville Road
Lawrenceville, NJ 08648}

\author{Jonathan Gagn\'e}
\affiliation{Department of Terrestrial Magnetism\\
Carnegie Institution of Washington Washington, DC 20015, USA}

\author{Olivia Nelson}
\affiliation{American Museum of Natural History \\
Department of Astrophysics\\ 
Central Park West at 79th Street\\ New York, NY 10034, USA}

\author{Kristina Coker}
\affiliation{American Museum of Natural History \\
Department of Astrophysics\\ 
Central Park West at 79th Street\\ New York, NY 10034, USA}

\author{Iliya Smithka}
\affiliation{American Museum of Natural History \\
Department of Astrophysics\\ 
Central Park West at 79th Street\\ New York, NY 10034, USA}

\author{Deion Desir}
\affiliation{American Museum of Natural History \\
Department of Astrophysics\\ 
Central Park West at 79th Street\\ New York, NY 10034, USA}

\author{Chelsea Vasquez}
\affiliation{American Museum of Natural History \\
Department of Astrophysics\\ 
Central Park West at 79th Street\\ New York, NY 10034, USA}

\begin{abstract}

We present a re-organization of the \citet{Oh17} (Oh17) wide, co-moving catalog of 4,555 groups of stars (10,606 individual objects) identified in the Tycho Gaia Astrometric Survey (TGAS) into new and known co evolving groups of stars in the Milky Way.  We use the BANYAN $\Sigma$  kinematic analysis tool to identify 1,015 individual stars in the Oh17 catalog that yielded a $>$80$\%$ probability in one of 27 known associations (e.g. the AB Doradus moving group, Columba, Upper Scorpius) in the vicinity of the Sun.  Among the 27 groups uncovered by Oh17 that had $>$ 10 connected components, we find that 4 are newly discovered.  We use a combination of Tycho, Gaia, 2MASS, WISE, Galex, and ROSAT photometry as well as Gaia parallaxes to determine that these new groups are likely older than the Pleiades but younger than $\sim$1 Gyr.  Using isochrone fitting we find that the majority of these new groups have solar type stars and solar type metallicity. Among the 35 Oh17 groups with 5 - 9 members, we find that 19 also appear new and co-moving, with Oh17 Group 30 being particularly exciting as it is well within 100pc (range of 77 - 90 pc) and also appears to be older than the Pleiades. For known star forming regions, open clusters, and moving groups identified by Oh17, we find that the majority were broken up into pieces over several Oh17 groups (e.g. Lower Centaurus Crux members are spread over 26 Oh17 groups) however we found no correlation with positions of the groups on color magnitude diagrams therefore no substructure of the association correlated with the Oh17 designated group.  We find that across the 27 groups tested by BANYAN $\Sigma$ there were 400 new members to 20 different associations uncovered by Oh17 that require further vetting.

\end{abstract}


\section{Introduction} \label{sec:intro}
Mapping the Milky Way with co-evolving stars can provide information on the dynamic history of our Galaxy.   Galactic position along with parallax, proper motion, and radial velocity measurements for individual stars helps us to create maps of the 6 dimensional spatial and velocity structure of the Milky Way.  Using such detailed maps of the Galaxy, investigations of the locations of stars with their kinematic clustering can be paired with parameters such as stellar separations, fundamental parameter estimates (log ($g$), metallicity, mass, radius) and chemical compositions to yield vital details about the past, present and future of the Milky Way.  

Widely separated companions are particularly important as they are easiest to study in detail given that individual sources can be resolved.  Studying binaries, triples, or hierarchical  systems with common origin, allows  much needed tests on age-calibration relations (e.g.  \citealt{Chaname12}) but also can lead to intriguing discoveries about the planetary formation history around Milky Way stars (e.g. \citealt{Oh17a}; \citealt{Teske15}). Furthermore, hierarchical systems at wide separations that have strong evidence for co-evolution are important objects for understanding the influence of the Galactic potential and large perturber structures (e.g. giant molecular clouds) in the Milky Way (e.g. \citealt{Weinberg87}).

 The Washington Double Star Catalog (WDS) maintained by the United States Naval Observatory is a principal resource for a list of double and multiple star information (\citealt{Mason18}) and contains over 140,000 entries with references for discovery as far back as 1895 (e.g. \citealt{Aitken95}).  In the current day, identifying widely separated companions is enabled by large or all-sky surveys such as the Sloan Digital Sky Survey (e.g. \citealt{Dhital10,Dhital15}), the Digitized Sky Surveys (e.g. \citealt{Lepine05,Lepine05a}), Tycho and Hipparcos (e.g. \citealt{Faherty10,Faherty11};\citealt{Kirkpatrick01} ).  Having precise parallax, proper motion and radial velocity are critical to identifying and confirming co-moving stars. That is why the European Space Agency's (ESA) Gaia space based observatory is a much-anticipated next major advancement in the science enabled by wide co-moving stars. By the time of the Gaia DR2 release (expected in April 2018), this mission will chart 10,000 times more stars than Hipparcos at a precision 100 times better.   In September 2016, the first data release of Gaia (DR1) which provided a subset of parallaxes and precise proper motions for Tycho stars (dubbed TGAS) led to several papers identifying new wide hierarchical systems in the Galaxy (e.g. \citealt{Oh17}, \citealt{Oelkers17}, \citealt{Andrews17}) .  

Aside from wide binaries in the Galaxy, important co-moving collections of moderate numbers of stars have also been discovered using astrometric surveys.  Using early, low astrometric precision catalogs, \citet{Kapteyn05aa} and \citet{Eggen65aa} identified large kinematically coherent associations of stars through their common proper motions and parallaxes.  With time, astrometric surveys became far more precise and these original associations were resolved into smaller age--coherent groups with the progression to milliarcsecond parallaxes and proper motions from the Tycho and Hipparcos surveys (\citealt{Perryman97aa}, \citealt{Hog00}).   

Within a few hundred parsecs of the Sun, there are numerous associations ranging in age from a few Myr (e.g.\ Rho Ophiuchus, 0.5--2~Myr, \citealt{Wilking08aa}; Taurus, 1--2~Myr, \citealt{Daemgen15aa}) through hundreds of Myr (e.g. Tucana Horologium, $\sim$45 Myr; AB Doradus, $\sim$150 Myr, \citealt{Bell15aa}; Pleiades, $\sim$112 Myr, \citealt{Dahm15aa}; Hyades, $\sim$750 Myr, \citealt{Brandt15aa}).  In--depth studies of the closest associations to the Sun ($<$ 200 pc) have revealed that they harbor large numbers of low--mass stars, brown dwarfs, and even objects whose mass falls below the deuterium burning boundary (so--called free-floating planetary--mass objects; \citealt{Gagne15aa}; \citealt{Faherty16aa}).  Moreover, given that moving groups harbor the closest young stars to the Sun, they are also the targeting ground for directly imaged exoplanets. Associations such as Tucana Horologium, TW Hya, and the AB Doradus moving group contain isolated objects that range in mass from a few solar masses down to a few Jupiter masses (\citealt{Gagne17aa}, \citealt{Faherty18aa}) as well as stars with planetary--mass companions. Observations of these associations enable investigations of the mass function, kinematics, and spatial distribution across the full range of objects generated through star formation processes in different isolated groups at young (1--2~Myr), medium (30--50 Myr), and older (100--700 Myr) ages.  \\

Given the expectation of a dramatic increase in high precision astrometry for stars in the Galaxy with Gaia, the DR2 catalog will re-organize our understanding of structures in the local Galactic neighborhood (e.g. \citealt{Kushniruk2017}). In this work, we examine the recent catalog of co-moving stars produced by \citet{Oh17} in search of new higher order structures in the Milky Way.  In Section ~\ref{sec:sample} we discuss the sample at large and in Section ~\ref{sec:data} we detail all the data we collected on individual sources.  In Section ~\ref{sec:Oh17reorg} we describe in detail or re-organization of the original catalog into known and new structures in the Galaxy using (primarily) the kinematic analysis tool called BANYAN $\Sigma$.  In Section ~\ref{sec:Activity} we focus on 5 new associations that appear to be newly identified and in Section ~\ref{sec:Isochrones} we review the age, mass, and metallicity parameters for each of them calculated using isochronal fitting.  Conclusions are presented in Section ~\ref{sec:conclusion}.

\section{The Sample} \label{sec:sample}
In September 2016, the European Space Agency (ESA) made public the first data release catalog (DR1) including a subset dubbed the Tycho Gaia Astrometric Survey (TGAS);  (\citealt{Lindegren16}).  The latter contains parallaxes and proper motions for 2,057,050 stars with astrometry grounded by positions in the Tycho-2 catalogue (\citealt{Hog00}). While there are numerous co-moving stars uncovered with the original Hipparcos and Tycho catalogs (e.g. \citealt{Lepine07}; \citealt{Faherty10}; \citealt{Shaya11}), the precision and photometric reach of Gaia lends itself to searches for wider and more distant pairs than has been previously possible. Indeed since the release of TGAS, several papers have conducted new analyses for wide co-moving pairs (\citealt{Oh17}; \citealt{Andrews17,Andrews18}; 
\citealt{Oelkers17}) that have pushed both the distance (separations $\gg$1 pc) and magnitude of proper motion for the discovered system.

In this paper we use the work of  \citet{Oh17} as our input sample.  \citet{Oh17} restricted the TGAS sample to those stars with parallax signal to noise values $\geq$ 8 (totaling 619,618) and searched within a 10~pc radius of each for a co-moving companion.  Using both parallax and proper motion values of this precise subsample of TGAS stars, \citet{Oh17} implemented a marginalized likelihood ratio test to discriminate candidate co-moving pairs from the field population. Unlike the standard method for identifying co-movers with a proper motion cut (e.g. recent work by \citealt{Oelkers17} which uncovered $\sim$ 1900 pairs in TGAS), the  \citet{Oh17} technique marginalized over the (unknown) true distances and velocities of the stars utilizing a probabilistic model for the assumptions on the 3D velocities of the two stars in a pair. 

The outcome of the \citet{Oh17} work is by no means a complete overview of wide, Gaia co-movers.  Indeed between the \citet{Oelkers17}, \citet{Andrews17}, and \citet{Oh17} wide companion catalogs from TGAS astrometry, there is overlap but unique collections in each.    Interesting for this work,  \citet{Oh17} uncovered 10,606 unique stars found to be co-moving with at least one but up to 151 others in TGAS (referred to as Oh17 sample from here-in). In the process and perhaps quite serendipitously for the project,  \citet{Oh17} uncovered parts of many known moving groups, open clusters, associations and star-forming regions. 
 
The 10,606 unique stars are organized into 4,555 groups.  Those are further broken down into 27 groups that have $\geq$ 10 connected components-- objects who passed the Oh17 kinematic association criterion--, 35 groups that have between 5 - 10 connected components, 39 groups that have four connected components, 218 that have three connected components, and 4,236 that have two connected components. 

As noted in \citet{Oh17}, radial velocities are required for each star to further verify that co-movers are not simply chance alignments, a prospect that becomes far more likely as the pairs have separations $>$ 1 pc (see \citealt{Price-Whelan2017}).

\section{Data on the Sample\label{sec:data}}
In order to examine the groups as a whole in the Oh17 sample, we supplemented the Gaia data for each unique star with catalog photometry and spectral information.  Using the Tool for OPerations on Catalogues And Tables (TOPCAT; \citealt{Taylor05}), we cross-matched with the Two micron All Sky catalog (2MASS; \citealt{Skrutskie06}), the Wide Field Infrared Survey Explorer Mission (WISE; \citealt{Wright10}), the Galaxy Evolution Explorer (GALEX; \citealt{Martin05}), and the Rontgen Satellite (ROSAT; \citealt{Voges00,Voges99})-- both bright and faint source catalogs.  For 2MASS and WISE we used a 1$\arcsec$ radius to match to ALLWISE which automatically had a cross-match with 2MASS at the 3$\arcsec$ level.  For GALEX we used 5 $\arcsec$ and ROSAT we used 30$\arcsec$ given their larger pixel sizes and greater positional uncertainties. We also used TOPCAT to cross-match all 10,606 unique stars with the Catalogue of Stellar Spectral Classifications compiled by \citet{Skiff14}.  We list the recovered 2MASS $JHK_{s}$, WISE $W1W2W3W4$, GALEX $FUV$ and $NUV$ photometry, ROSAT X-ray flux values as well as \citet{Skiff14} spectral type in Tables ~\ref{Tab:SpT}, ~\ref{Tab:Phot}, and ~\ref{Tab:Activity}.  We used the photometry in combination with the Gaia parallaxes in order to investigate color-magnitude diagrams and search for age-informative diagnostics in Section~\ref{sec:Activity} below. 

\section{The Oh17 Sample Reorganized\label{sec:Oh17reorg}}
As the Oh17 sample is organized into groups with only a few known collections of stars explicitly described, in this Section we re-organize the co-movers into their known associations, label them as new candidate members of groups, or recommend a collection of stars as an entirely new pair or co-moving association.  A detailed literature search of new co-moving pairs would normally be warranted to execute this task, however it is time consuming to sift through the numerous references to Tycho and Hipparcos stars (although see Gagne et al submitted for an analysis of new members across all of TGAS).  To expedite the re-organization into known associations, we utilized a kinematic tool called BANYAN $\Sigma$ (\citealt{Gagne18}). This tool uses a compiled list of bonafide members (see \citealt{Gagne18} for details on the bonafide definition) of 27 different associations within 150 pc of the Sun to determine the probability of a given star on the sky also being associated.

\subsection{Oh17 sample in BANYAN\label{sec:Oh17BANYAN}}
\label{sec:banyan}
To begin sorting which group corresponded to which known association of stars, we first applied the BANYAN $\Sigma$ kinematic code (\citealt{Gagne18}) to each of the 10,606 unique stars to ascertain the Bayesian probability that it belonged (new or known) to one of 27 co-moving collections. We chose a moderate membership probability threshold of $>$ 80$\%$ Bayesian likelihood for membership in a known group.  This number is arbitrary and moderately conservative (e.g. Gagne et al. submitted used $>$ 90$\%$ for their threshold) but given that these sources already have at least one connected component, we think it is justified.  Table ~\ref{Tab:Oh17Ban} shows that 1015 unique stars passed the probability criterion for one of the 27 collections tested by BANYAN.  Table ~\ref{Tab:Oh17Ban} gives the probability that an object was a member of a BANYAN tested association (column 10) but it also gives a breakdown of likelihood if it had a chance of being in more than one group (column 9).  

A few things of note came from the BANYAN results.  Given the 80$\%$ group membership probability that we employed to investigate the Oh17 co-moving stars , there were times when one object in a co-moving pair was found to be a part of a known group while it's  designated partner was not.  For instance, in the AB Doradus moving group, there were 24 individual stars identified by BANYAN to have a $>$ 80$\%$ probability of membership.  However there were four double-connected components in Oh17 where only one of the two stars were recovered as AB Doradus members (e.g. Oh17 group 2240 which contains a bonafide member).  This occurred across all of the 27 BANYAN groups.  Given that the \citet{Oh17} method requires matching kinematics for the pairs to emerge, we postulate that the majority of these broken up groups is due to the second component having a probability that simply did not exceed our 80$\%$ threshold.  Further investigation may prove that its connected partner(s) is also a candidate member, but with a lower likelihood.  All known information and BANYAN results are reported in Table ~\ref{Tab:Oh17Ban}.

\subsection{Known Associations from BANYAN\label{sec:BANYANKnown}}
Using the $>$80$\%$ probability threshold, members of each of the 27 BANYAN $\Sigma$ tested groups were identified. It is important to note that the \citet{Oh17} method was not designed to identify these large kinematic associations therefore it is not expected that it would or should recover all known members possible from Tycho and Hipparcos in a given association (see \citealt{Gagne2018C}).  As such, we emphasize that the Oh17 sample is not a complete look at known groups or even one where a significant portion of members should be identified.  

That stated, for our own purposes, we cross-matched all of the individual stars with the bonafide (BM), candidate (CM), high-likelihood (HM), low-likelihood (LM), ambiguous (AM) and rejected member (RM) lists used in \citet{Gagne18} to create BANYAN $\Sigma$.  Column (12) and column (13) of Table ~\ref{Tab:Oh17Ban} reflect whether an object might be a new candidate that requires further vetting  (NO) or whether it falls in one of the above listed BANYAN categories (BM, CM, LM, AM, or  RM).  Table~\ref{tab:summary} summarizes each of the 27 BANYAN tested groups.  We note that the NO objects may still be known literature sources however they are not in BANYAN $\Sigma$ and we only performed a cursory check in the literature for the most prominent associations.  In the case of the Hyades, the Pleiades, and Coma Berenices we verified that all of the NO sources were discussed as members in the literature therefore there are no new additions in the Oh17 catalog.  The remaining groups with NO objects listed, may have new members uncovered.  Specifically,  given that there are a significant number of Oh17 stars that are connected components to known or candidate members of associations, we suggest there are a significant number of new additions to moving groups uncovered.  Detailed vetting and literature searching is required to confirm new objects.  

Organizing the Oh17 groups/pairs by BANYAN results as laid out in Table ~\ref{Tab:Gaia} we find that Upper Centaurus Lupus (194 $>$ 80$\%$ probability members), and Lower Centaurus Crux (156 $>$ 80$\%$ probability members) were found in greatest number. Conversely, the groups 118TAU, $\rho$ Ophiucus, and Carina were not recovered at all. 
 
\begin{figure*}[!h]
\begin{center}
\epsscale{1.0}
\plottwo{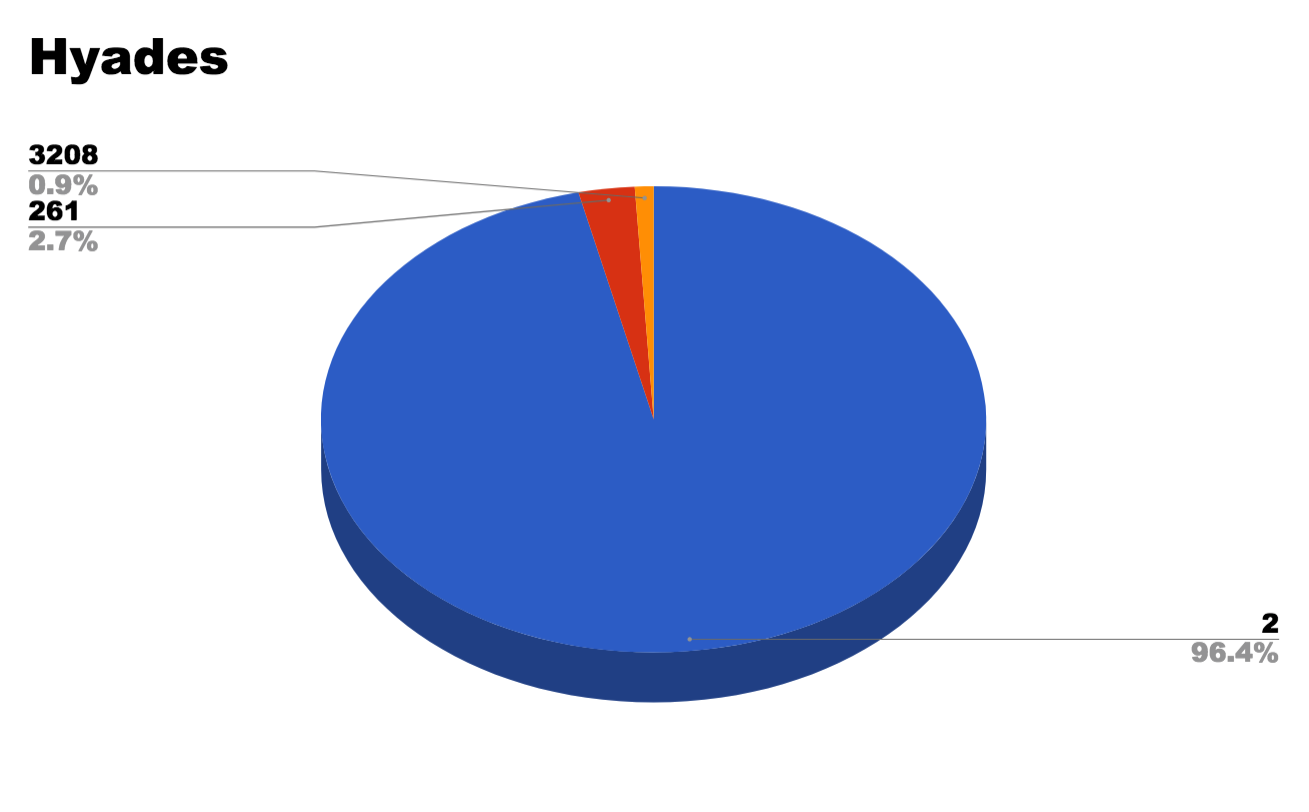}
{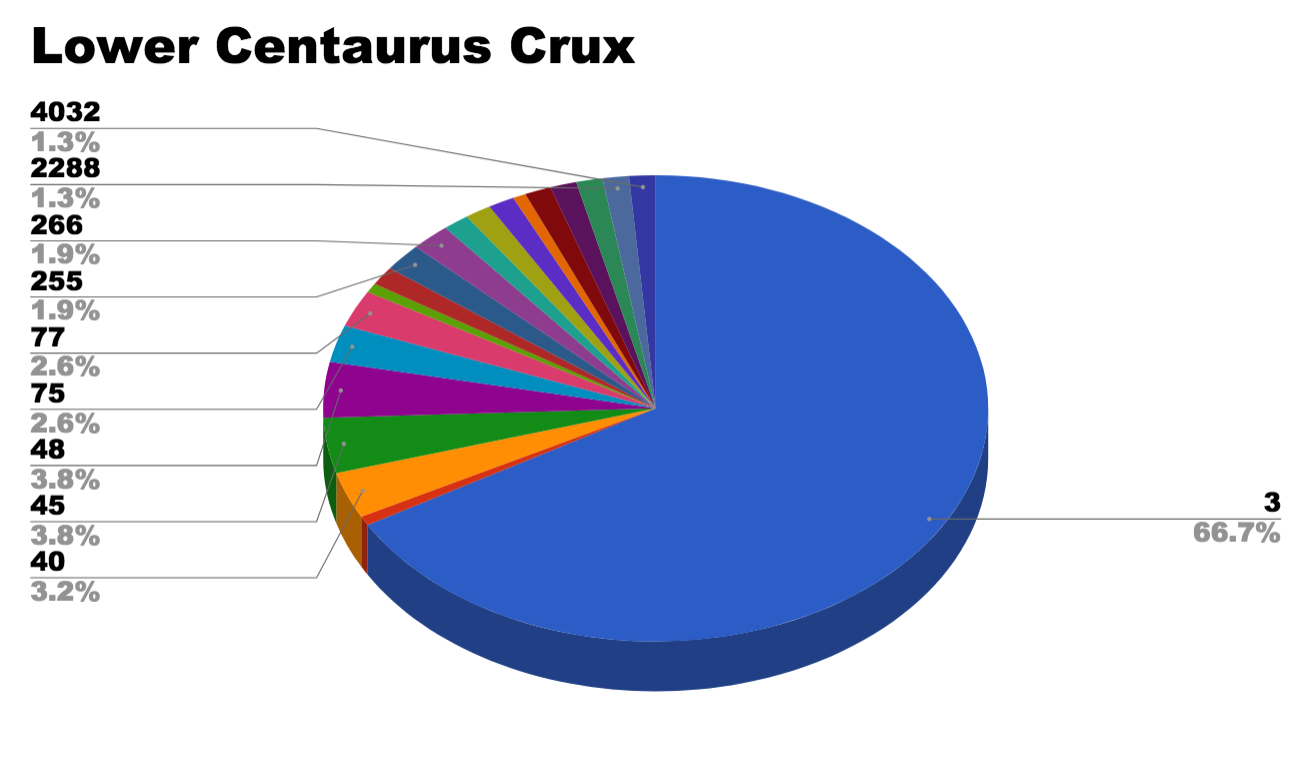}
\end{center}
\caption{A pie chart distribution of Oh17 designated groups with $>$ 80$\%$ membership in a BANYAN $\Sigma$ tested association. The percentages are based on the total number of objects that BANYAN found to be members and the contribution by an individually marked Oh17 group (labeled above the percentage).  At left we show the distribution of BANYAN $\Sigma$ predicted Hyades members and at right it is Lower Centaurus Crux members.  For each Oh17 group, we have color coded it in the pie chart.  We label the group number with percentage of the total at each pie slice.\label{fig:pie1}} 
\end{figure*}

\begin{figure*}[!h]
\begin{center}
\epsscale{1.0}
\plottwo{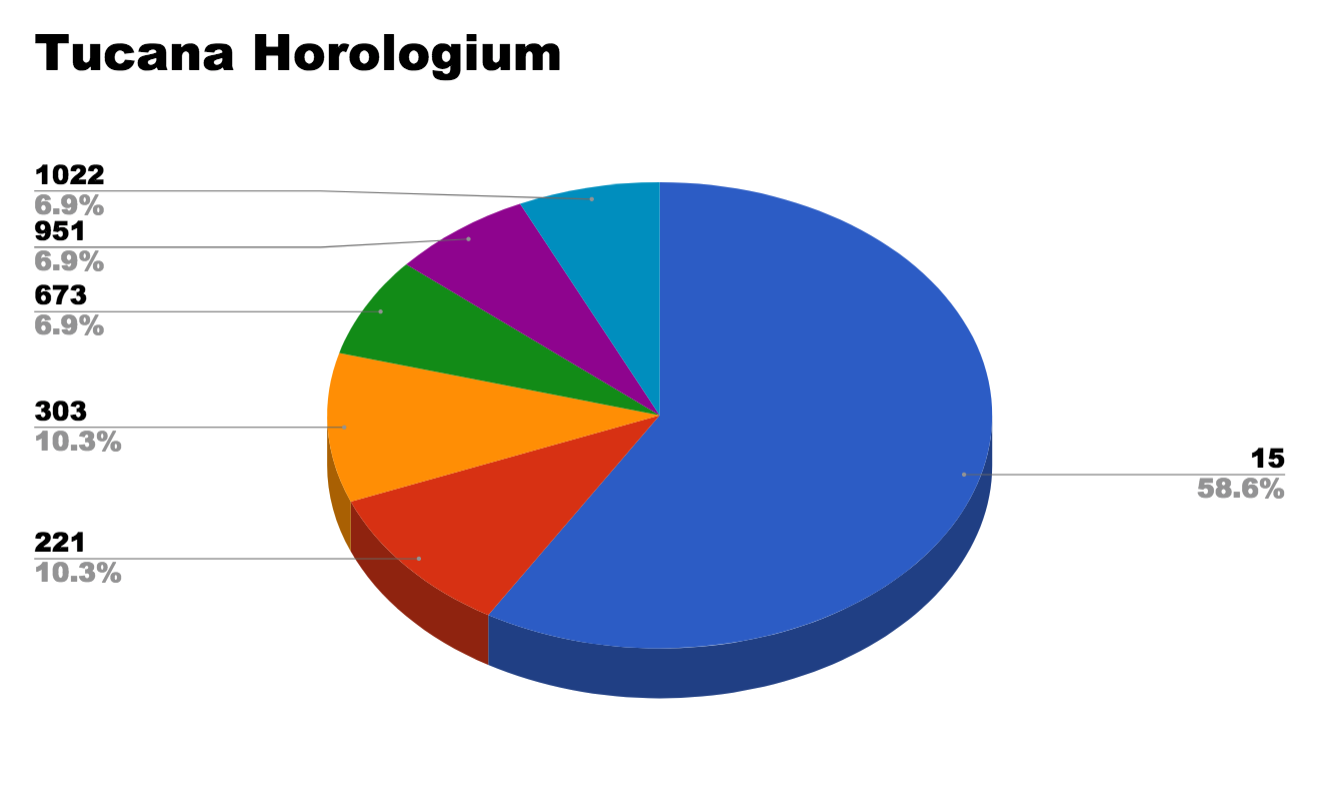}
{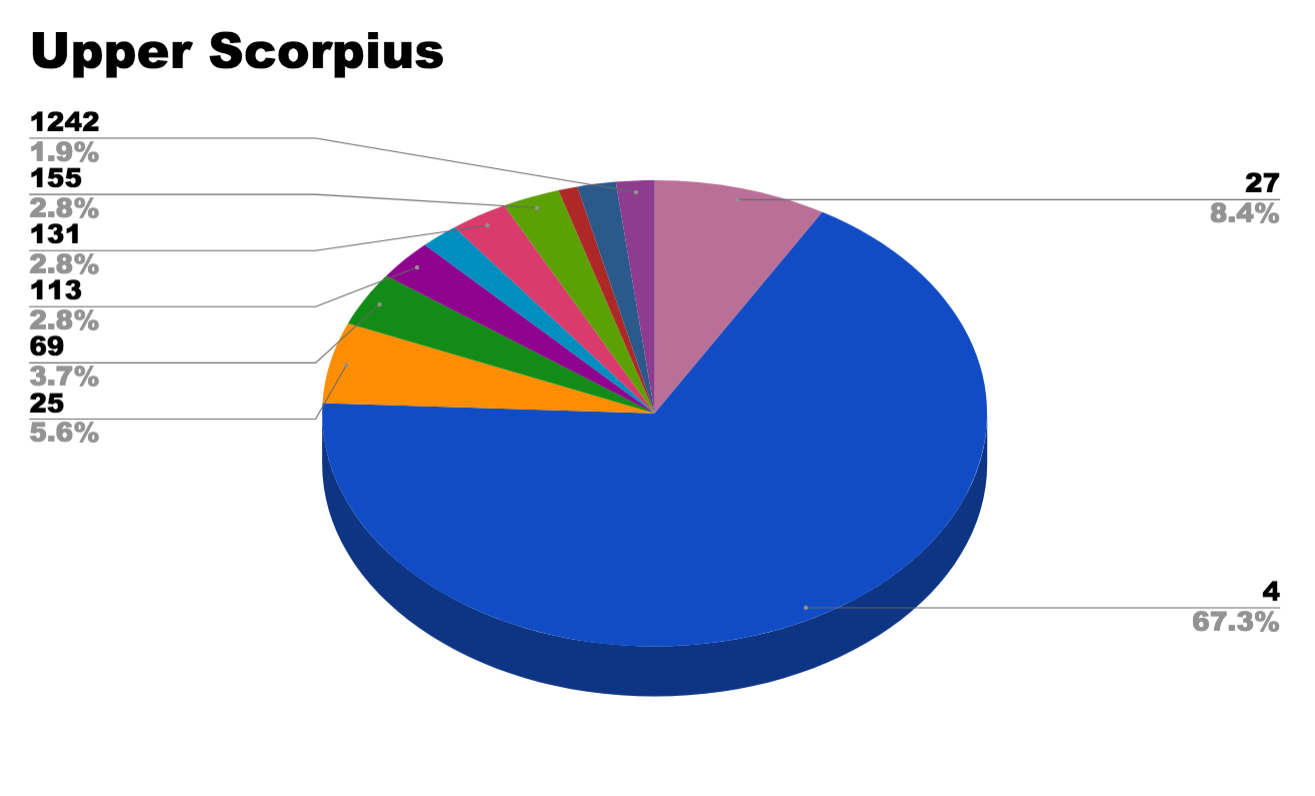}
\end{center}
\caption{See caption of Figure ~\ref{fig:pie1}.  At left we show the distribution of BANYAN $\Sigma$ predicted Tucana Horologium members and at right it is Upper Scorpius members.\label{fig:pie2}} 
\end{figure*}

\begin{figure*}[!h]
\begin{center}
\epsscale{1.0}
\plottwo{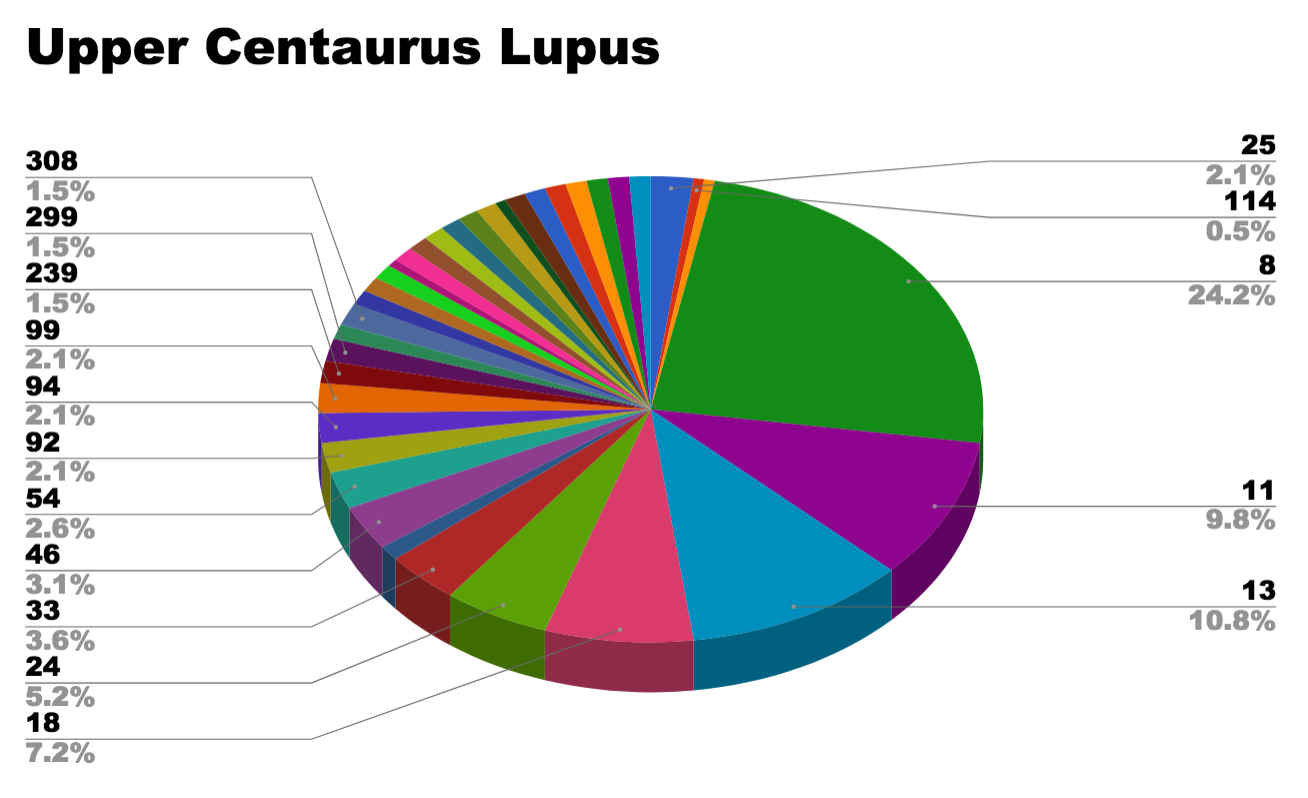}
{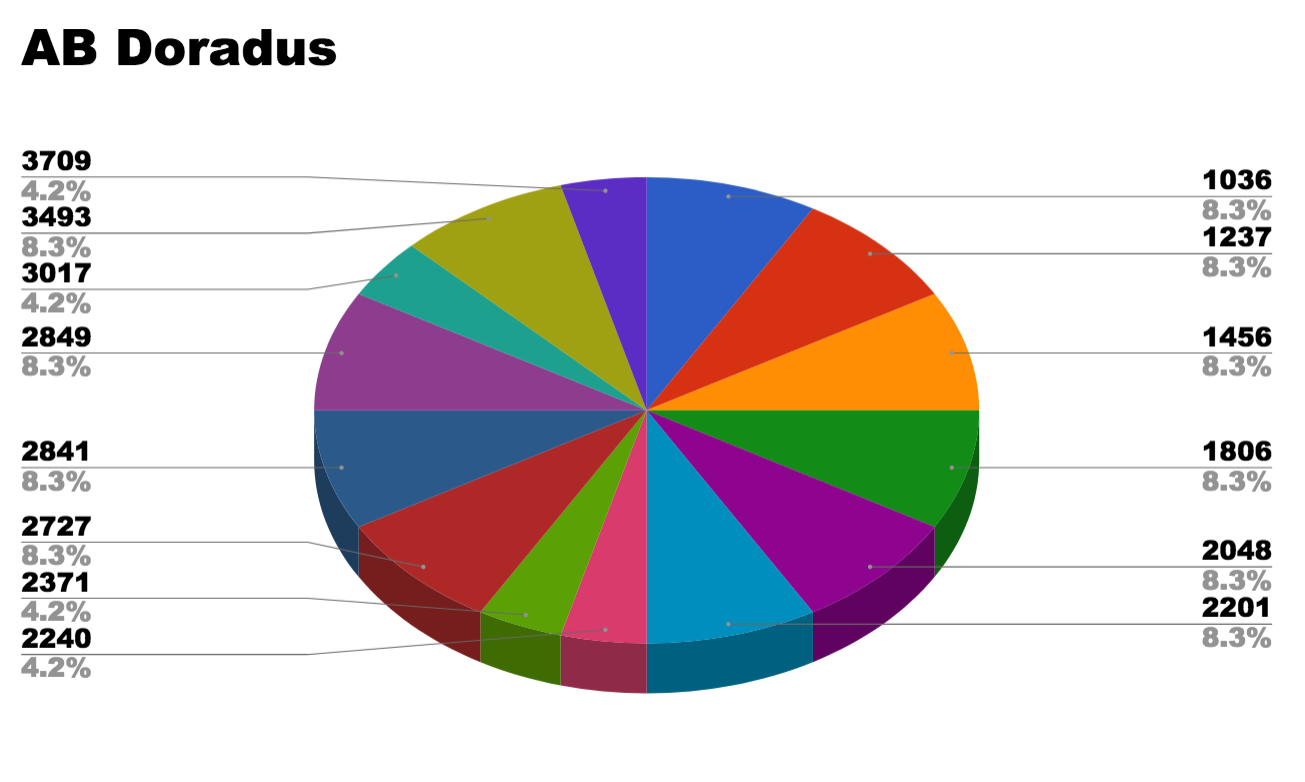}
\end{center}
\caption{See caption of Figure ~\ref{fig:pie1}.  At left we show the distribution of BANYAN $\Sigma$ predicted Upper Centaurus Lupus members and at right it is AB Doradus.\label{fig:pie3}} 
\end{figure*}

Members of the Pleiades (Oh17 group 0), Coma Berenices (Oh17 group 7), IC 2602  (Oh17 group 5) and Alessi 13 (Oh17 group 19) were recovered from BANYAN $\Sigma$ as all belonging to the same Oh17 defined group.  However-- with the exception of Corona Australis, $\epsilon$ Chamaeleontis, $\eta$ Chamaeleontis and Ursa Major where only one pair (or 1 object in a pair) was recovered-- the remaining 15 BANYAN $\Sigma$ groups were recovered across more than one Oh17 defined group (see Figures ~\ref{fig:pie1}, ~\ref{fig:pie2}, and ~\ref{fig:pie3} for pie charts illustrating the distributions).  For instance, BANYAN $\Sigma$ found 107 $>$ 80$\%$ membership probability Upper Scorpius stars.  However, as illustrated in Figure ~\ref{fig:pie2}, those are spread across 11 different Oh17 groups.  Among those, Oh17 group 4 (with 72 members), was the largest collection of Upper Scorpius objects, all of which came out as $>$ 80$\%$ membership BANYAN probability in the association. The 10 other Oh17 groups with the 35 remaining $>$ 80$\%$ membership probability Upper Scorpius stars ranged in size from 2 - 10.  

\begin{figure*}[!h]
\begin{center}
\epsscale{1.0}
\plotone{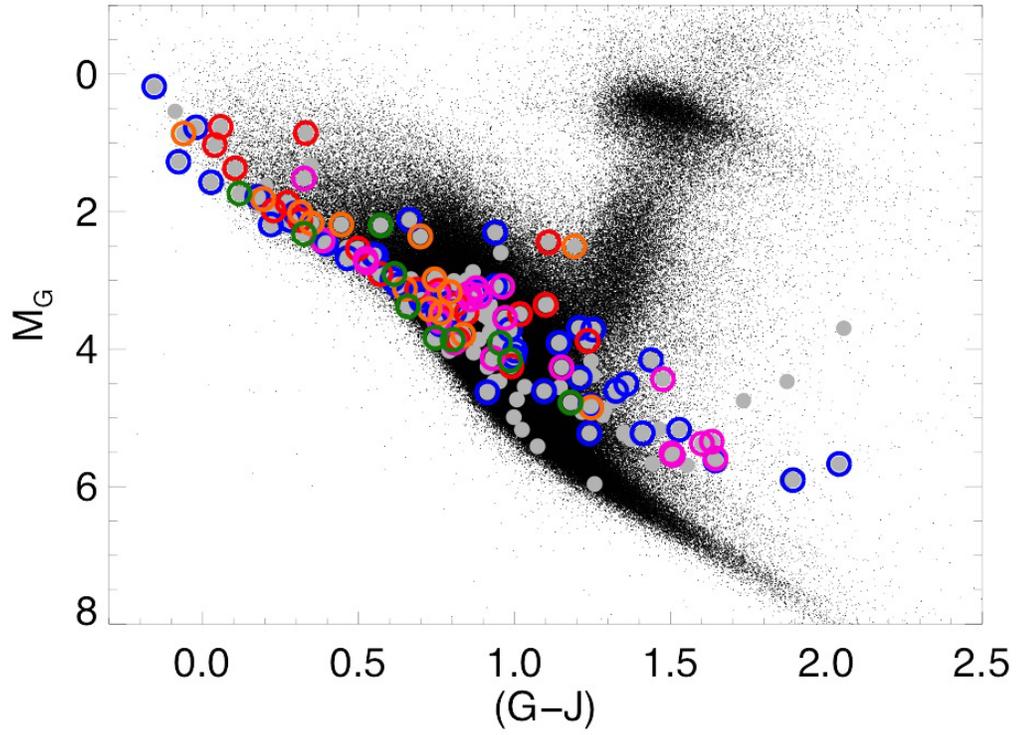}
\end{center}
\caption{The ($G-J$) versus $M_{G}$ color magnitude diagram for TGAS (black) and all BANYAN $\Sigma$ $>$ 80$\%$ membership probability of the Upper Centaurus Lupus association (UCL).  We highlight in different colored circles, the five largest Oh17 groups that were found to be co-moving members of UCL. \label{fig:UCL}} 
\end{figure*}

We investigated whether this substructure of Oh17 groups for the same BANYAN predicted group was significant but found no obvious evidence for a correlation with the members and their Oh17 group assignment.   For example, in Figure ~\ref{fig:UCL} we look at the 5 largest Oh17 groups that came out as Upper Centaurus Lupus in BANYAN on a ($G-J$) versus $M_{G}$ color magnitude diagram.  While we see nothing striking, \citet{Roser17} reports that using TGAS astrometry, spatial positions, and some follow-up radial velocities, they find that Group 11 is a part of a compact new moving group around V1062 Sco.   Further investigation is required for each of the individual groups to see if they end up mapping fine detailed kinematic structure or sub structures within a larger association.

Groups that were particularly close to the Sun (distance $\ll$ 100 pc) were almost entirely broken into pairs or triples by \citet{Oh17}.  For instance, as illustrated in Figure ~\ref{fig:pie3}, the AB Doradus moving group whose members span from 7 \textemdash 77 pc from the Sun and are scattered in right ascension and declination all over the sky, spanned 14 different Oh17 groups.  The \citet{Oh17} method probably breaks down for nearby groups given that the closer an association is to the Sun, the more important understanding the full kinematics becomes to deciphering membership.  With the \citet{Oh17} method, radial velocity measurements are not employed, making it difficult to differentiate the full kinematic signature of solar neighborhood groups.

We also found that several of the Oh17 groups with $>$ 10 sources had a mixture of objects with different BANYAN predicted associations.  Table ~\ref{tab:conflicts} lists each of the Oh17 groups that had a component in more than one of the BANYAN $\Sigma$ tested groups.  For instance, Oh17 group 25 has 6 BANYAN predicted members of Upper Scorpius and 4 BANYAN predicted members of Upper Centaurus Lupus. Group 13 has 1 BANYAN predicted members of Lower Centaurus Crux and 21 predicted members of Upper Centaurus Lupus.  These associations are known to have cross-contaminating kinematics as they occupy a similar part of XYZUVW space so it is not surprising to see the Oh17 method find them together.  Moreover, lacking a radial velocity means there is a crucial kinematic component missing for associations that are already very similar.

\subsection{Non-BANYAN but Known Groups Identified}\label{sec:out150}
As stated above, the Oh17 sample consists of 27 groups that have $\geq$ 10 connected components, which we choose as a cut-off number for what we investigated as a potentially new large association. We found that fifteen of those Oh17 groups contained BANYAN predicted members of nine different associations (as stated above some BANYAN groups were split among more than 1 Oh17 group).  The remaining 12 Oh17 groups with $>$ 10 members were split between (1) known associations that were simply not tested by BANYAN and (2) potential new groups.  For the former, we conducted a literature search including a cross match with the Gaia Open Cluster catalog (\citealt{Gaia-Collaboration17}) and found that Oh17 Group 1 is $\alpha$ Persei, Oh17 group 6 is Praesepe, Oh17 group 9 is IC2391, Oh17 group 17 is Blanco 1, Oh17 group 21 is NGC2451A,  Oh17 group 20 is Platais 3, and Oh17 group 16 is RSG2. Table ~\ref{Tab:Gaia} lists the membership (BANYAN or not) that we found through a literature search for any source in the Oh17 sample.

\subsection{New Moving Groups identified in Oh17\label{sec:NewMG}}
Of the 27 groups that have $\geq$ 10 connected components, five Oh17 groups appear to be newly identified. Those were Oh17 groups 10, 14, 23, and 26 (containing 29, 20, 10, and 10 connected components respectively).  The stellar members and associated kinematics for each new group are listed in Table ~\ref{Tab:Newgroups}.  Group 10 is perhaps the most exciting both because it has the largest number of members (29) and it falls within 100 pc (full range is 90 $-$ 115 pc).   We note that there is a reference to 3 of the stars (HIP 67005, HIP67231, and HIP66198) belonging to an unnamed open cluster by \citet{Latyshev77}-- along with 4 other stars that were not recovered by \citet{Oh17}.  However we found no other information aside from their coordinates in a literature search. 

For the remaining 3 groups, a quick search of the literature did not yield any indication that they were previously identified as co-moving associations.  Group 26 with 10 members is just a bit farther than Group 10 with a range of 105 $-$ 135 pc.  See Table~\ref{Tab:Newgroups} for the range of distances for all new groups as well as proper motion, radial velocity, and spectral types for their members.  

For the 35 Oh17 groups with 5 - 9 members, we find that 19 do not show any likelihood of membership in BANYAN tested groups, the Gaia Open Cluster catalog, or a very cursory literature search.  While we do not give these the same attention as the 5 new groups with  $\geq$ 10 connected components, we break out their individual components in Table ~\ref{Tab:Newgroups2}.  

\begin{figure*}[!h]
\begin{center}
\epsscale{2.0}
\plottwo{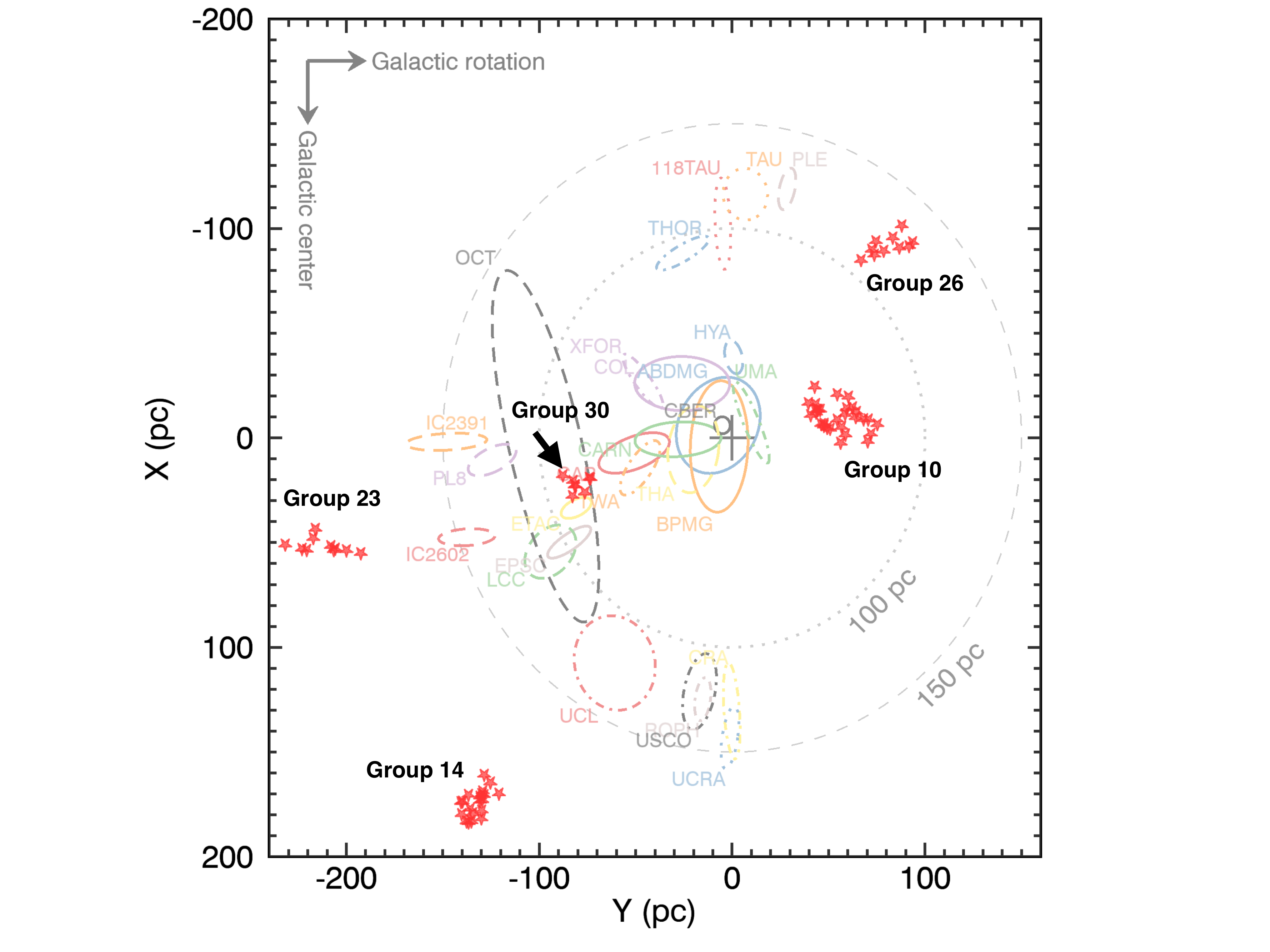}
{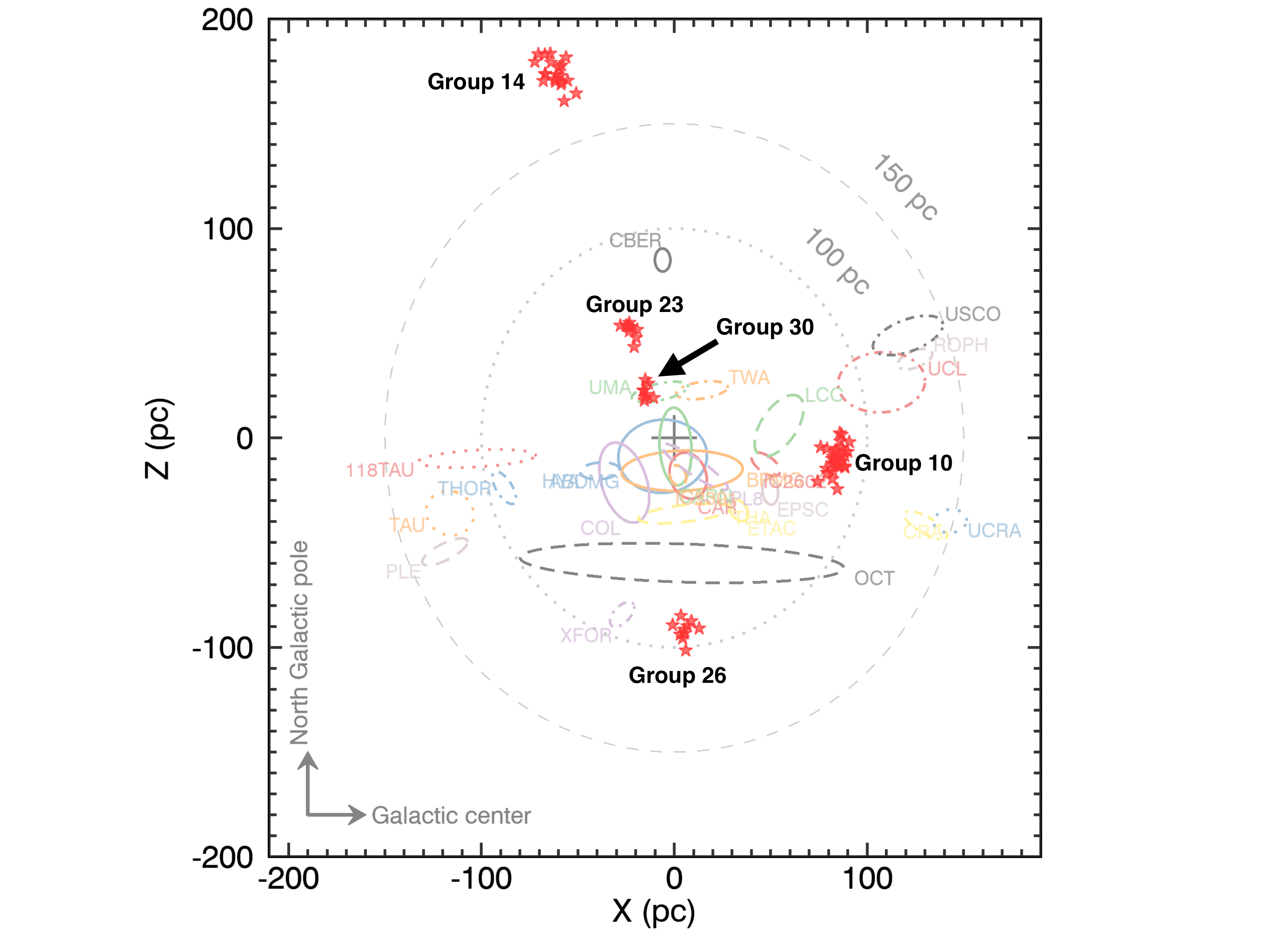}
\end{center}
\caption{The X versus Y and X versus Z positions of BANYAN $\Sigma$ tested groups as well as six (5 with $>$ 10 members and 1 with 8 members) new associations discussed in this work. \label{fig:XYXZ}} 
\end{figure*}

We note that Oh17 Group 30 is particularly exciting as it is within 100pc (range 77 - 90 pc; see Figure~\ref{fig:XYXZ}).  As such we provide a G-J color magnitude diagram for this 8 component group and note that there is 1 fairly strong X-ray active F5 star (see Figure ~\ref{fig:Group30-CMD}) indicating that it is Pleiades age or older.  This group will be the subject of a future study. 

\section{Age Estimates of the 5 New Moving Groups from X-ray and UV Activity\label{sec:Activity}}
We investigated the 5 new groups for age-indicative information among members. First, we plotted each in XYZ space to examine whether they overlapped in position with any of the BANYAN $\Sigma$ known groups.  Figure~\ref{fig:XYXZ} shows the results.  While each group occupies a tight portion in XY or XZ space, most are too distant for comparison and we find no obvious connection to known associations.  

There is a significant portion of stars in each group that have spectral information.  In Group 10, the closest of the new groups, there are AFG and K stars.  Group 14 only has A stars with literature spectral types while groups 23 and 26 have FG and A stars.  

\begin{figure*}
\begin{center}
\epsscale{1.2}
\plotone{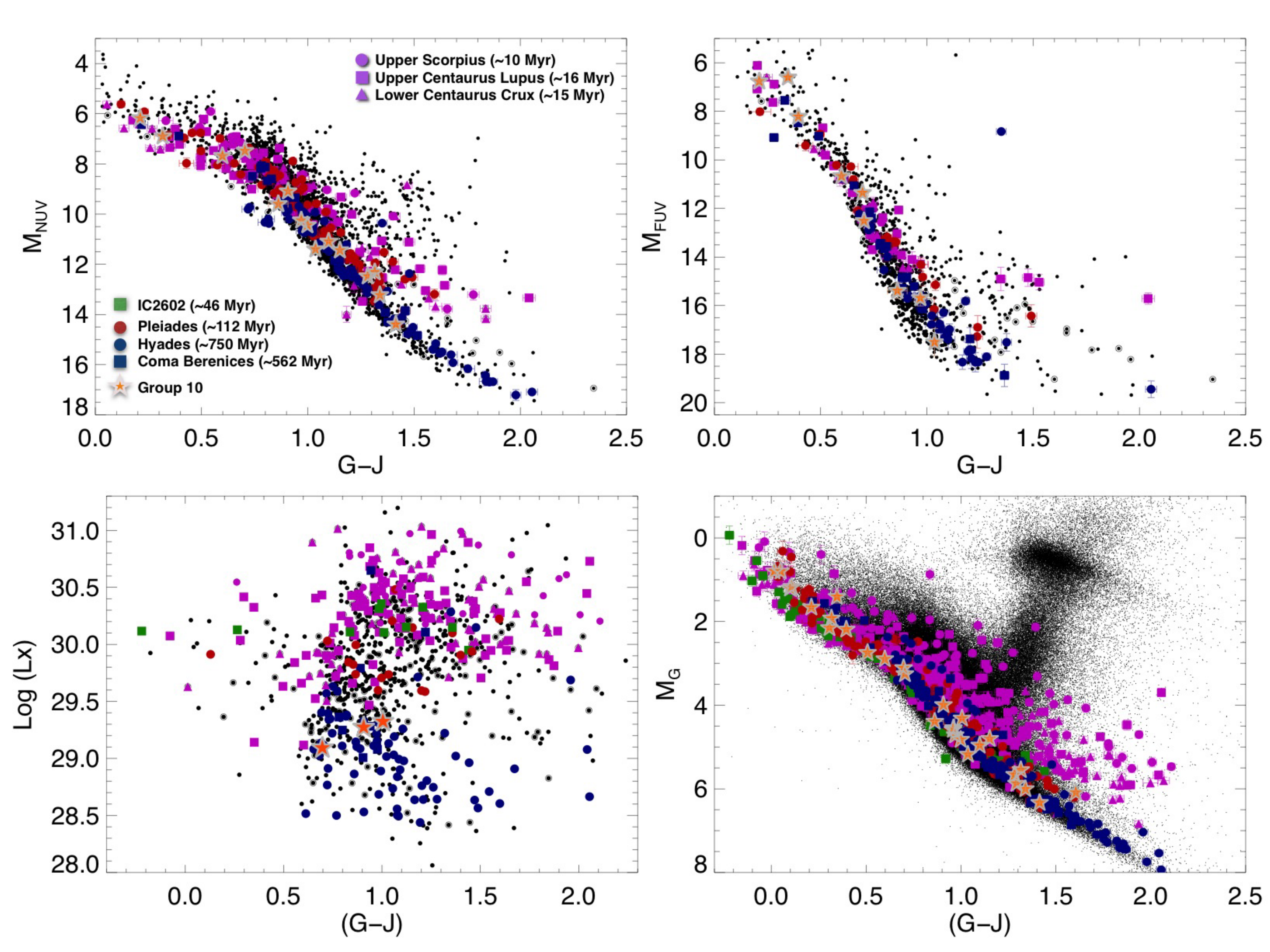}
\end{center}
\caption{A suite of color magnitude diagrams and an X-ray luminosity diagram highlighting the newly uncovered Group 10 from the Oh17 sample (orange five point stars).  At top left we show the ($G-J$) versus $M_{NUV}$ CMD for the full Oh17 sample (black) with select BANYAN $\Sigma$ selected groups highlighted to show different age bins.  Circled in black are Oh17 sources in known associations not otherwise color coded.  At top right we show the ($G-J$) versus $M_{FUV}$ CMD.  Bottom left is the ($G-J$) versus log$(L_{X})$.  Bottom right is the ($G-J$) versus $M_{G}$ CMD with all of TGAS stars with parallax signal to noise $>$ 10 shown as black points.  \label{fig:Group10-CMD}} 
\end{figure*}

\begin{figure*}
\begin{center}
\epsscale{1.2}
\plotone{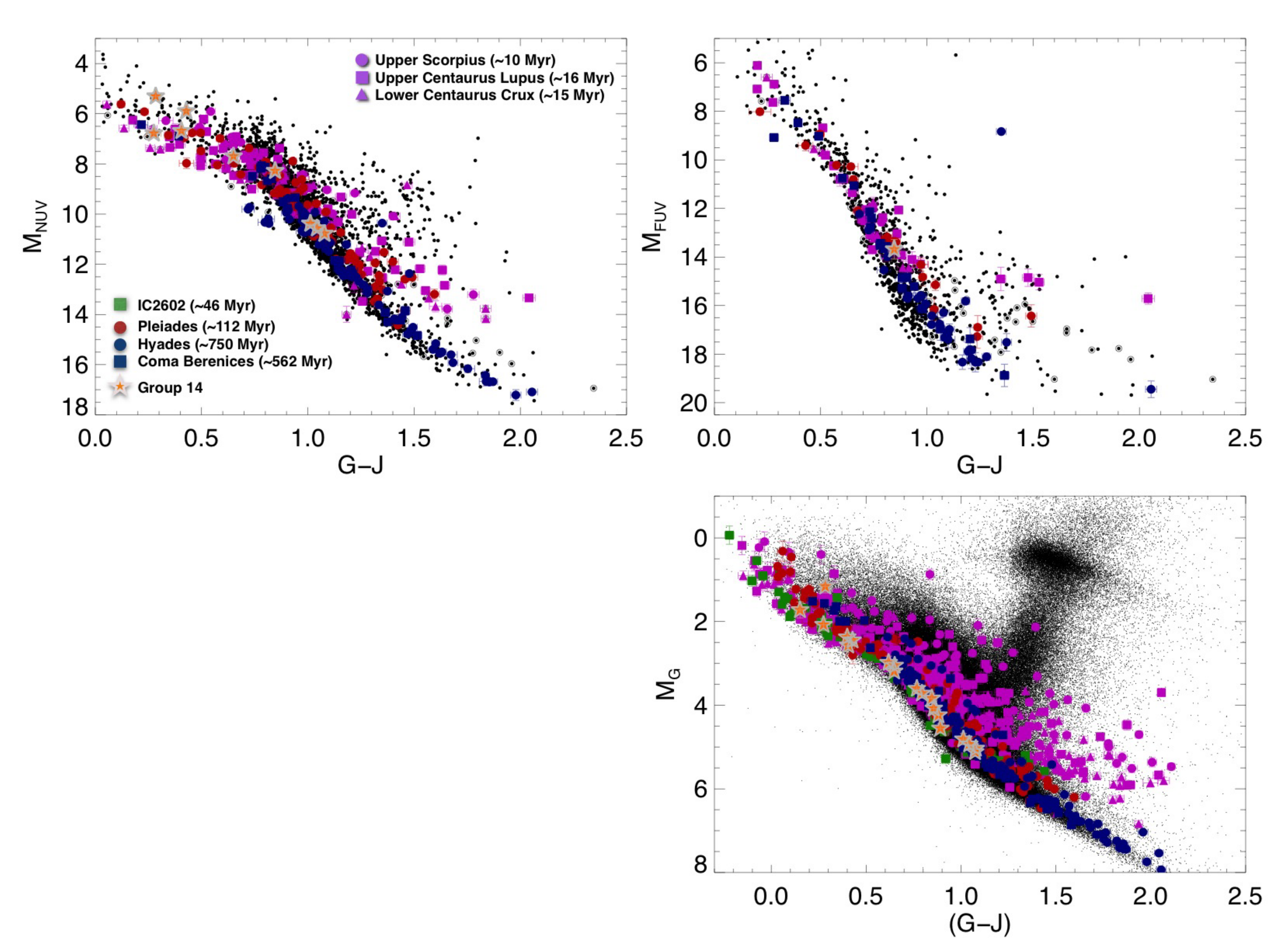}
\end{center}
\caption{See caption for ~\ref{fig:Group10-CMD}.  Group 14 from Oh17 is highlighted. No ROSAT detections were found for objects therefore this panel is blank in the figure. \label{fig:Group14-CMD}} 
\end{figure*}

\begin{figure*}[!h]
\begin{center}
\epsscale{1.2}
\plotone{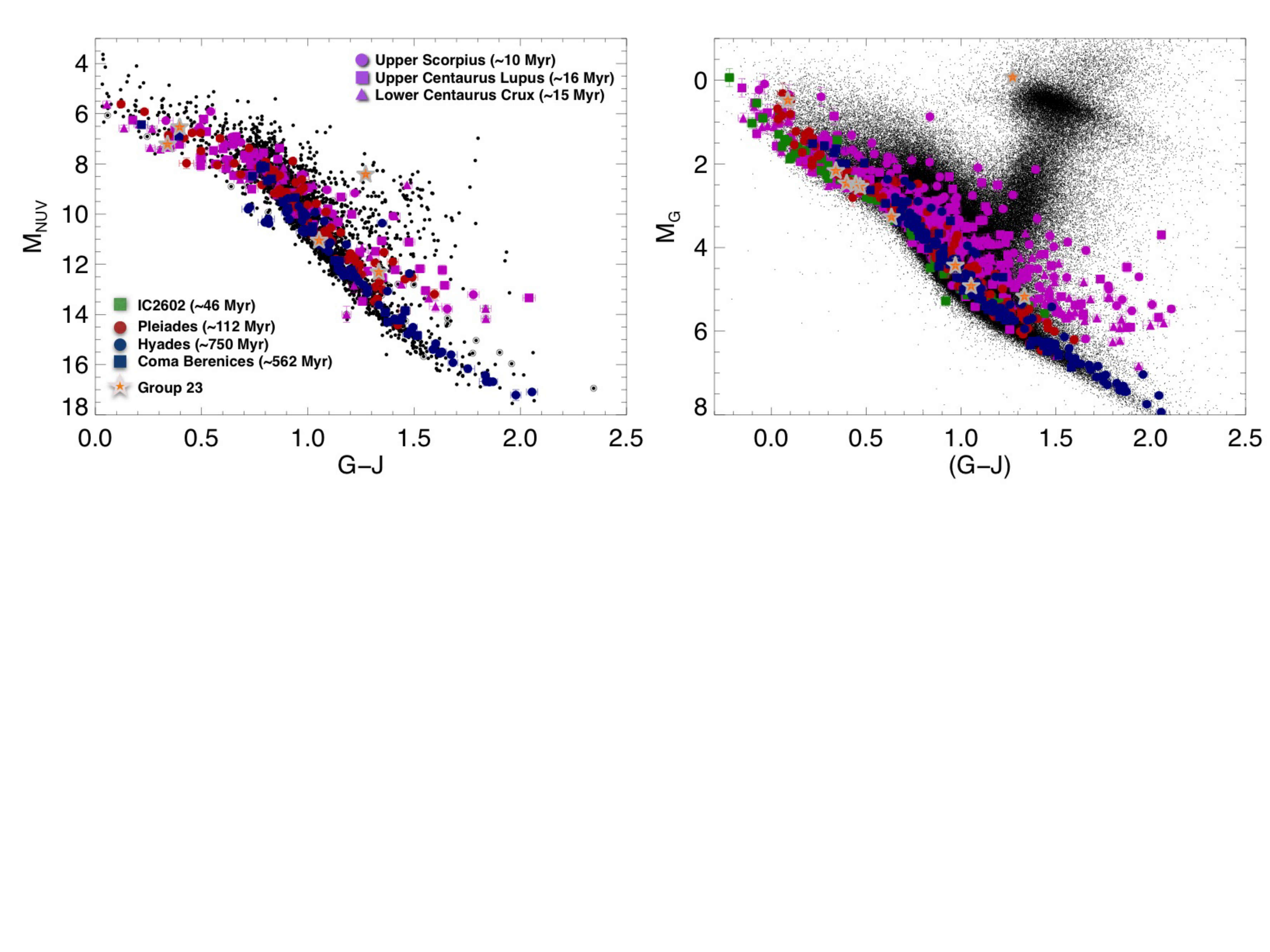}
\end{center}
\caption{See caption for ~\ref{fig:Group10-CMD}.  Group 23 from Oh17 is highlighted. No ROSAT  or $FUV$ detections were found for objects therefore those panels are blank in the figure.\label{fig:Group23-CMD}} 
\end{figure*}

\begin{figure*}[!h]
\begin{center}
\epsscale{1.2}
\plotone{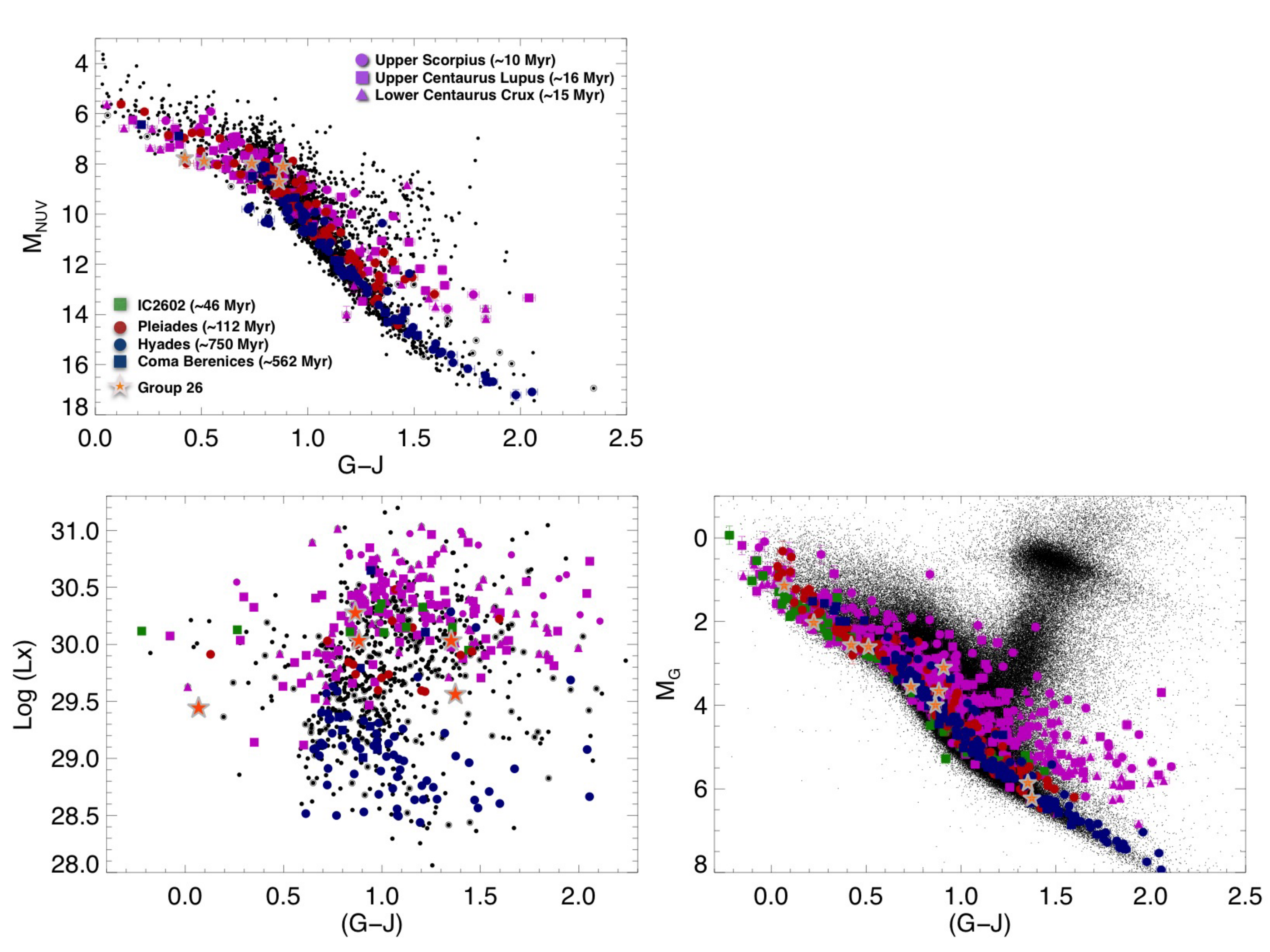}
\end{center}
\caption{See caption for ~\ref{fig:Group10-CMD}.  Group 26 from Oh17 is highlighted. No $FUV$ detections were found for objects therefore this panel is blank in the figure.\label{fig:Group26-CMD}} 
\end{figure*}

\begin{figure*}[!h]
\begin{center}
\epsscale{1.2}
\plotone{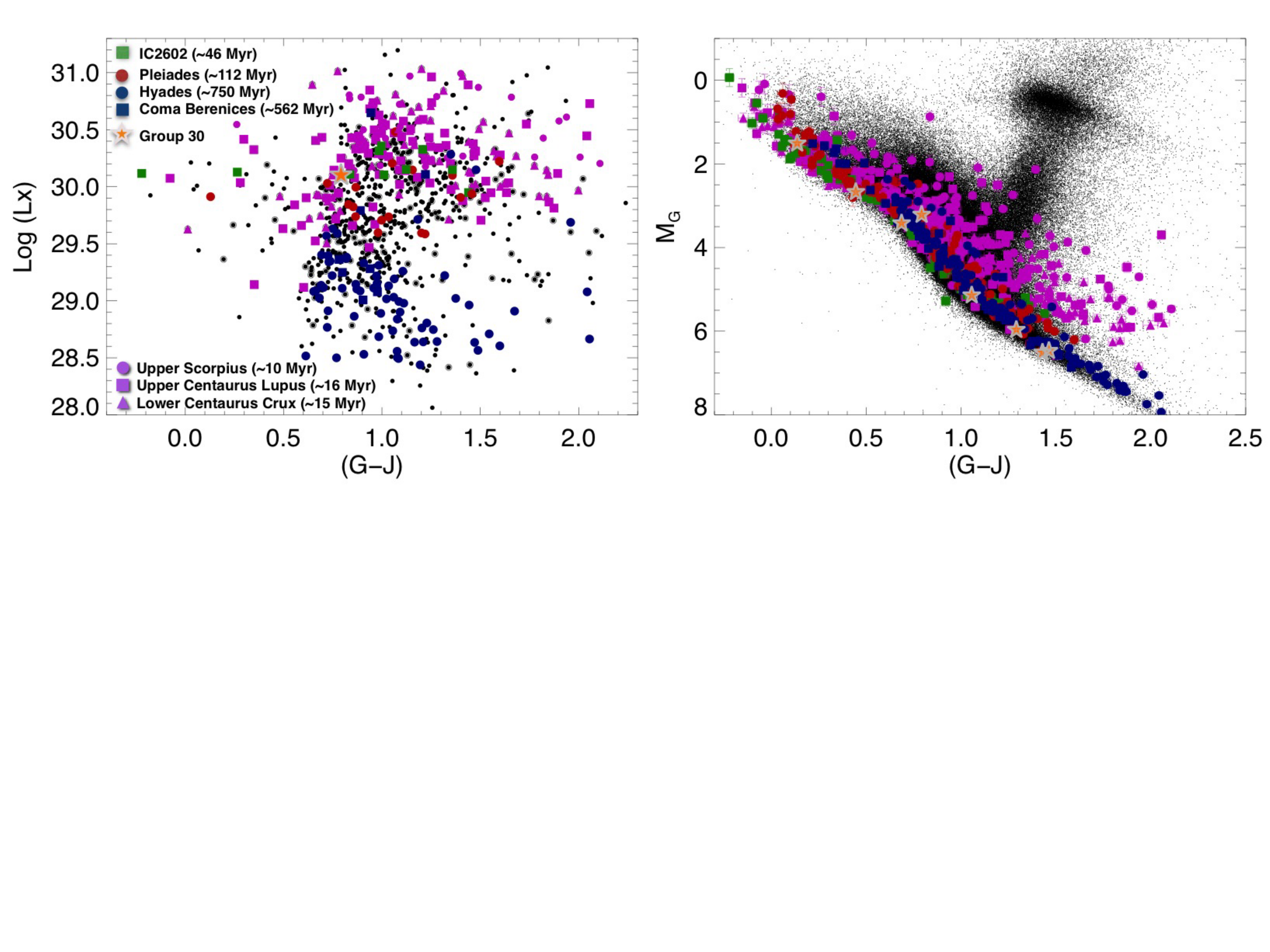}
\end{center}
\caption{See caption for ~\ref{fig:Group10-CMD}.  Group 30 from Oh17 is highlighted. No $FUV$ or $NUV$ detections were found for objects therefore those panels are blank in the figure.\label{fig:Group30-CMD}} 
\end{figure*}

To ascertain whether these new groups follow a logical temperature sequence, and how that sequence is related to known groups with ages, we examined a series of color magnitude diagrams.  As stated in Section ~\ref{sec:data}, we cross-matched all unique stars in the Oh17 sample with 2MASS, WISE, GALEX, and ROSAT. We used ($G-J$) as our color proxy for spectral type/effective temperature as this appeared to have a clean relationship after examining an array of photometric combinations.  The bottom right panels of Figures~\ref{fig:Group10-CMD},~\ref{fig:Group14-CMD},~\ref{fig:Group23-CMD}, and ~\ref{fig:Group26-CMD} show ($G-J$) versus $M_{G}$ for all of TGAS using a signal to noise cut off of 10.  Overplotted are the individual stars in each of the 4 uknown groups (respective to their own Figure) as well as BANYAN groups that were identified in Oh17 color coded to reflect different ages.  We chose to group objects into age bins of $\sim$ 15 Myr with Upper Scorpius, Upper Centaurus Lupus and Lower Centaurus Crux; $\sim$ 50 Myr with IC2602; $\sim$ 100 Myr with the Pleiades; and $\sim$ 500 - 800 Myr with the Hyades and Coma Berenices.  Moving from blue to red ($G-J$) color on the sequence we see that the younger groups shift redder and brighter than would be expected from field stars.   Each new group forms a fairly tight sequence across the range of ($G-J$) colors.  No group looks as young as the $\sim$15 Myr sequences.  While the groups show some scatter, they are all consistent with Pleiades age ($\sim$ 100 Myr) or older associations.  

Using X-ray and UV magnitudes we examined how the different groups (known and new) as well as the full sample measured on age-activity relations (e.g. \citealt{Nunez16}; \citealt{Preibisch05}; \citealt{Shkolnik11}; \citealt{Rodriguez13}).  We looked at the color magnitude diagrams of ($G-J$) versus $M_{NUV}$ and $M_{FUV}$ as well as ($G-J$)  versus X-ray luminosity.  The top panels of Figures~\ref{fig:Group10-CMD},~\ref{fig:Group14-CMD},~\ref{fig:Group23-CMD}, and ~\ref{fig:Group26-CMD} show $NUV$ and $FUV$ color magnitude diagrams for each group separately while the bottom left panel of each Figure shows the X-ray comparison.  Each group had several stars with $NUV$ detections.  Groups 23 and 26 had no $FUV$ detections and Group 14 had 1.    In X-ray,  Groups 14, and 23 had no detections in ROSAT.  Similar to the ($G-J$) versus $M_{G}$ color magnitude diagram comparisons, we find that the new groups follow logical sequences on the ultraviolet diagrams.  The $FUV$ and $NUV$ magnitude sequences are consistent with ages that are older than the Pleiades for each group.  The X-ray activity indicates that Group 10 is similar to Hyades members at ages of $\sim$ 750 Myr.  While Group 26 has stars that are more active and could be considered much younger with this diagnostic parameter.

\section{Age, Mass, and Metallicity From Isochronal Fitting\label{sec:Isochrones}}
\begin{figure*}[h]
\begin{center}
\epsscale{1.2}
\plotone{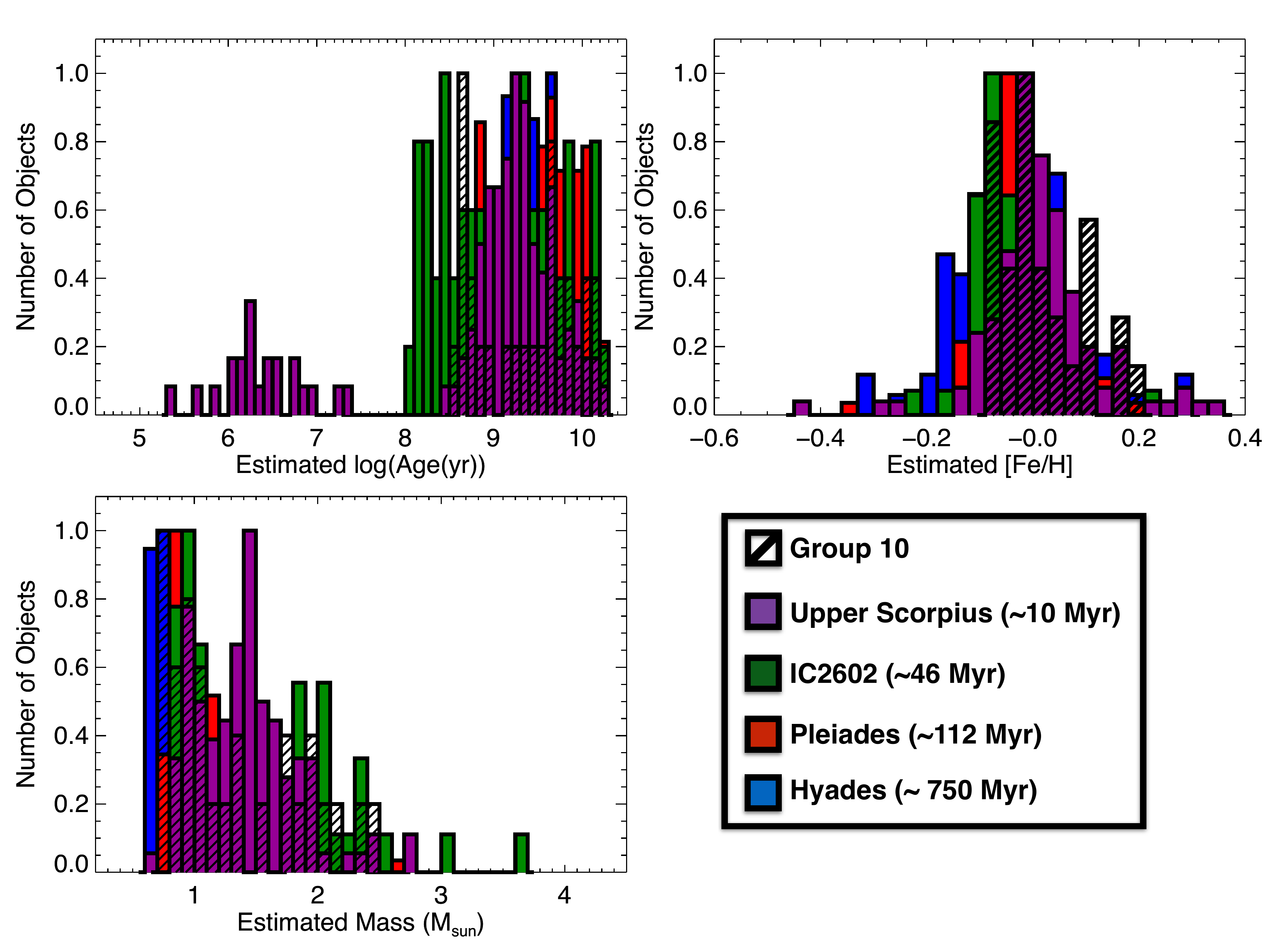}
\end{center}
\caption{The output of log (Age), [Fe/H], and Mass parameters from MIST isochrone fitting as described in \citet{Bochanski18aa}.  For context, we show both the group of interest (Group 10) as well as known associations at specific age bins:  Upper Scorpius ($\sim$10 Myr; purple), IC2602 ($\sim$46 Myr; green), Pleiades($\sim$112 Myr; red), Hyades ($\sim$750 Myr; Blue) \label{fig:Group10-Isochrone}} 
\end{figure*}

\begin{figure*}[ht]
\begin{center}
\epsscale{1.2}
\plotone{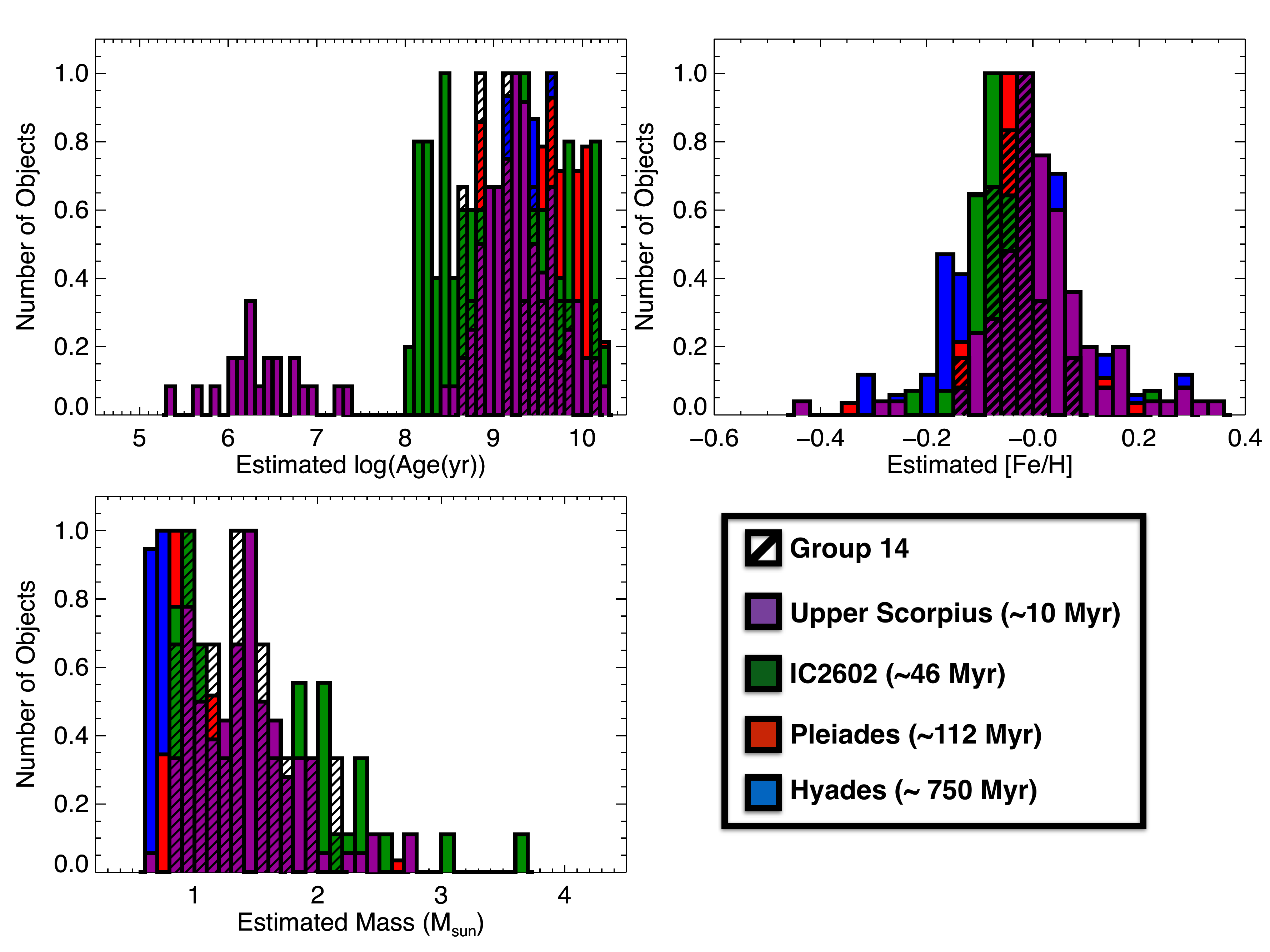}
\end{center}
\caption{See caption for Figure ~\ref{fig:Group10-Isochrone}.  Group 14 from Oh17 is highlighted. \label{fig:Group14-Isochrone}} 
\end{figure*}

\begin{figure*}[!ht]
\begin{center}
\epsscale{1.2}
\plotone{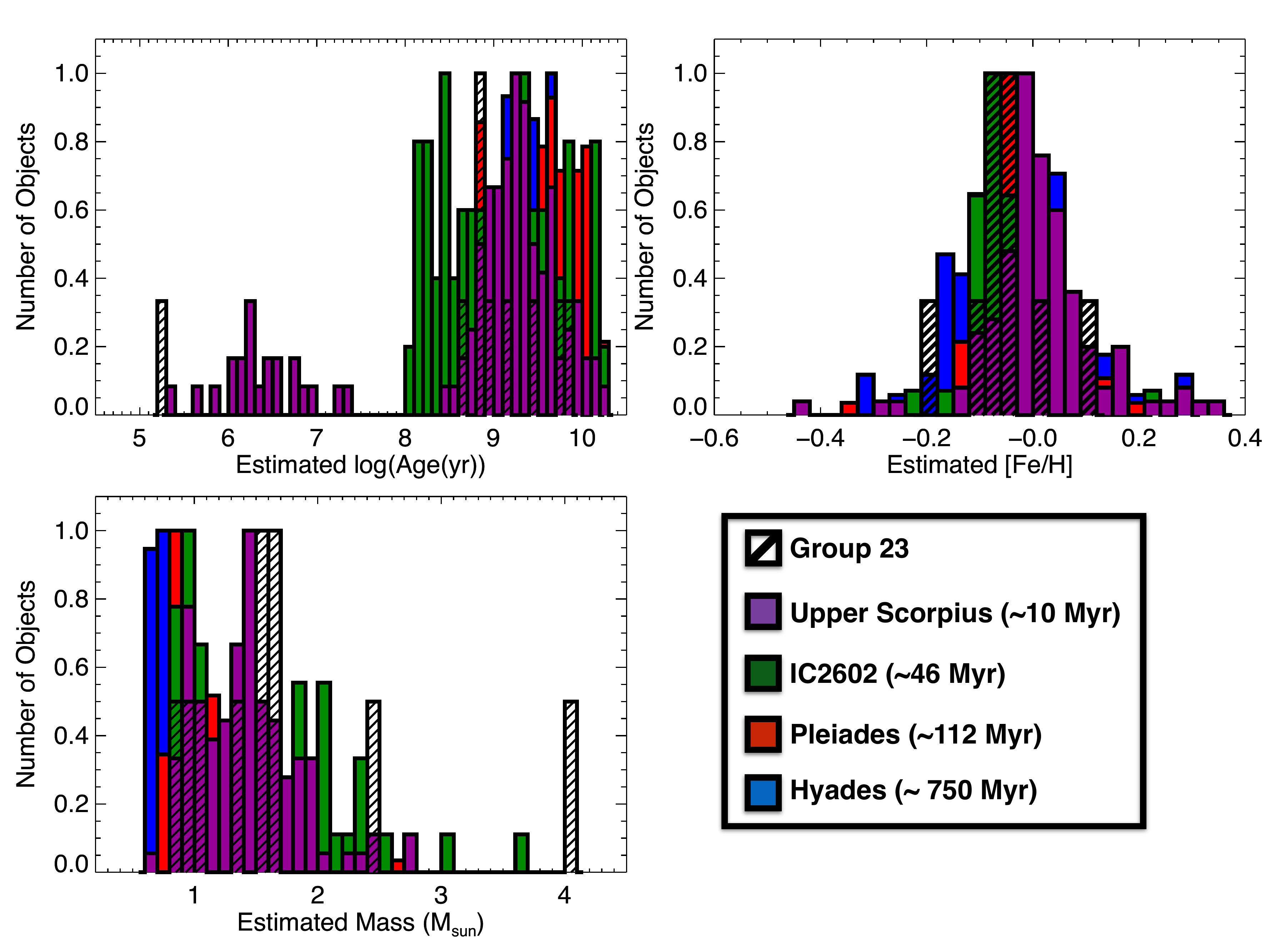}
\end{center}
\caption{See caption for Figure ~\ref{fig:Group10-Isochrone}.  Group 23 from Oh17 is highlighted.\label{fig:Group23-Isochrone}} 
\end{figure*}

\begin{figure*}[!ht]
\begin{center}
\epsscale{1.2}
\plotone{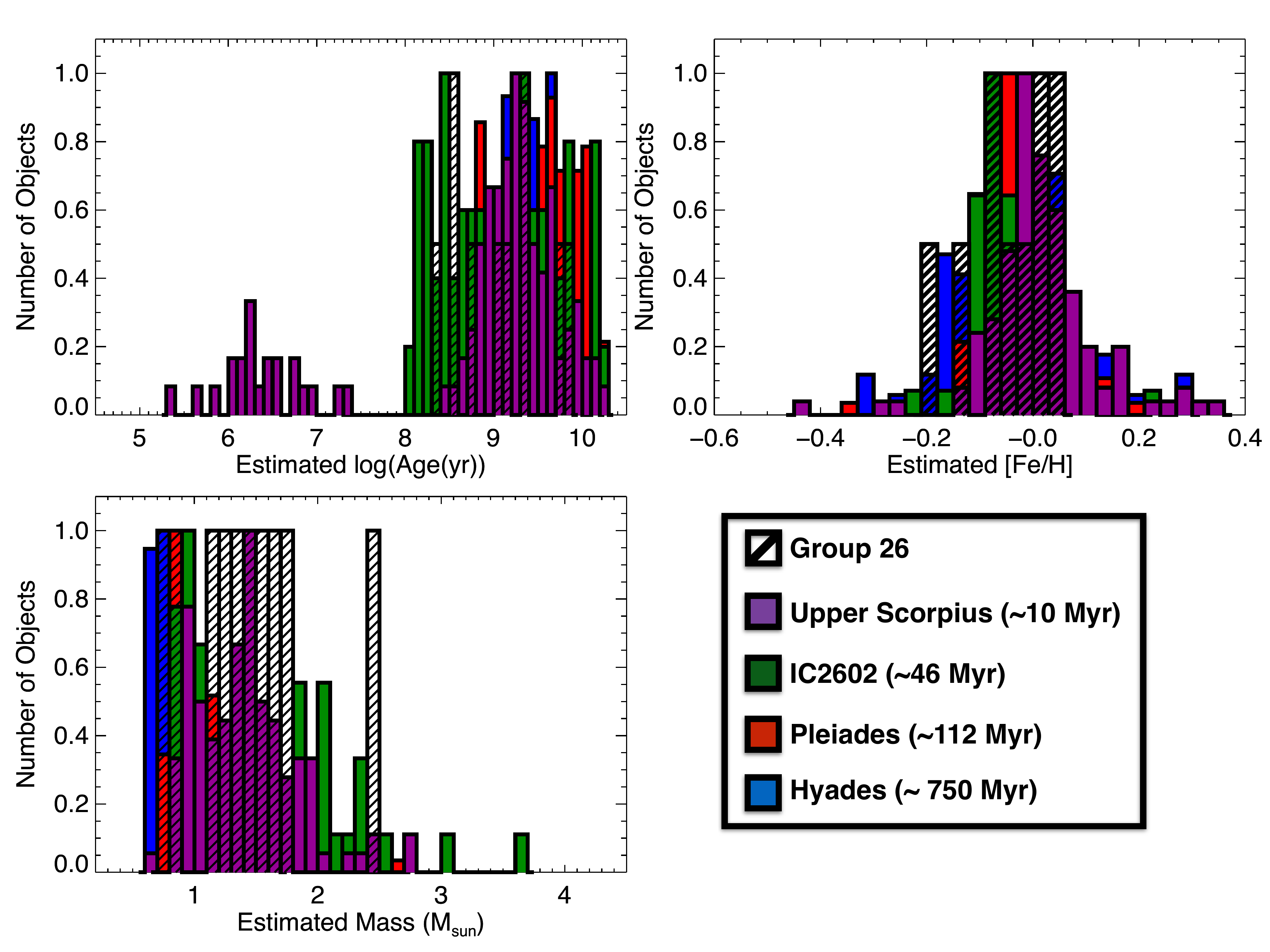}
\end{center}
\caption{See caption for Figure ~\ref{fig:Group10-Isochrone}.  Group 26 from Oh17 is highlighted. \label{fig:Group26-Isochrone}} 
\end{figure*}

We also turned to isochrone fitting to investigate the parameters of each new group.  Using the methodology of \citet{Bochanski18aa} where the Oh17 sample was supplemented with 2MASS and WISE photometry and then tested with posterior probabilities calculated using the trilinear interpolation
schemes within $isochrones$ and assumed priors described in \citet{Morton2015}, we look at the Mesa Isochrones and Stellar Track library (MIST; \citealt{Dotter2016}, \citealt{Choi2016}, \citealt{Paxton2011,Paxton2013,Paxton2015}) predictions for each star. We investigate both the age as well as the mass and [Fe/H] parameters.   Figures~\ref{fig:Group10-Isochrone}, ~\ref{fig:Group14-Isochrone},~\ref{fig:Group23-Isochrone}, and ~\ref{fig:Group26-Isochrone} show the results for each group.  For context, we overplotted one group in each of our age bins with MIST isochrone parameters from \citet{Bochanski18aa}:  Upper Scorpius ($\sim$ 10 Myr; purple), Pleiades  ($\sim$ 112 Myr; red), IC2602  ($\sim$ 50 Myr; green), and the Hyades  ($\sim$ 750 Myr; blue).  The histogram plots were normalized to 1 for ease of comparison and they were binned by 0.1 in Mass and Age and 0.03 in [Fe/H].  As stated in \citet{Bochanski18aa}, the age predictions from MIST isochrone fitting are scattered and can be significantly different than observable indications like Li depletion, gyrochronology, or UV/X-ray activity levels.  As such, we note the isochrone age predictions with skepticism and are more interested in the Mass and [Fe/H] distribution of the groups.

\subsection{Group 10\label{sec:isochrone10}}
Group 10 shows a slight bifurcation in age predictions from isochrones with half of the stars showing an indication of  ages younger than 1 Gyr and the other half falling older.   The metallicity distribution tends to near solar if not slightly metal rich compared to the known group predictions.  The stars identified in Group 10, the closest of the new groups examined, tend toward solar mass with a large number of higher mass objects.  There are several A stars in Group 10 spectroscopically identified so this is consistent with literature work on the members.

\subsection{Group 14, 26\label{sec:isochrone14}}
Groups 14 and 26 show isochrone age predictions for the stellar members that fall within the known associations plotted.  The metallicity predictions for stars in each group all fall slightly subsolar but are within the predictions of known groups.  The vast majority of stars identified in Groups 14 and 26 fall between 1 -2 solar masses, consistent with what we see on the color magnitude diagram sequences.

\subsection{Group 23\label{sec:isochrone23}}
Group 23 has 10 members, one of which falls clearly in the giant star area of the ($G-J$) versus $M_{G}$ color magnitude diagram.  That object, HD86703, also skews the age, mass, and metallicity plots by showing up as significantly younger, slightly metal rich, and high mass.  Otherwise, Group 23 is similar to Groups 14, 16, and 26 in its parameters.  Indeed HD86703 is the bright outlier on the $NUV$ color magnitude diagram of Figure~\ref{fig:Group23-CMD} and is classified as a G5/G6 giant star in the literature. 

\section{Conclusions\label{sec:conclusion}}
The Oh17 co-moving catalog is rich with discoveries for the local solar neighborhood and the nearby galactic substructure.  There were 10,606 individual stars in the Oh17 catalog split into 4,555 groups.  Those were further broken down into 27 groups with 10 or more connected components, 35 groups with 5 -9, 39 groups with 4,  218 that have three, and the remaining which have 2.  

The original Oh17 paper produced a rich and very useful catalog however it lacked a detailed literature search as to whether the groups were new, known, or parts of a whole. Given that future Gaia data releases will certainly uncover a wealth of previously unrecognized associations in the Galaxy, we looked to re-organize the sample of 4,555 groups into known or unknown collections of stars.  The BANYAN $\Sigma$ tool is one method for quickly parsing through the collection of pairs and hierarchical associations.  Using BANYAN $\Sigma$ we find that 1,015 individual stars in the Oh17 catalog have an 80$\%$ or larger probability of membership in one of the 27 groups analyzed.  Using those objects as a seed for interpretation of the overall Oh17 catalog,  we find that 24 of the 27 groups were uncovered in part (none in their entirety), and there are 400 new candidate members with Gaia astrometry across 20 different groups.  In fact a significant portion were uncovered as bonafide members in the literature and the Oh17 catalog found a co-moving companion or multiple connected companions with $>$80$\%$ probability in a group.   

We uncovered that a significant number of the large ($>$ 10 connected component) groups in the Oh17 sample were broken up parts of big, known associations like Upper Centaurus Lupus, Upper Centaurus Crux or Upper Scorpius.  We found no correlation with the individual Oh17 group and position on color magnitude diagrams therefore we do not identify traces of substructure in those known associations, however recent work suggests that further investigation is warranted (see \citealt{Roser17} and the discovery that Oh17 Group 11 was a compact new moving group around V1062 Sco).  

We investigated the Oh17 groups with $>$ 10 connected components in detail and found that  4 of those hierarchical associations are newly discovered co-moving collections of stars in the Milky Way.  Among those were Oh17 Groups 10, 14, 23, and 26 containing 29, 20, 10, and 10 connected components each.  Group 10 was the closest with a range of 105 - 135 pc making it a new candidate for searches of a co-evolving association with  directly imaged exoplanets and brown dwarfs.  Each group appears to be older than the Pleiades but with indications that they are younger than $\sim$ 1 Gyr.  Using the kinematics of these objects in updates of strong kinematic analysis tools like BANYAN $\Sigma$, one can uncover more members. Moreover, while we did not do a detailed search, we found that 19 of the 35 Oh17 groups with 5 - 9 members also appeared to be new co-moving associations in the Galaxy and warrant follow-up. Oh17 group 30 was particulary exciting given that it would be well within 100pc (range of 77 - 90 pc), has at least 1 X-ray active F5 star, and is in a compact area of young stars near the Sun.  Given that the Gaia DR2 release will occur in April of 2018, these new groups are likely just the tip of a large iceberg of discovery when it comes to the substructure of the Milky Way.  

\acknowledgments
We thank Dustin Lang for providing the $Gaia$-Tycho-2MASS cross-match data, and Seomyong Oh for useful conversations about the original catalog.  We also thank amateur astronomer Bruno Alessi for pointing out a recently discovered cluster.  This research has made use of: the Washington Double Star Catalog maintained at the U.S. Naval Observatory; the SIMBAD database and VizieR catalog access tool, operated at the Centre de Donnees astronomiques de Strasbourg, France (\citealt{Ochsenbein00}); data products from the Two Micron All Sky Survey (2MASS; \citealt{Skrutskie06}), which is a joint project of the University of Massachusetts and the Infrared Processing and Analysis Center (IPAC)/California Institute of Technology (Caltech), funded by the National Aeronautics and Space Administration (NASA) and the National Science Foundation; data products from the Wide-field Infrared Survey Explorer (WISE; and \citealt{Wright10}), which is a joint project of the University of California, Los Angeles, and the Jet Propulsion Laboratory (JPL)/Caltech, funded by NASA.  This project was developed in part at the 2017 Heidelberg Gaia Sprint, hosted by the Max-Planck-Institut fur Astronomie, Heidelberg. This work has made use of data from the European Space Agency(ESA) mission Gaia, processed by the Gaia Data Processing and Analysis Consortium. Funding for the DPAC has been provided by national institutions, in particular the institutions participating in the Gaia Multilateral Agreement. With deepest appreciation, we acknowledge Kathryn W. Davis for her generous founding support of the Master of Arts in Science Teaching (MAT) Program. Leadership support for the MAT program is provided by The Shelby Cullom Davis Charitable Fund.  The MAT program is supported in part by the National Science Foundation under Grant Number DUE-1340006 and the U.S. Department of Education under Grant Number U336S140026.

\software {BANYAN \citep{Gagne18}, isochrones \citep{Morton2015}, TOPCAT \citep{Taylor05}}

\newpage

\startlongtable
\tabletypesize{\scriptsize}
\clearpage
\begin{deluxetable*}{lllllccc}
\tablewidth{0pt}
\tablecolumns{13}
\tablecaption{Oh17 with spectral type and $T_{\rm eff}$\label{Tab:SpT}}
\tablewidth{0pt}
\tablehead{
\colhead{Name} &
\colhead{RA}&
\colhead{DEC}&
\colhead{SpT}&
\colhead{$T_{\rm eff}$}&
\colhead{BANYAN Grp}&
\colhead{BANYAN Prob\tablenotemark{a} (\%)}&
\colhead{Oh Grp}\\
&
&
&
&
\colhead{(K)}
&
&
\colhead{(\%)}\\
\colhead{(1)}&
\colhead{(2)}&
\colhead{(3)}&
\colhead{(4)}&
\colhead{(5)}&
\colhead{(6)}&
\colhead{(7)}&
\colhead{(8)}
}
\startdata
TYC 1253-388-1	&	59.45728	&	18.56219	&	$\cdots$	&	$\cdots$	&	FIELD	&	7.3	&	0	\\
TYC 1804-1924-1	&	57.07039	&	25.21493	&	F4V	&	6890	&	PLE	&	98.7	&	0	\\
HIP 18091	&	58.00344	&	19.59669	&	$\cdots$	&	$\cdots$	&	FIELD	&	36.6	&	0	\\
HIP 18544	&	59.50715	&	20.6766	&	F8	&	6200	&	PLE	&	56.9	&	0	\\
TYC 1261-1630-1	&	58.37032	&	20.90718	&	$\cdots$	&	$\cdots$	&	PLE	&	98.7	&	0	\\
TYC 1261-1415-1	&	58.88317	&	21.07928	&	$\cdots$	&	$\cdots$	&	PLE	&	79.1	&	0	\\
TYC 1261-24-1	&	59.59016	&	21.25748	&	$\cdots$	&	$\cdots$	&	PLE	&	95.2	&	0	\\
HIP 18266	&	58.61605	&	21.38955	&	$\cdots$	&	$\cdots$	&	PLE	&	98.6	&	0	\\
HIP 19367	&	62.23086	&	20.38579	&	F8	&	6200	&	FIELD	&	0.3	&	0	\\
HIP 18955	&	60.93413	&	22.9441	&	F5	&	6440	&	PLE	&	75.2	&	0	\\
HIP 16423	&	52.8682	&	21.82172	&	F2	&	6890	&	PLE	&	85.6	&	0	\\
HIP 15341	&	49.45743	&	22.832	&	A3	&	8720	&	FIELD	&	0.2	&	0	\\
TYC 1256-516-1	&	56.25702	&	19.55919	&	F5	&	6440	&	PLE	&	65.8	&	0	\\
HIP 17325	&	55.62454	&	20.1498	&	A2	&	8970	&	PLE	&	94.3	&	0	\\
HIP 17607	&	56.58068	&	20.8796	&	$\cdots$	&	$\cdots$	&	PLE	&	99.1	&	0	\\
TYC 1260-671-1	&	57.16395	&	21.92476	&	F0	&	7200	&	PLE	&	96.2	&	0	\\
HIP 17921	&	57.47955	&	22.24397	&	B8III	&	12400	&	PLE	&	99.8	&	0	\\
HIP 17892	&	57.40917	&	22.5333	&	B9	&	10500	&	PLE	&	99.8	&	0	\\
TYC 1800-118-1	&	57.297	&	22.60927	&	A0	&	9520	&	PLE	&	99.8	&	0	\\
TYC 1800-669-1	&	57.58886	&	23.09616	&	$\cdots$	&	$\cdots$	&	PLE	&	99.9	&	0	\\
HIP 17043	&	54.8051	&	21.84306	&	A0	&	9520	&	PLE	&	99.3	&	0	\\
HIP 17316	&	55.60007	&	21.47329	&	G0	&	6030	&	PLE	&	99.1	&	0	\\
TYC 1247-515-1	&	55.8798	&	22.15819	&	F8	&	6200	&	PLE	&	99.8	&	0	\\
HIP 17317	&	55.60019	&	22.42095	&	$\cdots$	&	$\cdots$	&	PLE	&	99.9	&	0	\\
HIP 17511	&	56.2456	&	22.03227	&	F5	&	6440	&	PLE	&	99.7	&	0	\\
TYC 1260-498-1	&	56.17434	&	22.46436	&	$\cdots$	&	$\cdots$	&	PLE	&	99.4	&	0	\\
TYC 1799-1102-1	&	56.41633	&	22.69428	&	A0	&	9520	&	PLE	&	99.9	&	0	\\
TYC 1800-1574-1	&	56.66006	&	22.9196	&	G0	&	6030	&	PLE	&	99.9	&	0	\\
HIP 17497	&	56.21357	&	23.26874	&	F3V	&	6890	&	PLE	&	99.9	&	0	\\
TYC 1800-1774-1	&	56.69618	&	22.91439	&	F8	&	6200	&	PLE	&	99.9	&	0	\\
TYC 1800-2170-1	&	56.95058	&	23.2179	&	$\cdots$	&	$\cdots$	&	PLE	&	99.9	&	0	\\
TYC 1800-471-1	&	57.48549	&	23.21844	&	F8	&	6200	&	PLE	&	99.7	&	0	\\
TYC 1800-496-1	&	57.41426	&	23.28992	&	$\cdots$	&	$\cdots$	&	PLE	&	99.9	&	0	\\
TYC 1800-628-1	&	57.4206	&	23.34138	&	A7V	&	7850	&	PLE	&	99.9	&	0	\\
TYC 1800-727-1	&	57.38645	&	23.38022	&	F3V	&	6890	&	PLE	&	99.9	&	0	\\
TYC 1800-2129-1	&	57.18299	&	23.25963	&	A8V	&	7850	&	PLE	&	100	&	0	\\
TYC 1800-1672-1	&	56.66674	&	23.11014	&	F5V	&	6440	&	PLE	&	99.9	&	0	\\
HIP 17572	&	56.45349	&	23.14696	&	A0	&	9520	&	PLE	&	99.9	&	0	\\
TYC 1800-1917-1	&	56.54194	&	23.3398	&	$\cdots$	&	$\cdots$	&	PLE	&	99.2	&	0	\\
TYC 1800-2027-1	&	56.756	&	23.49475	&	$\cdots$	&	$\cdots$	&	PLE	&	99.9	&	0	\\
TYC 1800-1616-1	&	56.72404	&	23.58337	&	$\cdots$	&	$\cdots$	&	PLE	&	99.9	&	0	\\
TYC 1800-1630-1	&	56.86188	&	23.67814	&	A3V	&	8720	&	PLE	&	99.8	&	0	\\
HIP 17692	&	56.83746	&	23.80315	&	A1V	&	9230	&	PLE	&	99.9	&	0	\\
\enddata 
\tablenotetext{a}{The BANYAN probability is applicable for the group noted.}
\tablecomments{This subtable is a preview of the entire sample, which will be available as a machine readable table.  Above we show the Tycho or Hipparcos name along with corresponding position, and the SpT and $T_{eff}$ compiled in \citet{Skiff14}.  We also show the BANYAN  $\Sigma$ probability of membership and the corresponding group as well as the Oh17 group number.}
\end{deluxetable*}

\startlongtable
\tabletypesize{\scriptsize}
\clearpage
\begin{deluxetable*}{lllllllllll}
\tablewidth{0pt}
\tablecolumns{13}
\tablecaption{Oh17 with photometry\label{Tab:Phot}}
\tablewidth{0pt}
\tablehead{
\colhead{Name} &
\colhead{SpT}&
\colhead{$B$}&
\colhead{$V$}&
\colhead{$J$}&
\colhead{$H$}&
\colhead{$K$}&
\colhead{$W1$}&
\colhead{$W2$}&
\colhead{$W3$}&
\colhead{$W4$}\\
\colhead{(1)}&
\colhead{(2)}&
\colhead{(3)}&
\colhead{(4)}&
\colhead{(5)}&
\colhead{(6)}&
\colhead{(7)}&
\colhead{(8)}&
\colhead{(9)}&
\colhead{(10)}&
\colhead{(11)}\\
}
\startdata
TYC 1253-388-1 	&	$\cdots$	&	12.45	&	11.32	&	      9.98$\pm$      0.03	&	      9.53$\pm$      0.02	&	      9.45$\pm$      0.02	&	      9.38$\pm$      0.02	&	      9.41$\pm$      0.02	&	      9.27$\pm$      0.05	&	      8.63$\pm$      $\cdots$\\	
TYC 1804-1924-1	&	F4V	&	9.62	&	9.22	&	      8.14$\pm$      0.02	&	      7.91$\pm$      0.02	&	      7.86$\pm$      0.02	&	      7.83$\pm$      0.03	&	      7.86$\pm$      0.02	&	      7.82$\pm$      0.02	&	      7.33$\pm$      0.12\\	
HIP 18091      	&	$\cdots$	&	11.13	&	10.45	&	      9.27$\pm$      0.02	&	      8.98$\pm$      0.02	&	      8.87$\pm$      0.02	&	      8.86$\pm$      0.02	&	      8.88$\pm$      0.02	&	      8.82$\pm$      0.03	&	      8.23$\pm$      0.33\\	
HIP 18544      	&	F8	&	9.88	&	9.38	&	      8.44$\pm$      0.02	&	      8.25$\pm$      0.02	&	      8.20$\pm$      0.03	&	      8.13$\pm$      0.02	&	      8.15$\pm$      0.02	&	      8.14$\pm$      0.02	&	      7.79$\pm$      0.20\\	
TYC 1261-1630-1	&	$\cdots$	&	12.86	&	11.68	&	     10.10$\pm$      0.02	&	      9.64$\pm$      0.02	&	      9.51$\pm$      0.02	&	      9.48$\pm$      0.02	&	      9.49$\pm$      0.02	&	      9.39$\pm$      0.04	&	      8.79$\pm$      $\cdots$\\	
TYC 1261-1415-1	&	$\cdots$	&	11.79	&	11.11	&	      9.62$\pm$      0.02	&	      9.28$\pm$      0.02	&	      9.17$\pm$      0.02	&	      9.14$\pm$      0.02	&	      9.19$\pm$      0.02	&	      9.16$\pm$      0.03	&	      8.27$\pm$      $\cdots$\\	
TYC 1261-24-1  	&	$\cdots$	&	12.35	&	11.42	&	      9.94$\pm$      0.03	&	      9.58$\pm$      0.03	&	      9.45$\pm$      0.02	&	      9.42$\pm$      0.02	&	      9.45$\pm$      0.02	&	      9.35$\pm$      0.04	&	      8.75$\pm$      $\cdots$\\	
HIP 18266      	&	$\cdots$	&	11.65	&	10.85	&	      9.60$\pm$      0.02	&	      9.26$\pm$      0.02	&	      9.18$\pm$      0.02	&	      9.16$\pm$      0.02	&	      9.20$\pm$      0.02	&	      9.16$\pm$      0.03	&	      8.26$\pm$      0.32\\	
HIP 19367      	&	F8	&	9.83	&	9.36	&	      8.38$\pm$      0.02	&	      8.19$\pm$      0.02	&	      8.11$\pm$      0.03	&	      8.11$\pm$      0.02	&	      8.13$\pm$      0.02	&	      8.10$\pm$      0.02	&	      8.16$\pm$      0.28\\	
HIP 18955      	&	F5	&	10.3	&	9.68	&	      8.52$\pm$      0.04	&	      8.26$\pm$      0.02	&	      8.18$\pm$      0.02	&	      8.16$\pm$      0.02	&	      8.17$\pm$      0.02	&	      8.17$\pm$      0.02	&	      8.13$\pm$      0.28\\	
HIP 16423      	&	F2	&	9.24	&	8.84	&	      8.07$\pm$      0.02	&	      7.89$\pm$      0.02	&	      7.82$\pm$      0.02	&	      7.81$\pm$      0.03	&	      7.84$\pm$      0.02	&	      7.79$\pm$      0.02	&	      7.71$\pm$      0.18\\	
HIP 15341      	&	A3	&	7.91	&	7.63	&	      7.07$\pm$      0.02	&	      6.98$\pm$      0.03	&	      6.89$\pm$      0.01	&	      6.90$\pm$      0.06	&	      6.89$\pm$      0.02	&	      6.92$\pm$      0.02	&	      6.78$\pm$      0.09\\	
TYC 1256-516-1 	&	F5	&	9.87	&	9.39	&	      8.40$\pm$      0.02	&	      8.23$\pm$      0.03	&	      8.15$\pm$      0.02	&	      8.10$\pm$      0.02	&	      8.11$\pm$      0.02	&	      8.09$\pm$      0.02	&	      7.85$\pm$      0.23\\	
HIP 17325      	&	A2	&	8.72	&	8.4	&	      7.75$\pm$      0.02	&	      7.67$\pm$      0.02	&	      7.65$\pm$      0.02	&	      7.74$\pm$      0.22	&	      7.61$\pm$      0.02	&	      7.62$\pm$      0.02	&	      7.81$\pm$      0.19\\	
HIP 17607      	&	$\cdots$	&	12.84	&	11.64	&	      9.99$\pm$      0.02	&	      9.61$\pm$      0.02	&	      9.52$\pm$      0.02	&	      9.47$\pm$      0.02	&	      9.50$\pm$      0.02	&	      9.46$\pm$      0.05	&	      8.56$\pm$      $\cdots$\\	
TYC 1260-671-1 	&	F0	&	8.79	&	8.41	&	      7.56$\pm$      0.03	&	      7.40$\pm$      0.02	&	      7.32$\pm$      0.02	&	      7.26$\pm$      0.04	&	      7.30$\pm$      0.02	&	      7.31$\pm$      0.02	&	      7.25$\pm$      0.12\\	
HIP 17921      	&	B8III	&	6.05	&	6.07	&	      5.97$\pm$      0.02	&	      6.05$\pm$      0.06	&	      5.98$\pm$      0.02	&	      6.06$\pm$      0.09	&	      5.99$\pm$      0.04	&	      6.11$\pm$      0.01	&	      5.70$\pm$      0.04\\	
HIP 17892      	&	B9	&	7.04	&	7	&	      6.85$\pm$      0.02	&	      6.92$\pm$      0.02	&	      6.88$\pm$      0.02	&	      6.89$\pm$      0.06	&	      6.92$\pm$      0.02	&	      6.95$\pm$      0.02	&	      6.80$\pm$      0.08\\	
TYC 1800-118-1 	&	A0	&	7.9	&	7.73	&	      7.31$\pm$      0.03	&	      7.27$\pm$      0.03	&	      7.24$\pm$      0.02	&	      7.21$\pm$      0.05	&	      7.24$\pm$      0.02	&	      7.29$\pm$      0.02	&	      6.92$\pm$      0.10\\	
TYC 1800-669-1 	&	$\cdots$	&	12.24	&	11.24	&	      9.86$\pm$      0.02	&	      9.51$\pm$      0.03	&	      9.39$\pm$      0.02	&	      9.35$\pm$      0.02	&	      9.39$\pm$      0.02	&	      9.31$\pm$      0.04	&	      8.25$\pm$      $\cdots$\\	
\enddata 
\tablecomments{This subtable is a preview of the entire sample, which will be available as a machine readable table.  Above we show the Tycho or Hipparcos name along with corresponding photometry from Tycho, 2MASS, and WISE as well as the SpT compiled from \citet{Skiff14}. Upper limits on photometry are shown as $\pm$ ---. }
\end{deluxetable*}

\tabletypesize{\scriptsize}
\clearpage
\begin{deluxetable*}{lcllllllllll}
\tablewidth{0pt}
\tablecolumns{13}
\tablecaption{Oh17 with photometry (cont)\label{Tab:Activity}}
\tablewidth{0pt}
\tablehead{
\colhead{Name} &
\colhead{SpT}&
\colhead{$NUV$}&
\colhead{$FUV$}&
\colhead{Name}&
\colhead{log($L_{x}$)}&
\colhead{Name}&
\colhead{log($L_{x}$)}&
\colhead{Grp}&
\colhead{BANYAN}\\
\colhead{(TGAS)}&
&
\colhead{(GALEX)}&
\colhead{(GALEX)}&
\colhead{(ROSAT Bright)}&
\colhead{(ROSAT)}&
\colhead{(ROSAT Faint)}&
\colhead{(ROSAT)}&
&
&
\\
\colhead{(1)}&
\colhead{(2)}&
\colhead{(3)}&
\colhead{(4)}&
\colhead{(5)}&
\colhead{(6)}&
\colhead{(7)}&
\colhead{(8)}&
\colhead{(9)}&
\colhead{(10)}\\
}
\startdata
TYC 1253-388-1	&	$\cdots$	&	17.86$\pm$0.04	&	$\cdots$	&	$\cdots$	&	$\cdots$	&	$\cdots$	&	$\cdots$	&	0	&	FIELD	~	7.3	\\
TYC 1804-1924-1	&	F4V	&	13.48$\pm$0.01	&	$\cdots$	&	$\cdots$	&	$\cdots$	&	$\cdots$	&	$\cdots$	&	0	&	PLE	~	98.7	\\
HIP 18091	&	$\cdots$	&	15.48$\pm$0.01	&	$\cdots$	&	$\cdots$	&	$\cdots$	&	$\cdots$	&	$\cdots$	&	0	&	FIELD	~	36.6	\\
HIP 18544	&	F8	&	13.69$\pm$0.01	&	18.75$\pm$0.10	&	$\cdots$	&	$\cdots$	&	$\cdots$	&	$\cdots$	&	0	&	PLE	~	56.9	\\
TYC 1261-1630-1	&	$\cdots$	&	18.22$\pm$0.03	&	21.79$\pm$0.48	&	$\cdots$	&	$\cdots$	&	035328.8+20544	&	29.94$\pm$0.38	&	0	&	PLE	~	98.7	\\
TYC 1261-1415-1	&	$\cdots$	&	17.18$\pm$0.03	&	$\cdots$	&	$\cdots$	&	$\cdots$	&	035531.7+21044	&	29.59$\pm$0.50	&	0	&	PLE	~	79.1	\\
TYC 1261-24-1	&	$\cdots$	&	17.70$\pm$0.03	&	$\cdots$	&	$\cdots$	&	$\cdots$	&	$\cdots$	&	$\cdots$	&	0	&	PLE	~	95.2	\\
HIP 18266	&	$\cdots$	&	16.72$\pm$0.03	&	$\cdots$	&	$\cdots$	&	$\cdots$	&	$\cdots$	&	$\cdots$	&	0	&	PLE	~	98.6	\\
HIP 19367	&	F8	&	14.01$\pm$0.00	&	18.87$\pm$0.07	&	$\cdots$	&	$\cdots$	&	040857.0+20232	&	29.82$\pm$0.40	&	0	&	FIELD	~	0.3	\\
HIP 18955	&	F5	&	14.32$\pm$0.01	&	20.02$\pm$0.17	&	$\cdots$	&	$\cdots$	&	$\cdots$	&	$\cdots$	&	0	&	PLE	~	75.2	\\
HIP 16423	&	F2	&	$\cdots$	&	17.64$\pm$0.08	&	$\cdots$	&	$\cdots$	&	$\cdots$	&	$\cdots$	&	0	&	PLE	~	85.6	\\
HIP 15341	&	A3	&	12.32$\pm$0.00	&	14.51$\pm$0.01	&	$\cdots$	&	$\cdots$	&	$\cdots$	&	$\cdots$	&	0	&	FIELD	~	0.2	\\
TYC 1256-516-1	&	F5	&	$\cdots$	&	$\cdots$	&	$\cdots$	&	$\cdots$	&	$\cdots$	&	$\cdots$	&	0	&	PLE	~	65.8	\\
HIP 17325	&	A2	&	12.53$\pm$0.00	&	$\cdots$	&	$\cdots$	&	$\cdots$	&	$\cdots$	&	$\cdots$	&	0	&	PLE	~	94.3	\\
HIP 17607	&	$\cdots$	&	$\cdots$	&	$\cdots$	&	$\cdots$	&	$\cdots$	&	$\cdots$	&	$\cdots$	&	0	&	PLE	~	99.1	\\
TYC 1260-671-1	&	F0	&	12.72$\pm$0.00	&	$\cdots$	&	034839.3+21553	&	30.03$\pm$0.26	&	$\cdots$	&	$\cdots$	&	0	&	PLE	~	96.2	\\
HIP 17921	&	B8III	&	$\cdots$	&	$\cdots$	&	$\cdots$	&	$\cdots$	&	$\cdots$	&	$\cdots$	&	0	&	PLE	~	99.8	\\
HIP 17892	&	B9	&	$\cdots$	&	$\cdots$	&	$\cdots$	&	$\cdots$	&	$\cdots$	&	$\cdots$	&	0	&	PLE	~	99.8	\\
TYC 1800-118-1	&	A0	&	$\cdots$	&	$\cdots$	&	$\cdots$	&	$\cdots$	&	$\cdots$	&	$\cdots$	&	0	&	PLE	~	99.8	\\
TYC 1800-669-1	&	$\cdots$	&	$\cdots$	&	$\cdots$	&	$\cdots$	&	$\cdots$	&	$\cdots$	&	$\cdots$	&	0	&	PLE	~	99.9	\\
HIP 17043	&	A0	&	11.64$\pm$0.00	&	$\cdots$	&	$\cdots$	&	$\cdots$	&	$\cdots$	&	$\cdots$	&	0	&	PLE	~	99.3	\\
HIP 17316	&	G0	&	14.54$\pm$0.01	&	$\cdots$	&	$\cdots$	&	$\cdots$	&	$\cdots$	&	$\cdots$	&	0	&	PLE	~	99.1	\\
TYC 1247-515-1	&	F8	&	15.56$\pm$0.01	&	$\cdots$	&	$\cdots$	&	$\cdots$	&	$\cdots$	&	$\cdots$	&	0	&	PLE	~	99.8	\\
HIP 17317	&	$\cdots$	&	15.25$\pm$0.01	&	20.82$\pm$0.16	&	$\cdots$	&	$\cdots$	&	$\cdots$	&	$\cdots$	&	0	&	PLE	~	99.9	\\
HIP 17511	&	F5	&	13.70$\pm$0.00	&	19.06$\pm$0.03	&	$\cdots$	&	$\cdots$	&	$\cdots$	&	$\cdots$	&	0	&	PLE	~	99.7	\\
TYC 1260-498-1	&	$\cdots$	&	16.44$\pm$0.00	&	21.84$\pm$0.28	&	$\cdots$	&	$\cdots$	&	034440.7+22274	&	29.74$\pm$0.57	&	0	&	PLE	~	99.4	\\
TYC 1799-1102-1	&	A0	&	$\cdots$	&	$\cdots$	&	$\cdots$	&	$\cdots$	&	$\cdots$	&	$\cdots$	&	0	&	PLE	~	99.9	\\
TYC 1800-1574-1	&	G0	&	$\cdots$	&	$\cdots$	&	$\cdots$	&	$\cdots$	&	$\cdots$	&	$\cdots$	&	0	&	PLE	~	99.9	\\
HIP 17497	&	F3V	&	$\cdots$	&	$\cdots$	&	$\cdots$	&	$\cdots$	&	$\cdots$	&	$\cdots$	&	0	&	PLE	~	99.9	\\
TYC 1800-1774-1	&	F8	&	$\cdots$	&	$\cdots$	&	$\cdots$	&	$\cdots$	&	$\cdots$	&	$\cdots$	&	0	&	PLE	~	99.9	\\
TYC 1800-2170-1	&	$\cdots$	&	$\cdots$	&	$\cdots$	&	$\cdots$	&	$\cdots$	&	$\cdots$	&	$\cdots$	&	0	&	PLE	~	99.9	\\
TYC 1800-471-1	&	F8	&	$\cdots$	&	$\cdots$	&	$\cdots$	&	$\cdots$	&	$\cdots$	&	$\cdots$	&	0	&	PLE	~	99.7	\\
TYC 1800-496-1	&	$\cdots$	&	$\cdots$	&	$\cdots$	&	$\cdots$	&	$\cdots$	&	$\cdots$	&	$\cdots$	&	0	&	PLE	~	99.9	\\
TYC 1800-628-1	&	A7V	&	$\cdots$	&	$\cdots$	&	$\cdots$	&	$\cdots$	&	$\cdots$	&	$\cdots$	&	0	&	PLE	~	99.9	\\
TYC 1800-727-1	&	F3V	&	$\cdots$	&	$\cdots$	&	$\cdots$	&	$\cdots$	&	$\cdots$	&	$\cdots$	&	0	&	PLE	~	99.9	\\
TYC 1800-2129-1	&	A8V	&	$\cdots$	&	$\cdots$	&	$\cdots$	&	$\cdots$	&	$\cdots$	&	$\cdots$	&	0	&	PLE	~	100	\\
TYC 1800-1672-1	&	F5V	&	$\cdots$	&	$\cdots$	&	$\cdots$	&	$\cdots$	&	$\cdots$	&	$\cdots$	&	0	&	PLE	~	99.9	\\
HIP 17572	&	A0	&	$\cdots$	&	$\cdots$	&	$\cdots$	&	$\cdots$	&	$\cdots$	&	$\cdots$	&	0	&	PLE	~	99.9	\\
TYC 1800-1917-1	&	$\cdots$	&	$\cdots$	&	$\cdots$	&	$\cdots$	&	$\cdots$	&	$\cdots$	&	$\cdots$	&	0	&	PLE	~	99.2	\\
TYC 1800-2027-1	&	$\cdots$	&	$\cdots$	&	$\cdots$	&	$\cdots$	&	$\cdots$	&	$\cdots$	&	$\cdots$	&	0	&	PLE	~	99.9	\\
\enddata 
\tablecomments{This subtable is a preview of the entire sample, which will be available as a machine readable table.  Above we show the Tycho or Hipparcos name along with corresponding photometry from Galex and X-ray luminosity from ROSAT bright and/or faint (with corresponding name).  We also include the Oh17 group number and the BANYAN $\Sigma$ predicted association.}
\end{deluxetable*}

\tabletypesize{\scriptsize}
\clearpage
\begin{longrotatetable}
\begin{deluxetable}{lllllllllllllllllllllll}
\tablecaption{Oh17 sample with BANYAN $\Sigma$ input of known members\label{Tab:Oh17Ban}}
\tablehead{
\colhead{Name} &
\colhead{RA}&
\colhead{DEC}&
\colhead{ $\mu_{\alpha}$ cos ($\delta$) } &
\colhead{$\mu_{\delta}$ }&
\colhead{$\pi$}&
\colhead{v$_{\rm rad}$} & 
\colhead{Group-BANYAN}&
\colhead{Prob}&
\colhead{Group-Oh} &
\colhead{Known}&
\colhead{Known}\\
\colhead{}&
\colhead{}&
\colhead{}&
\colhead{(mas $^{-1}$) }&
\colhead{(mas $^{-1}$)}&
\colhead{(mas)}&
\colhead{(km $s ^{-1}$)}&
\colhead{}\\
\colhead{(1)}     &
\colhead{(2)}     &
\colhead{(3)}     &
\colhead{(4)}     &
\colhead{(5)}     &
\colhead{(6)}    &
\colhead{(7)}    &
\colhead{(8)}     &
\colhead{(9)}     &
\colhead{(10)}&
\colhead{(11)}&
\colhead{(12)}\\
}
\startdata
TYC2899-1744-1	&	76.17226	&	40.39956	&	12.147$\pm$2.195	&	-111.141$\pm$1.942	&	16.238$\pm$0.262	&	$\cdots$	&	ABDMG	&	92.6	&	1036	&	No	&	No\\
HIP23399	&	75.43895	&	42.34373	&	17.984$\pm$0.076	&	-123.319$\pm$0.052	&	17.337$\pm$0.285	&	$\cdots$	&	ABDMG	&	94.1	&	1036	&	No	&	No\\
HIP17405	&	55.93921	&	16.66598	&	156.198$\pm$0.146	&	-310.030$\pm$0.092	&	58.125$\pm$0.363	&	$\cdots$	&	ABDMG	&	98.4	&	1237	&	No	&	No\\
HIP17414	&	55.9697	&	16.67071	&	157.960$\pm$0.181	&	-316.306$\pm$0.112	&	58.046$\pm$0.235	&	$\cdots$	&	ABDMG	&	99.5	&	1237	&	No	&	No\\
HIP41181	&	126.05475	&	44.94519	&	-60.136$\pm$0.393	&	-176.593$\pm$0.270	&	27.097$\pm$0.245	&	$\cdots$	&	ABDMG	&	96.7	&	1456	&	No	&	No\\
HIP41184	&	126.06486	&	44.94897	&	-63.222$\pm$0.103	&	-177.913$\pm$0.064	&	27.102$\pm$0.244	&	$\cdots$	&	ABDMG	&	97.1	&	1456	&	No	&	No\\
HIP29964	&	94.61743	&	-72.04454	&	-7.670$\pm$0.122	&	74.367$\pm$0.137	&	25.612$\pm$0.220	&	$\cdots$	&	ABDMG(56);BPMG(44)	&	99.6	&	1806	&	BF	&	BPMG\\
TYC9162-379-1	&	79.22423	&	-68.35218	&	14.406$\pm$1.465	&	58.566$\pm$1.503	&	22.114$\pm$0.455	&	$\cdots$	&	ABDMG(75);BPMG(25)	&	97.1	&	1806	&	No	&	No\\
HIP14809	&	47.80793	&	22.41534	&	55.656$\pm$0.110	&	-125.193$\pm$0.090	&	19.711$\pm$0.240	&	$\cdots$	&	ABDMG	&	93.3	&	2048	&	BF	&	ABDMG\\
TYC1807-46-1	&	56.49168	&	27.55932	&	43.262$\pm$1.190	&	-118.850$\pm$0.468	&	18.245$\pm$0.275	&	$\cdots$	&	ABDMG	&	96.6	&	2048	&	No	&	No\\
HIP2981	&	9.48814	&	47.40737	&	110.257$\pm$0.039	&	-82.865$\pm$0.028	&	22.101$\pm$0.311	&	$\cdots$	&	ABDMG	&	88.3	&	2201	&	No	&	No\\
HIP3589	&	11.46274	&	54.97752	&	96.401$\pm$0.030	&	-73.969$\pm$0.042	&	19.878$\pm$0.340	&	$\cdots$	&	ABDMG	&	88.1	&	2201	&	BF	&	ABDMG\\
HIP118008	&	359.04558	&	-39.05311	&	206.231$\pm$0.056	&	-185.819$\pm$0.064	&	45.471$\pm$0.228	&	$\cdots$	&	ABDMG	&	99.8	&	2240	&	BF	&	ABDMG\\
HIP79578	&	243.54936	&	-31.66473	&	-75.560$\pm$0.035	&	-256.211$\pm$0.027	&	41.194$\pm$0.476	&	$\cdots$	&	ABDMG	&	98.8	&	2371	&	No	&	No\\
TYC4718-894-1	&	58.8352	&	-1.7296	&	42.085$\pm$0.903	&	-91.433$\pm$0.588	&	18.102$\pm$0.250	&	$\cdots$	&	ABDMG	&	97.6	&	2727	&	No	&	No\\
HIP19183	&	61.67321	&	1.68352	&	36.571$\pm$0.063	&	-94.590$\pm$0.037	&	17.547$\pm$0.336	&	$\cdots$	&	ABDMG	&	95.9	&	2727	&	BF	&	ABDMG\\
HIP31878	&	99.9582	&	-61.47789	&	-26.981$\pm$0.104	&	74.960$\pm$0.093	&	45.328$\pm$0.236	&	$\cdots$	&	ABDMG(75);BPMG(25)	&	99.8	&	2841	&	BF	&	ABDMG\\
HIP30314	&	95.62882	&	-60.21838	&	-11.418$\pm$0.027	&	64.559$\pm$0.028	&	41.973$\pm$0.276	&	$\cdots$	&	ABDMG(83);BPMG(17)	&	99.8	&	2841	&	BF	&	ABDMG\\
TYC5899-26-1	&	73.10226	&	-16.82363	&	123.265$\pm$1.133	&	-212.593$\pm$1.008	&	63.400$\pm$0.366	&	$\cdots$	&	ABDMG	&	99.6	&	2849	&	BF	&	ABDMG\\
HIP17695	&	56.84801	&	-1.97335	&	180.433$\pm$0.170	&	-274.096$\pm$0.127	&	59.266$\pm$0.332	&	$\cdots$	&	ABDMG	&	99.7	&	2849	&	BF	&	ABDMG\\
HIP19422	&	62.39687	&	69.54015	&	73.004$\pm$0.037	&	-298.653$\pm$0.055	&	53.358$\pm$0.242	&	$\cdots$	&	ABDMG	&	93.3	&	3017	&	No	&	No\\
HIP40910	&	125.23018	&	14.07024	&	-83.698$\pm$0.137	&	-261.809$\pm$0.080	&	44.351$\pm$0.274	&	$\cdots$	&	ABDMG	&	98.5	&	3493	&	No	&	No\\
HIP44295	&	135.32228	&	15.26443	&	-125.739$\pm$0.122	&	-320.354$\pm$0.082	&	54.658$\pm$0.306	&	$\cdots$	&	ABDMG	&	94.7	&	3493	&	No	&	No\\
\enddata 
\tablecomments{This subtable is a preview of the entire sample, which will be available as a machine readable table.  Above we show the Tycho or Hipparcos name along with corresponding position, proper motion, parallax, and radial velocity (where available).  Columns (9) and (10) show the detailed results from BANYAN $\Sigma$ of probability of membership in a nearby association (10) and the best group fit.  Column (12) describes whether an object was used as a bonafide member (BF) in BANYAN $\Sigma$ or whether it was a candidate member (CM), high likelihood member (HM), ambiguous member (AM), rejected member (RM), or not investigated (NO).  Column (13) reflects the corresponding group of investigation to Column (12).  The Oh17 Group number is listed in Column (11). An object marked as (NO) may be a newly discovered candidate of a known group.  However we note that we did not do a detailed literature search as to whether some of these sources were investigated by others therefore further vetting is required for each source.
}
\end{deluxetable} 
\end{longrotatetable}

\tabletypesize{\scriptsize}
\clearpage
\begin{deluxetable*}{lccccccccc}
\tablewidth{0pt}
\tablecolumns{13}
\tablecaption{BANYAN $\Sigma$ summary of Oh17\label{tab:summary}}
\tablewidth{0pt}
\tablehead{
\colhead{BANYAN Name} &
\colhead{Total}&
\colhead{\# BF}&
\colhead{\# CM}&
\colhead{ \# HM } &
\colhead{\# LM}&
\colhead{\# RM}    &
\colhead{\# AM}    &
\colhead{\# NO (NEW?)}    \\
\colhead{(1)}     &
\colhead{(2)}     &
\colhead{(3)}     &
\colhead{(4)}     &
\colhead{(5)}     &
\colhead{(6)}     &
\colhead{(7)}     &
\colhead{(8)}     &
\colhead{(9)}\\
}
\startdata
118TAU	&	$\cdots$	&	$\cdots$	&	$\cdots$	&	$\cdots$	&	$\cdots$	&	$\cdots$	&	$\cdots$	&	$\cdots$	\\
ABDMG	&	24	&	8	&	1	&	$\cdots$	&	$\cdots$	&	$\cdots$	&	$\cdots$	&	15	\\
BPMG	&	6	&	3	&	$\cdots$	&	$\cdots$	&	$\cdots$	&	$\cdots$	&	$\cdots$	&	3	\\
CAR	&	0	&	$\cdots$	&	$\cdots$	&	$\cdots$	&	$\cdots$	&	$\cdots$	&	$\cdots$	&	$\cdots$	\\
CARN	&	9	&	5	&	$\cdots$	&	$\cdots$	&	$\cdots$	&	$\cdots$	&	$\cdots$	&	4	\\
CBER	&	45	&	33	&	2	&	$\cdots$	&	$\cdots$	&	$\cdots$	&	$\cdots$	&	10	\\
COL	&	18	&	4	&	$\cdots$	&	$\cdots$	&	$\cdots$	&	$\cdots$	&	$\cdots$	&	14	\\
CRA	&	1	&	$\cdots$	&	$\cdots$	&	1	&	$\cdots$	&	$\cdots$	&	$\cdots$	&	$\cdots$	\\
EPSC	&	10	&	5	&	2	&	1	&	$\cdots$	&	$\cdots$	&	1	&	1	\\
ETAC	&	2	&	1	&	$\cdots$	&	1	&	$\cdots$	&	$\cdots$	&	$\cdots$	&	$\cdots$	\\
HYA	&	110	&	88	&	14	&	$\cdots$	&	$\cdots$	&	$\cdots$	&	$\cdots$	&	8	\\
IC2391 	&	28	&	18	&	$\cdots$	&	8	&	$\cdots$	&	$\cdots$	&	$\cdots$	&	2	\\
IC2602	&	31	&	11	&	1	&	$\cdots$	&	11	&	$\cdots$	&	1	&	7	\\
LCC	&	156	&	37	&	9	&	41	&	$\cdots$	&	$\cdots$	&	1	&	68	\\
OCT	&	34	&	$\cdots$	&	$\cdots$	&	$\cdots$	&	$\cdots$	&	$\cdots$	&	$\cdots$	&	34	\\
PL8	&	22	&	$\cdots$	&	$\cdots$	&	5	&	$\cdots$	&	$\cdots$	&	$\cdots$	&	17	\\
PLE	&	136	&	117	&	6	&	$\cdots$	&	$\cdots$	&	1	&	$\cdots$	&	12	\\
ROPH	&	$\cdots$	&	$\cdots$	&	$\cdots$	&	$\cdots$	&	$\cdots$	&	$\cdots$	&	$\cdots$	&	$\cdots$	\\
TAU	&	33	&	9	&	$\cdots$	&	2	&	$\cdots$	&	$\cdots$	&	$\cdots$	&	22	\\
THA	&	29	&	24	&	$\cdots$	&	$\cdots$	&	$\cdots$	&	$\cdots$	&	$\cdots$	&	5	\\
THOR	&	5	&	2	&	2	&	$\cdots$	&	$\cdots$	&	$\cdots$	&	$\cdots$	&	1	\\
TWA	&	3	&	$\cdots$	&	$\cdots$	&	$\cdots$	&	$\cdots$	&	$\cdots$	&	$\cdots$	&	3	\\
UCL	&	194	&	27	&	8	&	41	&	$\cdots$	&	$\cdots$	&	2	&	116	\\
UCRA	&	2	&	$\cdots$	&	$\cdots$	&	$\cdots$	&	$\cdots$	&	$\cdots$	&	$\cdots$	&	2	\\
UMA	&	1	&	$\cdots$	&	1	&	$\cdots$	&	$\cdots$	&	$\cdots$	&	$\cdots$	&	$\cdots$	\\
USCO	&	107	&	23	&	4	&	24	&	$\cdots$	&	$\cdots$	&	$\cdots$	&	56	\\
XFOR	&	9	&	3	&	$\cdots$	&	6	&	$\cdots$	&	$\cdots$	&	$\cdots$	&	$\cdots$	\\
\enddata 
\tablecomments{The number of Oh17 objects for each BANYAN selected group that are either: Bonafide members (BF), Candidate Members (CM), High Likely Members (HM), Low Likely Members (LM), Ambiguous Members (AM), Rejected Members (RM), not in BANYAN $\Sigma$ (NO).\\
The full names of BANYAN $\Sigma$ groups are:  118 Tau (118TAU), AB Doradus (ABDMG), $\beta$ Pictoris (BPMG), Carina (CAR), Carina-Near (CARN), Coma Berenices (CBER), Columba (COL), Corona Australis (CRA), $\epsilon$ Chamaeleontis (EPSC),$\eta$ Chamaeleontis (ETAC), the Hyades cluster (HYA), Lower Centaurus Crux (LCC), Octans (OCT), Platais 8 (PL8), the Pleiades cluster (PLE), $\rho$ Ophiucus (ROPH), the Tucana-Horologium association (THA), 32 Orionis (THOR), TW Hya (TWA), Upper Centaurus Lupus (UCL), Upper CrA (UCRA), the core of the Ursa Major cluster (UMA), Upper Scorpius (USCO),  Taurus (TAU), and $\chi$ For (XFOR)\\}
\end{deluxetable*}

\startlongtable
\clearpage
\begin{deluxetable*}{llllllll}
\tabletypesize{\scriptsize}
\tablewidth{0pt}
\tablecolumns{13}
\tablecaption{BANYAN Groups Matched to Oh17\label{Tab:Gaia}}
\tablewidth{0pt}
\tablehead{
\colhead{BANYAN Name} &
\colhead{\# Oh17 Groups}&
\colhead{Oh17 Group Number}&
\colhead{ \# in BANYAN } &
\colhead{\# in Oh17 Groups}&
\colhead{Age (Myr)}    &
\colhead{Age ref}\\
\colhead{(1)}     &
\colhead{(2)}     &
\colhead{(3)}     &
\colhead{(4)}     &
\colhead{(5)}     &
\colhead{(6)}     &
\colhead{(7)}\\
}
\startdata
118TAU	&	$\cdots$	&	$\cdots$	&	$\cdots$	&	$\cdots$	&$\sim$10&14\\
\hline
\hline
ABDMG	&	14	&	1036, 1237, 1456, 1806, 2048,  	&	24	&	28	& 149$^{+51}_{-19}$&1\\
& & 2201, 2240, 2371, 2727, 2841, \\
& & 2849, 3017, 3493, 3709 \\
\hline
\hline
BPMG	&	3	&	241, 2230, 3062	&	6	&	7	& 24$\pm$3& 1\\
\hline
\hline
CAR	&	$\cdots$	&	$\cdots$ &	$\cdots$	&	$\cdots$& 45$^{+11}_{-7}$&1\\
\hline
\hline
CARN	&	5	&	215, 272, 2347, 2785, 3746 &	9	&	12	&$\sim$200&2 \\
\hline
\hline 
CBER	&	1	&	7	&	45	&	47	&562$^{+98}_{-84}$&3\\
\hline
\hline 
COL	&	7	&	355, 93, 241, 690, 1460, &	18	&	23&42$^{+6}_{-4}$&1	\\
& & 2495,4103 \\
\hline
\hline 
CRA	&	1	&	3261	&	1	&	2&	4-5&16\\
\hline
\hline 
EPSC	&	1	&	3	&	10	&	114	&3.7$^{+4.6}_{-1.4}$&4\\
\hline
\hline 
ETAC	&	1	&	2071	&	2	&	2	&11$\pm$3&1\\
\hline
\hline 
HYA	&	3	&	2, 261, 3208	&	110& 123&750$\pm$100&5	\\
\hline
\hline 
IC2391	&	2	&	9, 2520 & 28	&	38	&	50$\pm$5 &18	\\
\hline
\hline 
IC2602	&	1	&	5	&	31	&	59&46$^{+6}_{-5}$&6	\\
\hline
\hline 
LCC	&	21	&	3, 13, 40, 45, 48, 	&	156	&	197	&15$\pm$3&7\\
& & 75, 77, 100, 183, 239, \\
& & 255, 266, 937, 1340, 2035,  \\
& & 2091, 2162, 2206, 2283, 2288,  \\
& & 4032 \\
\hline
\hline 
OCT	&	14	&	31,37, 143, 670, 2351,   &	34	&	38&35$\pm$5&8	\\
& & 2974, 3508, 3521, 3562, 3730,    \\
& & 3994, 3995, 4187, 4289, 4340 \\
\hline
\hline 
PL8	&	4	&	12, 269, 574, 4289	&	22	&	30	&$\sim$60&9\\
\hline
\hline 
PLE	&	1	&	0	&	136	&	151	&112$\pm$5&10\\
\hline
\hline 
ROPH	&	$\cdots$	&	$\cdots$	&	$\cdots$	&	$\cdots$&	$<$2&13\\
\hline
\hline 
TAU	&	9	&	28, 29, 286, 300, 1348,	&	33	&	34&	1-2&15\\
	&		&	3774, 3824, 3856, 3870	&		&	  &	   &  \\
\hline
\hline 
THA	&	6	&	15, 221, 303, 673, &	29	&	35	&45$\pm$4&1\\
& &  951, 1022	\\
\hline
\hline 
THOR	&	1	&	44	&	5	&	6	&22$^{+4}_{-3}$&1\\
\hline
\hline 
TWA	&	2	&	172, 3001	&	3	&	5	&10$\pm$3&1\\
\hline
\hline 
UCL	&	38	&	8, 11, 13, 18, 24,  	&	194	&	215&16$\pm$2&7	\\
& & 25, 33, 40, 45, 46 \\
& & 54, 92, 94, 99,114 \\
& & 239, 299, 304, 308, 408, \\
& & 668, 672, 1443, 2239, 2668,   \\
& & 2828, 3019, 3155, 3238, 3245,   \\
& & 3362, 3908, 4027, 4132, 4218,   \\
& & 4300, 4404, 4490 \\
\hline
\hline 
UCRA	&	2	&	1308, 3261	&	2	&	4&	10&17\\
\hline
\hline 
UMA	&	1	&	1058	&	1	&	2&414$\pm$23&11	\\
\hline
\hline 
USCO	&	11	&	4, 25, 27, 69, 113,  &	107	&	115	&10$\pm$3&7\\
& & 114, 131, 155, 447,  \\
& & 831, 1242 \\
\hline
\hline 
XFOR	&	1	&	19	&	9	&	13	&$\sim$500&12\\
\hline
\hline 
& & {\bf NON-BANYAN}\\
\hline
Alpha Perseus	&	---	&	1	&	---	&	125	\\
RSG 2\tablenotemark{a}	&	---	&	16	&	---	&	18	\\
Blanco1	&	---	&	17	&	---	&	16	\\
Praesepe 	&	---	&	6, 4141	&	---	&	60	\\
NGC2451A &	---	&	21, 233, 236	&	---	&	18	\\
Platais 9	&	---	&	22			&	---	&	18	\\									
NGC2516	&	---	&	3967	&	---	&	2	\\
NGC3532	&	---	&	3116	&	---	&	2	\\
NGC6475	&	---	&	57, 2870, 2891, 2983	&	---	&	11	\\
NGC6633	&	---	&	1335	&	---	&	2	\\
NGC7092	&	---	&	3101	&	---	&	2	\\
\enddata 
\tablenotetext{a}{A new open cluster discovered recently by \citet{Roser16}}
\tablecomments{All 21 of the BANYAN $\Sigma$ tested groups compared to the Oh17 catalog group numbers.  Non-BANYAN tested groups are also listed with corresponding Oh17 group numbers. \\ 
\\
The full names of BANYAN $\Sigma$ groups are:  118 Tau (118TAU), AB Doradus (ABDMG), $\beta$ Pictoris (BPMG), Carina (CAR), Carina-Near (CARN), Coma Berenices (CBER), Columba (COL), Corona Australis (CRA), $\epsilon$ Chamaeleontis  (EPSC),$\eta$ Chamaeleontis (ETAC), the Hyades cluster (HYA), Lower Centaurus Crux (LCC), Octans (OCT), Platais 8 (PL8), the Pleiades cluster (PLE), $\rho$ Ophiucus (ROPH), the Tucana-Horologium association (THA), 32 Orionis (THOR), TW Hya (TWA), Upper Centaurus Lupus (UCL), Upper CrA (UCRA), the core of the Ursa Major cluster (UMA), Upper Scorpius (USCO),  Taurus (TAU), and $\chi$ For (XFOR)\\
\\
{\bf References:} (1)  \citet{Bell15aa}, (2) \citet{Zuckerman06}, (3) \citet{Silaj14}, (4) \citet{Murphy13}, (5) \citet{Brandt15}, (6) \citet{Dobbie10}, (7) \citet{Pecaut16}, (8)\citet{Murphy15}, (9) \citet{Platais98} , (10) \citet{Dahm15}, (11) \citet{Jones15}, (12) \citet{Pohnl10}, (13) \citet{Wilking2008}, (14) \citet{Mamajek2016}, (15) \citet{Kenyon1995}, (16) \citet{Gennaro2012}, (17) \citet{Gagne18}, (18) \citet{Barrado2004}} 

\end{deluxetable*}

\tabletypesize{\scriptsize}
\clearpage
\begin{deluxetable*}{lllll}[htbp]
\tablewidth{0pt}
\tablecolumns{13}
\tablecaption{Oh17 Group with Conflicting BANYAN Prediction\label{tab:conflicts}}
\tablewidth{0pt}
\tablehead{
\colhead{Oh17 Group \#} &
\colhead{BANYAN Group}\\
\colhead{(1)}     &
\colhead{(2)}\\
}
\startdata
3	& EPSC, LCC\\
13	& UCL, LCC\\
25	& UCL, USCO\\
40	& LCC, UCL\\
45	& UCL, LCC\\
114	& USCO, UCL\\
239	& LCC, UCL\\
241	& BPMG, COL\\
3261 & CRA, UCRA\\
\enddata 
\tablecomments{A list of the Oh17 groups that had members with $>$80$\%$ probability in more than 1 BANYAN $\Sigma$ tested group }
\end{deluxetable*}

\startlongtable
\tabletypesize{\scriptsize}
\clearpage
\begin{deluxetable*}{lccccccccc}
\tablewidth{0pt}
\tablecolumns{13}
\tablecaption{Potentially new associations from Oh17 with $>$ 10 connected components\label{Tab:Newgroups}}
\tablewidth{0pt}
\tablehead{
\colhead{Name} &
\colhead{RA}&
\colhead{DEC}&
\colhead{SpT}&
\colhead{ $\mu_{\alpha}$ } &
\colhead{$\mu_{\delta}$ }&
\colhead{$\pi$}&
\colhead{Group} \\
\colhead{}&
\colhead{}&
\colhead{}&
\colhead{}&
\colhead{(mas $^{-1}$) }&
\colhead{(mas $^{-1}$)}&
\colhead{(mas)}\\
\colhead{(1)}     &
\colhead{(2)}     &
\colhead{(3)}     &
\colhead{(4)}     &
\colhead{(5)}     &
\colhead{(6)}    &
\colhead{(7)}    &
\colhead{(8)}\\
}
\startdata
HIP69721&214.07256&58.38940&F5&-16.254$\pm$0.039&-2.903$\pm$0.049&9.341$\pm$0.293&10\\
HIP67005&205.97812&52.06439&A1V&-18.270$\pm$0.018&-5.605$\pm$0.021&10.737$\pm$0.325&10\\
TYC3851-600-1&207.11420&54.04270&$\cdots$&-18.288$\pm$0.297&-3.934$\pm$0.783&10.709$\pm$0.255&10\\
HIP67231&206.64844&54.43266&A2P&-18.533$\pm$0.018&-4.750$\pm$0.021&10.333$\pm$0.503&10\\
TYC3851-336-1&205.40212&53.33751&G0&-18.006$\pm$0.617&-3.350$\pm$0.743&10.040$\pm$0.261&10\\
TYC3851-369-1&205.78236&54.02590&G5&-19.004$\pm$0.298&-2.867$\pm$0.663&10.431$\pm$0.297&10\\
HIP66198&203.53030&55.34841&$\cdots$&-19.079$\pm$0.022&-6.070$\pm$0.023&10.704$\pm$0.570&10\\
TYC3850-257-1&201.21590&54.89743&A5&-19.011$\pm$0.347&-6.271$\pm$0.396&11.125$\pm$0.264&10\\
HIP63702&195.81947&57.31521&F8&-17.103$\pm$0.063&-8.196$\pm$0.070&10.246$\pm$0.261&10\\
TYC3480-1209-1&223.27004&51.26115&K2&-14.193$\pm$0.392&-0.685$\pm$0.682&9.772$\pm$0.238&10\\
TYC3868-177-1&230.81618&54.84823&$\cdots$&-13.792$\pm$0.398&-1.234$\pm$0.845&8.986$\pm$0.262&10\\
HIP74458&228.24117&56.04643&A2&-13.157$\pm$0.041&-1.189$\pm$0.039&8.672$\pm$0.300&10\\
TYC3861-1374-1&222.52355&53.63483&$\cdots$&-14.564$\pm$0.914&-1.915$\pm$0.748&9.633$\pm$0.295&10\\
TYC3860-1483-1&219.85980&54.77406&$\cdots$&-17.561$\pm$0.458&-2.798$\pm$0.653&10.964$\pm$0.261&10\\
HIP72389&222.01173&56.15920&G5&-15.865$\pm$0.117&-1.522$\pm$0.123&10.266$\pm$0.221&10\\
HIP69917&214.62966&52.03331&A2&-17.194$\pm$0.026&-3.122$\pm$0.030&10.059$\pm$0.271&10\\
HIP69650&213.82070&52.53591&A4V&-17.603$\pm$0.022&-3.474$\pm$0.027&10.402$\pm$0.280&10\\
HIP69958&214.73284&54.86376&A5Vn&-16.565$\pm$0.029&-2.063$\pm$0.029&9.790$\pm$0.681&10\\
TYC3865-934-1&216.29629&57.63321&G0&-15.381$\pm$0.367&-2.485$\pm$0.645&9.571$\pm$0.284&10\\
HIP73730&226.07328&59.53505&A2&-13.661$\pm$0.027&-0.164$\pm$0.029&9.000$\pm$0.275&10\\
TYC3875-762-1&231.92341&59.98704&$\cdots$&-13.297$\pm$0.289&0.094$\pm$1.020&8.928$\pm$0.281&10\\
TYC3867-281-1&226.10718&59.88078&K2&-13.399$\pm$0.353&-0.068$\pm$1.138&9.351$\pm$0.280&10\\
HIP71911&220.63149&60.23096&F0&-16.235$\pm$0.065&-3.840$\pm$0.066&9.428$\pm$0.219&10\\
TYC3867-1373-1&222.87595&59.53208&$\cdots$&-15.362$\pm$0.456&-1.742$\pm$1.371&9.638$\pm$0.392&10\\
TYC4173-609-1&219.82002&61.93126&$\cdots$&-17.035$\pm$0.363&-3.995$\pm$0.976&9.892$\pm$0.300&10\\
HIP69275&212.72088&62.52220&F2IV&-17.166$\pm$0.043&-3.034$\pm$0.050&9.678$\pm$0.247&10\\
TYC4174-1117-1&209.67123&63.68876&$\cdots$&-18.907$\pm$0.583&-4.208$\pm$0.691&10.613$\pm$0.260&10\\
TYC3471-233-1&211.95507&51.95266&$\cdots$&-16.804$\pm$0.351&-4.588$\pm$0.894&9.982$\pm$0.242&10\\
HIP68637&210.74889&50.97178&A0IV&-16.444$\pm$0.018&-6.210$\pm$0.019&9.936$\pm$0.418&10\\
\hline
\hline
\\
TYC9280-112-1&258.43252&-69.98264&$\cdots$&-11.735$\pm$0.607&-12.739$\pm$0.774&4.244$\pm$0.253&14\\
TYC9279-1700-1&256.44526&-70.39820&A(?)&-13.393$\pm$0.421&-13.273$\pm$0.472&4.424$\pm$0.241&14\\
TYC9279-2048-1&254.64074&-70.24192&$\cdots$&-13.924$\pm$0.544&-12.111$\pm$0.681&4.330$\pm$0.219&14\\
TYC9279-1772-1&254.96898&-70.17474&$\cdots$&-13.458$\pm$0.493&-12.506$\pm$0.712&4.361$\pm$0.264&14\\
TYC9275-1592-1&252.40693&-69.30057&$\cdots$&-14.436$\pm$0.554&-13.326$\pm$0.713&4.734$\pm$0.252&14\\
TYC9275-963-1&255.52187&-68.19909&$\cdots$&-13.093$\pm$0.456&-12.835$\pm$0.652&4.478$\pm$0.222&14\\
TYC9276-2997-1&258.71396&-68.85196&$\cdots$&-11.928$\pm$0.512&-13.501$\pm$0.699&4.222$\pm$0.231&14\\
TYC9275-2648-1&255.75727&-68.62989&A1V&-13.721$\pm$0.306&-13.050$\pm$0.296&4.639$\pm$0.406&14\\
TYC9275-2499-1&255.66161&-68.61236&$\cdots$&-13.545$\pm$0.483&-13.917$\pm$0.716&4.516$\pm$0.247&14\\
TYC9275-3434-1&257.82869&-68.11085&$\cdots$&-12.428$\pm$0.365&-12.500$\pm$0.628&4.284$\pm$0.269&14\\
TYC9275-1819-1&256.02271&-68.19380&$\cdots$&-12.999$\pm$0.438&-12.991$\pm$0.577&4.348$\pm$0.246&14\\
TYC9275-1067-1&255.89374&-67.88703&A2Vs&-12.990$\pm$0.369&-12.302$\pm$0.419&4.280$\pm$0.275&14\\
TYC9275-2142-1&254.16395&-68.35217&A6III&-13.716$\pm$0.446&-12.913$\pm$0.547&4.601$\pm$0.275&14\\
TYC9275-251-1&252.96875&-68.07482&$\cdots$&-13.832$\pm$0.544&-11.963$\pm$0.576&4.422$\pm$0.317&14\\
TYC9050-754-1&252.84677&-67.47804&$\cdots$&-14.366$\pm$0.434&-13.203$\pm$0.572&4.583$\pm$0.244&14\\
TYC9275-1107-1&254.72914&-67.78803&A5IV&-14.559$\pm$0.445&-13.242$\pm$0.475&4.544$\pm$0.299&14\\
TYC9064-2249-1&257.50549&-66.88909&$\cdots$&-13.776$\pm$0.452&-14.554$\pm$0.734&4.672$\pm$0.246&14\\
TYC9051-124-1&254.90482&-66.98483&$\cdots$&-13.290$\pm$0.440&-12.357$\pm$0.637&4.457$\pm$0.276&14\\
HIP82908&254.13678&-66.10902&A0V&-14.133$\pm$0.039&-12.773$\pm$0.037&4.436$\pm$0.321&14\\
TYC9050-901-1&250.97105&-67.24941&$\cdots$&-15.309$\pm$0.528&-13.511$\pm$0.715&4.755$\pm$0.265&14\\
\hline
\hline
\\
TYC8950-174-1&148.39804&-64.19227&A(?)&-28.902$\pm$0.707&23.115$\pm$0.660&4.874$\pm$0.323&23\\
HIP48707&149.02216&-63.11015&A4V&-29.082$\pm$0.071&23.529$\pm$0.067&4.717$\pm$0.269&23\\
TYC8951-88-1&151.54021&-63.92761&A2V&-29.790$\pm$0.707&22.781$\pm$0.628&5.008$\pm$0.242&23\\
HIP48873&149.52952&-62.58526&G5/6III&-29.149$\pm$0.030&23.367$\pm$0.028&4.735$\pm$0.254&23\\
TYC8946-285-1&148.20684&-62.10954&$\cdots$&-29.946$\pm$0.915&23.556$\pm$0.656&4.446$\pm$0.267&23\\
TYC8942-2165-1&148.16318&-61.72158&$\cdots$&-29.677$\pm$0.903&23.039$\pm$0.632&4.406$\pm$0.267&23\\
HIP48281&147.65756&-60.52759&A1IV&-28.084$\pm$0.027&23.504$\pm$0.028&4.580$\pm$0.336&23\\
TYC8941-397-1&145.63238&-60.09132&A5IV&-26.126$\pm$0.730&23.833$\pm$0.752&4.611$\pm$0.326&23\\
TYC8943-2975-1&150.07020&-61.51121&$\cdots$&-31.134$\pm$0.803&23.904$\pm$0.569&4.696$\pm$0.242&23\\
TYC8942-2267-1&146.93234&-60.93701&$\cdots$&-28.471$\pm$1.132&23.109$\pm$0.985&4.311$\pm$0.350&23\\
\hline
\hline
\\
TYC3713-616-1&44.21811&58.58569&G0&32.929$\pm$0.829&-29.927$\pm$0.451&8.431$\pm$0.239&26\\
TYC3715-742-1&50.82901&58.72919&A5&31.259$\pm$0.507&-34.520$\pm$0.375&8.369$\pm$0.247&26\\
TYC4062-755-1&52.27789&60.56034&G5&32.722$\pm$0.834&-38.415$\pm$0.547&8.684$\pm$0.257&26\\
TYC4048-1560-1&46.98214&60.52351&$\cdots$&29.288$\pm$1.353&-29.686$\pm$0.622&7.930$\pm$0.322&26\\
TYC4049-648-1&47.75587&60.95448&A3II&29.963$\pm$0.561&-30.319$\pm$0.393&7.470$\pm$0.231&26\\
TYC4053-1110-1&51.31316&62.12718&$\cdots$&32.027$\pm$1.392&-36.279$\pm$0.756&8.752$\pm$0.278&26\\
HIP12355&39.76509&62.57544&A0&33.219$\pm$0.034&-25.385$\pm$0.039&7.596$\pm$0.264&26\\
HIP12346&39.74336&62.59140&B9&33.483$\pm$0.021&-25.348$\pm$0.024&7.808$\pm$0.434&26\\
TYC4056-654-1&46.26338&65.18254&F8&33.095$\pm$0.390&-29.208$\pm$0.386&7.969$\pm$0.318&26\\
TYC3715-701-1&51.58515&58.88690&F5&35.325$\pm$0.750&-37.785$\pm$0.514&9.274$\pm$0.243&26\\
\enddata 
\tablecomments{Collections of stars in Groups 10, 14, 23, and 26 from the Oh17 sample that appear to be newly discovered. }
\end{deluxetable*}

\startlongtable
\tabletypesize{\scriptsize}
\clearpage
\begin{deluxetable*}{lccccccccc}
\tablewidth{0pt}
\tablecolumns{13}
\tablecaption{Potentially new associations from Oh17 with 5 - 9 connected components\label{Tab:Newgroups2}}
\tablewidth{0pt}
\tablehead{
\colhead{Name} &
\colhead{RA}&
\colhead{DEC}&
\colhead{SpT}&
\colhead{ $\mu_{\alpha}$ } &
\colhead{$\mu_{\delta}$ }&
\colhead{$\pi$}&
\colhead{Group} \\
\colhead{}&
\colhead{}&
\colhead{}&
\colhead{}&
\colhead{mas $^{-1}$ }&
\colhead{mas $^{-1}$}&
\colhead{mas}\\
\colhead{(1)}     &
\colhead{(2)}     &
\colhead{(3)}     &
\colhead{(4)}     &
\colhead{(5)}     &
\colhead{(6)}    &
\colhead{(7)}    &
\colhead{(8)}\\
}
\startdata
TYC8950-1447-1&147.07979&-64.05587&$\cdots$&-43.466$\pm$1.072&46.020$\pm$0.687&12.982$\pm$0.268&30\\
HIP45594&139.39198&-63.38719&F3V&-28.931$\pm$0.050&42.406$\pm$0.050&11.025$\pm$0.246&30\\
TYC9210-1730-1&153.12405&-67.87577&$\cdots$&-47.205$\pm$0.715&42.869$\pm$0.510&12.245$\pm$0.232&30\\
TYC9210-1818-1&152.09121&-67.94434&$\cdots$&-39.815$\pm$1.079&38.142$\pm$0.717&11.325$\pm$0.315&30\\
HIP47335&144.68804&-66.85889&A9IV/V&-36.597$\pm$0.052&45.317$\pm$0.054&11.671$\pm$0.232&30\\
TYC8953-1289-1&144.79319&-66.77079&$\cdots$&-37.709$\pm$3.199&45.094$\pm$2.661&11.653$\pm$0.376&30\\
HIP46460&142.12690&-66.70166&A0V&-35.443$\pm$0.025&49.410$\pm$0.026&12.977$\pm$0.461&30\\
HIP47017&143.73490&-64.99927&F5V&-34.670$\pm$0.068&44.232$\pm$0.061&11.620$\pm$0.234&30\\
\hline
\hline
TYC7700-1309-1&144.25993&-42.63547&$\cdots$&-32.821$\pm$0.973&14.060$\pm$0.463&6.968$\pm$0.272&34\\
HIP48403&148.03021&-43.79756&A9IV/V(m)&-35.664$\pm$0.041&13.527$\pm$0.044&7.488$\pm$0.390&34\\
TYC7702-1556-1&148.57572&-41.70651&F8&-33.243$\pm$0.925&12.140$\pm$0.441&7.324$\pm$0.321&34\\
TYC7690-1513-1&139.75253&-43.40861&F5V&-31.488$\pm$0.879&16.398$\pm$0.897&7.181$\pm$0.312&34\\
HIP47161&144.15275&-42.08476&F3V&-33.935$\pm$0.102&14.598$\pm$0.099&6.982$\pm$0.222&34\\
HIP48234&147.50611&-40.16895&A1/2V&-34.784$\pm$0.033&12.641$\pm$0.034&7.230$\pm$0.274&34\\
TYC7700-2419-1&144.21453&-41.97493&$\cdots$&-30.378$\pm$1.091&12.498$\pm$0.502&6.833$\pm$0.330&34\\
\hline
\hline
TYC3698-2538-1&34.17816&58.21182&F2IV&36.798$\pm$0.673&-27.865$\pm$0.575&8.674$\pm$0.274&35\\
HIP9690&31.16681&65.10339&A0V&43.445$\pm$0.021&-26.549$\pm$0.025&9.598$\pm$0.506&35\\
TYC3697-428-1&30.91825&59.76027&F0II&41.841$\pm$0.576&-28.335$\pm$0.468&9.832$\pm$0.271&35\\
TYC4036-884-1&29.99388&62.69480&$\cdots$&45.049$\pm$0.445&-27.465$\pm$0.763&9.433$\pm$0.300&35\\
HIP11156&35.87925&61.77354&F5V&39.542$\pm$0.065&-29.346$\pm$0.060&9.169$\pm$0.270&35\\
TYC4046-788-1&35.85626&61.78297&$\cdots$&40.546$\pm$1.380&-29.693$\pm$0.507&8.966$\pm$0.240&35\\
TYC4037-1304-1&32.35060&63.22518&$\cdots$&37.965$\pm$0.544&-24.123$\pm$0.740&8.351$\pm$0.232&35\\
\hline
\hline
TYC3698-1416-1&34.05663&58.66893&$\cdots$&16.484$\pm$1.081&-13.733$\pm$0.410&2.631$\pm$0.235&36\\
HIP10844&34.89591&59.30818&A2V&15.517$\pm$0.071&-13.776$\pm$0.063&2.649$\pm$0.253&36\\
TYC3698-475-1&35.90293&59.92108&A3V&15.083$\pm$0.785&-13.633$\pm$0.484&2.661$\pm$0.228&36\\
TYC3698-985-1&34.16233&59.83703&$\cdots$&15.798$\pm$0.972&-13.631$\pm$0.377&2.677$\pm$0.228&36\\
TYC3698-495-1&33.51432&59.79895&$\cdots$&15.683$\pm$1.094&-13.389$\pm$0.398&2.678$\pm$0.225&36\\
TYC3698-3123-1&36.63970&58.36576&$\cdots$&15.503$\pm$1.030&-14.365$\pm$0.453&2.673$\pm$0.258&36\\
TYC3698-747-1&33.46726&59.75090&A0V&15.850$\pm$1.151&-13.708$\pm$0.437&2.628$\pm$0.239&36\\
\hline
\hline
HIP117376&356.99239&78.37609&F8&22.344$\pm$0.090&1.349$\pm$0.084&6.359$\pm$0.228&38\\
TYC4500-124-1&0.17226&79.67775&G&22.601$\pm$0.637&0.630$\pm$0.602&6.372$\pm$0.278&38\\
TYC4500-310-1&9.91810&79.09186&$\cdots$&23.259$\pm$0.888&-3.678$\pm$0.837&6.681$\pm$0.259&38\\
TYC4500-1478-1&9.52596&79.05572&$\cdots$&22.385$\pm$0.644&-2.306$\pm$0.722&6.549$\pm$0.249&38\\
TYC4501-1813-1&11.34654&79.73042&F5&23.907$\pm$1.174&-3.876$\pm$0.654&6.687$\pm$0.250&38\\
TYC4500-616-1&4.36837&79.79943&$\cdots$&23.831$\pm$1.096&-1.806$\pm$1.129&6.291$\pm$0.282&38\\
HIP115764&351.80796&79.54195&A2&21.916$\pm$0.038&4.080$\pm$0.040&6.364$\pm$0.700&38\\
\hline
\hline
HIP23819&76.79954&-3.49642&A2/3V&11.312$\pm$0.067&-13.707$\pm$0.054&5.741$\pm$0.281&39\\
TYC4745-475-1&73.30027&-3.81951&$\cdots$&13.013$\pm$0.818&-14.544$\pm$0.693&6.125$\pm$0.237&39\\
HIP23386&75.41076&-2.72076&A1/2V&11.985$\pm$0.065&-14.462$\pm$0.050&5.481$\pm$0.330&39\\
TYC4741-307-1&74.07619&-1.89251&$\cdots$&11.062$\pm$1.004&-14.083$\pm$0.754&5.690$\pm$0.244&39\\
HIP22716&73.27011&-1.27584&A7/F0&12.649$\pm$0.061&-15.773$\pm$0.041&5.867$\pm$0.344&39\\
HIP22689&73.19139&0.68718&F8&11.499$\pm$0.214&-15.130$\pm$0.124&5.707$\pm$0.237&39\\
HIP23661&76.27642&-3.67021&A2V&11.557$\pm$0.056&-13.612$\pm$0.046&6.091$\pm$0.336&39\\
\hline
\hline
TYC9233-1754-1&170.35634&-72.84477&A2V&-25.955$\pm$0.878&-1.928$\pm$0.613&4.120$\pm$0.267&41\\
HIP54740&168.12389&-71.74939&B8V&-25.582$\pm$0.036&-1.247$\pm$0.039&4.222$\pm$0.244&41\\
TYC9220-3213-1&166.17019&-71.67039&$\cdots$&-26.136$\pm$0.754&-0.681$\pm$0.592&4.076$\pm$0.245&41\\
HIP54712&168.02257&-71.21747&B7Vn&-26.348$\pm$0.032&-1.295$\pm$0.031&4.420$\pm$0.274&41\\
TYC9216-1951-1&167.26498&-70.49593&$\cdots$&-26.590$\pm$0.812&-1.286$\pm$0.672&4.080$\pm$0.266&41\\
TYC9220-3429-1&164.25354&-71.35402&$\cdots$&-25.284$\pm$0.472&0.389$\pm$0.466&4.278$\pm$0.266&41\\
TYC9233-453-1&170.14833&-72.02462&$\cdots$&-24.680$\pm$0.999&-1.998$\pm$0.563&4.107$\pm$0.367&41\\
\hline
\hline
TYC3709-701-1&45.22227&56.76031&$\cdots$&28.772$\pm$1.063&-27.059$\pm$0.491&6.725$\pm$0.257&42\\
TYC3700-400-1&40.67272&54.15157&$\cdots$&27.166$\pm$1.150&-21.516$\pm$0.634&6.275$\pm$0.279&42\\
TYC3709-588-1&46.63326&56.33607&$\cdots$&28.689$\pm$1.578&-26.658$\pm$0.792&6.524$\pm$0.345&42\\
HIP14047&45.22317&52.35193&B9V&27.129$\pm$0.065&-24.901$\pm$0.068&6.770$\pm$0.437&42\\
TYC3309-1348-1&42.21890&52.47538&$\cdots$&28.137$\pm$0.914&-24.381$\pm$0.668&6.647$\pm$0.347&42\\
HIP13488&43.41998&53.80268&A2&28.509$\pm$0.070&-26.244$\pm$0.051&6.812$\pm$0.250&42\\
\hline
\hline
TYC661-692-1&57.71066&11.00143&F8&23.839$\pm$0.716&-23.964$\pm$0.409&6.476$\pm$0.241&43\\
HIP18033&57.81623&13.04598&B9II-III&23.194$\pm$0.030&-22.986$\pm$0.013&6.483$\pm$0.546&43\\
TYC664-136-1&57.91530&14.79663&$\cdots$&25.105$\pm$0.741&-24.219$\pm$0.428&6.280$\pm$0.252&43\\
HIP18778&60.33910&9.33362&F8&25.241$\pm$0.122&-26.186$\pm$0.068&6.548$\pm$0.256&43\\
TYC662-820-1&59.08010&11.41968&$\cdots$&25.069$\pm$1.335&-24.482$\pm$0.556&6.660$\pm$0.412&43\\
TYC662-217-1&59.92566&12.16893&$\cdots$&23.671$\pm$0.693&-24.894$\pm$0.424&6.824$\pm$0.317&43\\
\hline
\hline
TYC3698-121-1&33.52886&59.73304&A0V&15.448$\pm$1.083&-13.620$\pm$0.420&2.882$\pm$0.241&47\\
HIP9795&31.49397&60.27831&G2II&16.715$\pm$0.096&-13.524$\pm$0.088&2.914$\pm$0.231&47\\
TYC3698-1347-1&33.38928&59.42791&$\cdots$&15.654$\pm$0.960&-13.609$\pm$0.345&2.873$\pm$0.232&47\\
TYC3698-501-1&33.37689&59.45773&$\cdots$&16.153$\pm$0.827&-13.464$\pm$0.589&2.954$\pm$0.282&47\\
TYC3698-1731-1&33.82643&59.81540&$\cdots$&15.228$\pm$1.096&-13.427$\pm$0.462&2.887$\pm$0.223&47\\
TYC4033-2479-1&32.66574&60.07869&K0&15.567$\pm$0.965&-13.498$\pm$0.524&2.836$\pm$0.238&47\\
\hline
\hline
HIP43909&134.17590&-63.04840&B8/9Vn&-27.470$\pm$0.028&14.054$\pm$0.027&4.914$\pm$0.335&49\\
HIP43135&131.79257&-63.81253&A0V&-26.131$\pm$0.054&15.481$\pm$0.040&5.154$\pm$0.263&49\\
TYC8930-1190-1&132.04977&-63.35331&$\cdots$&-26.729$\pm$0.851&15.076$\pm$0.682&4.987$\pm$0.248&49\\
TYC8930-2088-1&132.27606&-63.06794&$\cdots$&-27.764$\pm$1.167&15.406$\pm$1.033&5.327$\pm$0.413&49\\
TYC8931-646-1&135.02120&-63.28649&$\cdots$&-26.964$\pm$1.219&12.878$\pm$1.097&5.076$\pm$0.264&49\\
TYC8930-1213-1&131.06627&-62.51399&$\cdots$&-25.408$\pm$0.911&15.105$\pm$0.805&5.184$\pm$0.294&49\\
\hline
\hline
TYC5314-259-1&68.38328&-7.97834&$\cdots$&-4.724$\pm$1.069&-2.278$\pm$0.815&6.333$\pm$0.229&51\\
TYC4746-535-1&68.21826&-5.72537&F3V&-3.875$\pm$0.516&-2.017$\pm$0.713&6.516$\pm$0.639&51\\
TYC4743-981-1&69.40904&-5.51202&A9V&-4.823$\pm$0.500&-2.146$\pm$0.455&6.487$\pm$0.321&51\\
TYC5317-617-1&68.64190&-10.59108&F3V&-4.531$\pm$0.479&-0.393$\pm$0.471&6.669$\pm$0.309&51\\
HIP21484&69.20860&-8.46049&B9V&-4.241$\pm$0.032&-1.386$\pm$0.025&6.540$\pm$0.449&51\\
\hline
\hline
HIP16609&53.44380&8.29048&A5&26.983$\pm$0.071&-23.388$\pm$0.043&6.916$\pm$0.264&52\\
TYC72-816-1&58.44870&5.70640&$\cdots$&28.956$\pm$0.742&-25.953$\pm$0.467&7.503$\pm$0.263&52\\
HIP17512&56.24565&8.31947&G5&26.671$\pm$0.097&-24.288$\pm$0.064&7.111$\pm$0.440&52\\
TYC658-828-1&56.46727&8.54080&$\cdots$&28.188$\pm$0.826&-25.445$\pm$0.488&7.533$\pm$0.279&52\\
HIP17907&57.44383&9.40739&B9&25.379$\pm$0.037&-24.417$\pm$0.018&7.162$\pm$0.509&52\\
\hline
\hline
TYC8180-844-1&145.55105&-51.05258&$\cdots$&-19.043$\pm$1.318&8.059$\pm$0.775&8.162$\pm$0.302&53\\
HIP48338&147.79501&-53.18296&F0/2IV&-19.648$\pm$0.059&6.854$\pm$0.057&7.817$\pm$0.258&53\\
HIP46740&142.89615&-51.25188&A4/5IV/V&-21.168$\pm$0.047&9.718$\pm$0.051&8.252$\pm$0.242&53\\
TYC8175-288-1&140.82228&-50.22978&$\cdots$&-19.665$\pm$1.131&10.253$\pm$1.107&8.116$\pm$0.380&53\\
TYC8584-2682-1&143.10856&-52.62763&$\cdots$&-17.960$\pm$1.958&8.918$\pm$0.916&8.057$\pm$0.347&53\\
\hline
\hline
TYC8534-396-1&94.22929&-52.87385&$\cdots$&1.378$\pm$0.731&8.411$\pm$0.659&8.348$\pm$0.232&56\\
TYC8542-1617-1&96.52613&-56.52116&G0&1.148$\pm$0.820&12.801$\pm$0.726&8.437$\pm$0.269&56\\
HIP30685&96.72389&-53.58192&F0V&1.288$\pm$0.075&11.312$\pm$0.083&8.927$\pm$0.227&56\\
TYC8534-211-1&95.47952&-52.73230&$\cdots$&1.949$\pm$0.922&9.413$\pm$0.690&8.532$\pm$0.244&56\\
TYC8115-252-1&96.89736&-50.77373&$\cdots$&2.073$\pm$1.563&8.170$\pm$0.624&8.196$\pm$0.280&56\\
\hline
\hline
HIP34706&107.82022&-70.11906&K0III&-20.228$\pm$0.077&90.907$\pm$0.066&5.774$\pm$0.213&58\\
TYC9183-1267-1&110.73158&-69.43246&$\cdots$&-23.350$\pm$0.766&93.447$\pm$0.706&5.836$\pm$0.237&58\\
TYC8922-1022-1&113.48855&-66.31245&$\cdots$&-25.490$\pm$0.840&82.590$\pm$0.797&5.844$\pm$0.316&58\\
TYC8919-2129-1&117.55586&-65.35849&$\cdots$&-36.701$\pm$0.797&87.584$\pm$0.624&5.862$\pm$0.318&58\\
TYC8918-990-1&113.69697&-63.95839&$\cdots$&-26.029$\pm$1.078&81.378$\pm$0.866&5.914$\pm$0.378&58\\
\hline
\hline
HIP105282&319.86989&49.51030&B6V&14.414$\pm$0.025&2.174$\pm$0.024&6.276$\pm$0.513&59\\
HIP103196&313.60852&48.92983&AM&13.083$\pm$0.032&2.685$\pm$0.030&6.263$\pm$0.297&59\\
TYC3579-1214-1&313.08423&48.72394&$\cdots$&12.447$\pm$1.589&3.931$\pm$0.859&6.251$\pm$0.274&59\\
HIP103658&315.02763&48.67945&B9P&13.870$\pm$0.031&3.278$\pm$0.038&6.216$\pm$0.302&59\\
TYC3592-755-1&315.13891&48.71559&F2&14.353$\pm$0.890&2.638$\pm$0.817&6.207$\pm$0.266&59\\
\hline
\hline
TYC81-1439-1&67.13638&6.09780&A3&19.980$\pm$0.663&-21.486$\pm$0.487&6.215$\pm$0.553&60\\
TYC81-986-1&65.60088&6.52921&A3&19.850$\pm$0.643&-21.099$\pm$0.542&6.263$\pm$0.320&60\\
TYC80-202-1&63.96299&7.11771&$\cdots$&23.279$\pm$0.975&-25.185$\pm$0.566&6.134$\pm$0.268&60\\
TYC668-737-1&65.35161&8.89843&$\cdots$&21.998$\pm$1.326&-24.095$\pm$0.707&6.166$\pm$0.378&60\\
HIP20425&65.63759&5.69412&F5&20.631$\pm$0.107&-21.217$\pm$0.081&6.377$\pm$0.265&60\\
\hline
\hline
HIP35588&110.16249&-52.19871&F3/5V&-9.239$\pm$0.030&0.645$\pm$0.030&6.734$\pm$0.391&61\\
TYC8131-363-1&109.54213&-52.03277&A1V&-9.722$\pm$0.514&1.429$\pm$0.476&6.234$\pm$0.254&61\\
TYC8131-1097-1&109.21415&-51.59885&F2V&-8.141$\pm$0.863&0.824$\pm$0.610&6.326$\pm$0.260&61\\
TYC8127-223-1&109.63125&-50.21998&$\cdots$&-9.479$\pm$0.876&1.219$\pm$0.535&6.642$\pm$0.220&61\\
TYC8128-923-1&111.16376&-49.76533&F0IV/V&-9.642$\pm$1.025&0.925$\pm$0.675&6.469$\pm$0.251&61\\
\enddata 
\tablecomments{Collections of stars in Groups with 5 -9 connected components from the Oh17 sample that appear to be newly discovered. }
\end{deluxetable*} 

\bibliographystyle{yahapj}
\bibliography{maincopy}

\begin{thebibliography}{76}
\expandafter\ifx\csname natexlab\endcsname\relax\def\natexlab#1{#1}\fi

\bibitem[{{Aitken}(1895)}]{Aitken95}
{Aitken}, R.~G. 1895, \href{http://dx.doi.org/10.1086/121036}{\pasp, 7, 305}

\bibitem[{{Andrews} {et~al.}(2017){Andrews}, {Chanam{\'e}}, \&
  {Ag{\"u}eros}}]{Andrews17}
{Andrews}, J.~J., {Chanam{\'e}}, J., \& {Ag{\"u}eros}, M.~A. 2017,
  \href{http://dx.doi.org/10.1093/mnras/stx2000}{\mnras, 472, 675}

\bibitem[{{Andrews} {et~al.}(2018){Andrews}, {Chanam{\'e}}, \&
  {Ag{\"u}eros}}]{Andrews18}
---. 2018, \href{http://dx.doi.org/10.1093/mnras/stx2685}{\mnras, 473, 5393}

\bibitem[{{Barrado y Navascu{\'e}s} {et~al.}(2004){Barrado y Navascu{\'e}s},
  {Stauffer}, \& {Jayawardhana}}]{Barrado2004}
{Barrado y Navascu{\'e}s}, D., {Stauffer}, J.~R., \& {Jayawardhana}, R. 2004,
  \href{http://dx.doi.org/10.1086/423485}{\apj, 614, 386}

\bibitem[{{Bell} {et~al.}(2015){Bell}, {Mamajek}, \& {Naylor}}]{Bell15aa}
{Bell}, C.~P.~M., {Mamajek}, E.~E., \& {Naylor}, T. 2015,
  \href{http://dx.doi.org/10.1093/mnras/stv1981}{\mnras, 454, 593}

\bibitem[{{Bochanski} {et~al.}(2018){Bochanski}, {Faherty}, {Gagn{\'e}},
  {Nelson}, {Coker}, {Smithka}, {Desir}, \& {Vasquez}}]{Bochanski18aa}
{Bochanski}, J.~J., {Faherty}, J.~K., {Gagn{\'e}}, J., {et~al.} 2018, ArXiv
  e-prints, \href{http://arxiv.org/abs/1801.00537}{{\sffamily arXiv:1801.00537
  [astro-ph.SR]}}

\bibitem[{{Brandt} \& {Huang}(2015{\natexlab{a}})}]{Brandt15aa}
{Brandt}, T.~D., \& {Huang}, C.~X. 2015{\natexlab{a}},
  \href{http://dx.doi.org/10.1088/0004-637X/807/1/58}{\apj, 807, 58}

\bibitem[{{Brandt} \& {Huang}(2015{\natexlab{b}})}]{Brandt15}
---. 2015{\natexlab{b}},
  \href{http://dx.doi.org/10.1088/0004-637X/807/1/24}{\apj, 807, 24}

\bibitem[{{Chanam{\'e}} \& {Ram{\'{\i}}rez}(2012)}]{Chaname12}
{Chanam{\'e}}, J., \& {Ram{\'{\i}}rez}, I. 2012,
  \href{http://dx.doi.org/10.1088/0004-637X/746/1/102}{\apj, 746, 102}

\bibitem[{{Choi} {et~al.}(2016){Choi}, {Dotter}, {Conroy}, {Cantiello},
  {Paxton}, \& {Johnson}}]{Choi2016}
{Choi}, J., {Dotter}, A., {Conroy}, C., {et~al.} 2016,
  \href{http://dx.doi.org/10.3847/0004-637X/823/2/102}{\apj, 823, 102}

\bibitem[{{Daemgen} {et~al.}(2015){Daemgen}, {Bonavita}, {Jayawardhana},
  {Lafreni{\`e}re}, \& {Janson}}]{Daemgen15aa}
{Daemgen}, S., {Bonavita}, M., {Jayawardhana}, R., {Lafreni{\`e}re}, D., \&
  {Janson}, M. 2015, \href{http://dx.doi.org/10.1088/0004-637X/799/2/155}{\apj,
  799, 155}

\bibitem[{{Dahm}(2015{\natexlab{a}})}]{Dahm15aa}
{Dahm}, S.~E. 2015{\natexlab{a}},
  \href{http://dx.doi.org/10.1088/0004-637X/813/2/108}{\apj, 813, 108}

\bibitem[{{Dahm}(2015{\natexlab{b}})}]{Dahm15}
---. 2015{\natexlab{b}},
  \href{http://dx.doi.org/10.1088/0004-637X/813/2/108}{\apj, 813, 108}

\bibitem[{{Dhital} {et~al.}(2010){Dhital}, {West}, {Stassun}, \&
  {Bochanski}}]{Dhital10}
{Dhital}, S., {West}, A.~A., {Stassun}, K.~G., \& {Bochanski}, J.~J. 2010,
  \href{http://dx.doi.org/10.1088/0004-6256/139/6/2566}{\aj, 139, 2566}

\bibitem[{{Dhital} {et~al.}(2015){Dhital}, {West}, {Stassun}, {Schluns}, \&
  {Massey}}]{Dhital15}
{Dhital}, S., {West}, A.~A., {Stassun}, K.~G., {Schluns}, K.~J., \& {Massey},
  A.~P. 2015, \href{http://dx.doi.org/10.1088/0004-6256/150/2/57}{\aj, 150, 57}

\bibitem[{{Dobbie} {et~al.}(2010){Dobbie}, {Lodieu}, \& {Sharp}}]{Dobbie10}
{Dobbie}, P.~D., {Lodieu}, N., \& {Sharp}, R.~G. 2010,
  \href{http://dx.doi.org/10.1111/j.1365-2966.2010.17355.x}{\mnras, 409, 1002}

\bibitem[{{Dotter}(2016)}]{Dotter2016}
{Dotter}, A. 2016, \href{http://dx.doi.org/10.3847/0067-0049/222/1/8}{\apjs,
  222, 8}

\bibitem[{{Eggen}(1965)}]{Eggen65aa}
{Eggen}, O.~J. 1965, {Moving Groups of Stars}, ed. A.~{Blaauw} \& M.~{Schmidt}
  (the University of Chicago Press), 111

\bibitem[{{Faherty}(in prep)}]{Faherty18aa}
{Faherty}, J.~K. in prep, ApJ

\bibitem[{{Faherty} {et~al.}(2011){Faherty}, {Burgasser}, {Bochanski},
  {Looper}, {West}, \& {van der Bliek}}]{Faherty11}
{Faherty}, J.~K., {Burgasser}, A.~J., {Bochanski}, J.~J., {et~al.} 2011,
  \href{http://dx.doi.org/10.1088/0004-6256/141/3/71}{\aj, 141, 71}

\bibitem[{{Faherty} {et~al.}(2010){Faherty}, {Burgasser}, {West}, {Bochanski},
  {Cruz}, {Shara}, \& {Walter}}]{Faherty10}
{Faherty}, J.~K., {Burgasser}, A.~J., {West}, A.~A., {et~al.} 2010,
  \href{http://dx.doi.org/10.1088/0004-6256/139/1/176}{\aj, 139, 176}

\bibitem[{{Faherty} {et~al.}(2016){Faherty}, {Riedel}, {Cruz}, {Gagne},
  {Filippazzo}, {Lambrides}, {Fica}, {Weinberger}, {Thorstensen}, {Tinney},
  {Baldassare}, {Lemonier}, \& {Rice}}]{Faherty16aa}
{Faherty}, J.~K., {Riedel}, A.~R., {Cruz}, K.~L., {et~al.} 2016,
  \href{http://dx.doi.org/10.3847/0067-0049/225/1/10}{\apjs, 225, 10}

\bibitem[{{Gagn{\'e}} {et~al.}(2018{\natexlab{a}}){Gagn{\'e}}, {Roy-Loubier},
  {Faherty}, {Doyon}, \& {Malo}}]{Gagne2018C}
{Gagn{\'e}}, J., {Roy-Loubier}, O., {Faherty}, J.~K., {Doyon}, R., \& {Malo},
  L. 2018{\natexlab{a}}, ArXiv e-prints,
  \href{http://arxiv.org/abs/1804.03093}{{\sffamily arXiv:1804.03093
  [astro-ph.SR]}}

\bibitem[{{Gagn{\'e}} {et~al.}(2015){Gagn{\'e}}, {Faherty}, {Cruz},
  {Lafreni{\'e}re}, {Doyon}, {Malo}, {Burgasser}, {Naud}, {Artigau},
  {Bouchard}, {Gizis}, \& {Albert}}]{Gagne15aa}
{Gagn{\'e}}, J., {Faherty}, J.~K., {Cruz}, K.~L., {et~al.} 2015,
  \href{http://dx.doi.org/10.1088/0067-0049/219/2/33}{\apjs, 219, 33}

\bibitem[{{Gagn{\'e}} {et~al.}(2017){Gagn{\'e}}, {Faherty}, {Mamajek}, {Malo},
  {Doyon}, {Filippazzo}, {Weinberger}, {Donaldson}, {L{\'e}pine},
  {Lafreni{\`e}re}, {Artigau}, {Burgasser}, {Looper}, {Boucher}, {Beletsky},
  {Camnasio}, {Brunette}, \& {Arboit}}]{Gagne17aa}
{Gagn{\'e}}, J., {Faherty}, J.~K., {Mamajek}, E.~E., {et~al.} 2017,
  \href{http://dx.doi.org/10.3847/1538-4365/228/2/18}{\apjs, 228, 18}

\bibitem[{{Gagn{\'e}} {et~al.}(2018{\natexlab{b}}){Gagn{\'e}}, {Mamajek},
  {Malo}, {Riedel}, {Rodriguez}, {Lafreni{\`e}re}, {Faherty}, {Roy-Loubier},
  {Pueyo}, {Robin}, \& {Doyon}}]{Gagne18}
{Gagn{\'e}}, J., {Mamajek}, E.~E., {Malo}, L., {et~al.} 2018{\natexlab{b}},
  ArXiv e-prints, \href{http://arxiv.org/abs/1801.09051}{{\sffamily
  arXiv:1801.09051 [astro-ph.SR]}}

\bibitem[{{Gaia Collaboration} {et~al.}(2017){Gaia Collaboration}, {van Leeuwen
  F.}, {Vallenari}, {Jordi}, {Lindegren}, {Bastian}, {Prusti}, {de Bruijne},
  {Brown}, {Babusiaux}, \& et~al.}]{Gaia-Collaboration17}
{Gaia Collaboration}, {van Leeuwen F.}, {Vallenari}, A., {et~al.} 2017, VizieR
  Online Data Catalog, 360

\bibitem[{{Gennaro} {et~al.}(2012){Gennaro}, {Prada Moroni}, \&
  {Tognelli}}]{Gennaro2012}
{Gennaro}, M., {Prada Moroni}, P.~G., \& {Tognelli}, E. 2012,
  \href{http://dx.doi.org/10.1111/j.1365-2966.2011.19945.x}{\mnras, 420, 986}

\bibitem[{{H{\o}g} {et~al.}(2000){H{\o}g}, {Fabricius}, {Makarov}, {Urban},
  {Corbin}, {Wycoff}, {Bastian}, {Schwekendiek}, \& {Wicenec}}]{Hog00}
{H{\o}g}, E., {Fabricius}, C., {Makarov}, V.~V., {et~al.} 2000, \aap, 355, L27

\bibitem[{{Jones} {et~al.}(2015){Jones}, {White}, {Boyajian}, {Schaefer},
  {Baines}, {Ireland}, {Patience}, {ten Brummelaar}, {McAlister}, {Ridgway},
  {Sturmann}, {Sturmann}, {Turner}, {Farrington}, \& {Goldfinger}}]{Jones15}
{Jones}, J., {White}, R.~J., {Boyajian}, T., {et~al.} 2015,
  \href{http://dx.doi.org/10.1088/0004-637X/813/1/58}{\apj, 813, 58}

\bibitem[{Kapteyn(1905)}]{Kapteyn05aa}
Kapteyn, J.~C. 1905, British Assoc. Adv. Sci. Rep., 257

\bibitem[{{Kenyon} \& {Hartmann}(1995)}]{Kenyon1995}
{Kenyon}, S.~J., \& {Hartmann}, L. 1995,
  \href{http://dx.doi.org/10.1086/192235}{\apjs, 101, 117}

\bibitem[{{Kirkpatrick} {et~al.}(2001){Kirkpatrick}, {Dahn}, {Monet}, {Reid},
  {Gizis}, {Liebert}, \& {Burgasser}}]{Kirkpatrick01}
{Kirkpatrick}, J.~D., {Dahn}, C.~C., {Monet}, D.~G., {et~al.} 2001,
  \href{http://dx.doi.org/10.1086/321085}{\aj, 121, 3235}

\bibitem[{{Kushniruk} {et~al.}(2017){Kushniruk}, {Schirmer}, \&
  {Bensby}}]{Kushniruk2017}
{Kushniruk}, I., {Schirmer}, T., \& {Bensby}, T. 2017,
  \href{http://dx.doi.org/10.1051/0004-6361/201731147}{\aap, 608, A73}

\bibitem[{{Latyshev}(1977)}]{Latyshev77}
{Latyshev}, I.~N. 1977, Astronomicheskij Tsirkulyar, 969, 7

\bibitem[{{L{\'e}pine}(2005)}]{Lepine05a}
{L{\'e}pine}, S. 2005, \href{http://dx.doi.org/10.1086/432161}{\aj, 130, 1247}

\bibitem[{{L{\'e}pine} \& {Bongiorno}(2007)}]{Lepine07}
{L{\'e}pine}, S., \& {Bongiorno}, B. 2007,
  \href{http://dx.doi.org/10.1086/510333}{\aj, 133, 889}

\bibitem[{{L{\'e}pine} \& {Shara}(2005)}]{Lepine05}
{L{\'e}pine}, S., \& {Shara}, M.~M. 2005,
  \href{http://dx.doi.org/10.1086/427854}{\aj, 129, 1483}

\bibitem[{{Lindegren} {et~al.}(2016){Lindegren}, {Lammers}, {Bastian},
  {Hern{\'a}ndez}, {Klioner}, {Hobbs}, {Bombrun}, {Michalik}, {Ramos-Lerate},
  {Butkevich}, {Comoretto}, {Joliet}, {Holl}, {Hutton}, {Parsons},
  {Steidelm{\"u}ller}, {Abbas}, {Altmann}, {Andrei}, {Anton}, {Bach},
  {Barache}, {Becciani}, {Berthier}, {Bianchi}, {Biermann}, {Bouquillon},
  {Bourda}, {Br{\"u}semeister}, {Bucciarelli}, {Busonero}, {Carlucci},
  {Casta{\~n}eda}, {Charlot}, {Clotet}, {Crosta}, {Davidson}, {de Felice},
  {Drimmel}, {Fabricius}, {Fienga}, {Figueras}, {Fraile}, {Gai}, {Garralda},
  {Geyer}, {Gonz{\'a}lez-Vidal}, {Guerra}, {Hambly}, {Hauser}, {Jordan},
  {Lattanzi}, {Lenhardt}, {Liao}, {L{\"o}ffler}, {McMillan}, {Mignard}, {Mora},
  {Morbidelli}, {Portell}, {Riva}, {Sarasso}, {Serraller}, {Siddiqui}, {Smart},
  {Spagna}, {Stampa}, {Steele}, {Taris}, {Torra}, {van Reeven}, {Vecchiato},
  {Zschocke}, {de Bruijne}, {Gracia}, {Raison}, {Lister}, {Marchant},
  {Messineo}, {Soffel}, {Osorio}, {de Torres}, \& {O'Mullane}}]{Lindegren16}
{Lindegren}, L., {Lammers}, U., {Bastian}, U., {et~al.} 2016,
  \href{http://dx.doi.org/10.1051/0004-6361/201628714}{\aap, 595, A4}

\bibitem[{Mamajek(2016)}]{Mamajek2016}
Mamajek, E. 2016, A New Candidate Young Stellar Group at d=121 pc Associated
  with 118 Tauri, Figshare

\bibitem[{{Martin} {et~al.}(2005){Martin}, {Fanson}, {Schiminovich},
  {Morrissey}, {Friedman}, {Barlow}, {Conrow}, {Grange}, {Jelinsky},
  {Milliard}, {Siegmund}, {Bianchi}, {Byun}, {Donas}, {Forster}, {Heckman},
  {Lee}, {Madore}, {Malina}, {Neff}, {Rich}, {Small}, {Surber}, {Szalay},
  {Welsh}, \& {Wyder}}]{Martin05}
{Martin}, D.~C., {Fanson}, J., {Schiminovich}, D., {et~al.} 2005,
  \href{http://dx.doi.org/10.1086/426387}{\apjl, 619, L1}

\bibitem[{{Mason} {et~al.}(2018){Mason}, {Wycoff}, {Hartkopf}, {Douglass}, \&
  {Worley}}]{Mason18}
{Mason}, B.~D., {Wycoff}, G.~L., {Hartkopf}, W.~I., {Douglass}, G.~G., \&
  {Worley}, C.~E. 2018, VizieR Online Data Catalog, 1

\bibitem[{{Morton}(2015)}]{Morton2015}
{Morton}, T.~D. 2015, {isochrones: Stellar model grid package}, Astrophysics
  Source Code Library, \href{http://arxiv.org/abs/1503.010}{{\sffamily
  ascl:1503.010}}

\bibitem[{{Murphy} \& {Lawson}(2015)}]{Murphy15}
{Murphy}, S.~J., \& {Lawson}, W.~A. 2015,
  \href{http://dx.doi.org/10.1093/mnras/stu2450}{\mnras, 447, 1267}

\bibitem[{{Murphy} {et~al.}(2013){Murphy}, {Lawson}, \& {Bessell}}]{Murphy13}
{Murphy}, S.~J., {Lawson}, W.~A., \& {Bessell}, M.~S. 2013,
  \href{http://dx.doi.org/10.1093/mnras/stt1375}{\mnras, 435, 1325}

\bibitem[{{N{\'u}{\~n}ez} \& {Ag{\"u}eros}(2016)}]{Nunez16}
{N{\'u}{\~n}ez}, A., \& {Ag{\"u}eros}, M.~A. 2016,
  \href{http://dx.doi.org/10.3847/0004-637X/830/1/44}{\apj, 830, 44}

\bibitem[{{Ochsenbein} {et~al.}(2000){Ochsenbein}, {Bauer}, \&
  {Marcout}}]{Ochsenbein00}
{Ochsenbein}, F., {Bauer}, P., \& {Marcout}, J. 2000,
  \href{http://dx.doi.org/10.1051/aas:2000169}{\aaps, 143, 23}

\bibitem[{{Oelkers} {et~al.}(2017){Oelkers}, {Stassun}, \&
  {Dhital}}]{Oelkers17}
{Oelkers}, R.~J., {Stassun}, K.~G., \& {Dhital}, S. 2017,
  \href{http://dx.doi.org/10.3847/1538-3881/aa6d55}{\aj, 153, 259}

\bibitem[{{Oh} {et~al.}(2017{\natexlab{a}}){Oh}, {Price-Whelan}, {Brewer},
  {Hogg}, {Spergel}, \& {Myles}}]{Oh17a}
{Oh}, S., {Price-Whelan}, A.~M., {Brewer}, J.~M., {et~al.} 2017{\natexlab{a}},
  ArXiv e-prints, \href{http://arxiv.org/abs/1709.05344}{{\sffamily
  arXiv:1709.05344 [astro-ph.SR]}}

\bibitem[{{Oh} {et~al.}(2017{\natexlab{b}}){Oh}, {Price-Whelan}, {Hogg},
  {Morton}, \& {Spergel}}]{Oh17}
{Oh}, S., {Price-Whelan}, A.~M., {Hogg}, D.~W., {Morton}, T.~D., \& {Spergel},
  D.~N. 2017{\natexlab{b}},
  \href{http://dx.doi.org/10.3847/1538-3881/aa6ffd}{\aj, 153, 257}

\bibitem[{{Paxton} {et~al.}(2011){Paxton}, {Bildsten}, {Dotter}, {Herwig},
  {Lesaffre}, \& {Timmes}}]{Paxton2011}
{Paxton}, B., {Bildsten}, L., {Dotter}, A., {et~al.} 2011,
  \href{http://dx.doi.org/10.1088/0067-0049/192/1/3}{\apjs, 192, 3}

\bibitem[{{Paxton} {et~al.}(2013){Paxton}, {Cantiello}, {Arras}, {Bildsten},
  {Brown}, {Dotter}, {Mankovich}, {Montgomery}, {Stello}, {Timmes}, \&
  {Townsend}}]{Paxton2013}
{Paxton}, B., {Cantiello}, M., {Arras}, P., {et~al.} 2013,
  \href{http://dx.doi.org/10.1088/0067-0049/208/1/4}{\apjs, 208, 4}

\bibitem[{{Paxton} {et~al.}(2015){Paxton}, {Marchant}, {Schwab}, {Bauer},
  {Bildsten}, {Cantiello}, {Dessart}, {Farmer}, {Hu}, {Langer}, {Townsend},
  {Townsley}, \& {Timmes}}]{Paxton2015}
{Paxton}, B., {Marchant}, P., {Schwab}, J., {et~al.} 2015,
  \href{http://dx.doi.org/10.1088/0067-0049/220/1/15}{\apjs, 220, 15}

\bibitem[{{Pecaut} \& {Mamajek}(2016)}]{Pecaut16}
{Pecaut}, M.~J., \& {Mamajek}, E.~E. 2016,
  \href{http://dx.doi.org/10.1093/mnras/stw1300}{\mnras, 461, 794}

\bibitem[{{Perryman} {et~al.}(1997){Perryman}, {Lindegren}, {Kovalevsky},
  {Hoeg}, {Bastian}, {Bernacca}, {Cr{\'e}z{\'e}}, {Donati}, {Grenon},
  {Grewing}, {van Leeuwen}, {van der Marel}, {Mignard}, {Murray}, {Le Poole},
  {Schrijver}, {Turon}, {Arenou}, {Froeschl{\'e}}, \&
  {Petersen}}]{Perryman97aa}
{Perryman}, M.~A.~C., {Lindegren}, L., {Kovalevsky}, J., {et~al.} 1997, \aap,
  323, L49

\bibitem[{{Platais} {et~al.}(1998){Platais}, {Kozhurina-Platais}, \& {van
  Leeuwen}}]{Platais98}
{Platais}, I., {Kozhurina-Platais}, V., \& {van Leeuwen}, F. 1998,
  \href{http://dx.doi.org/10.1086/300606}{\aj, 116, 2423}

\bibitem[{{P{\"o}hnl} \& {Paunzen}(2010)}]{Pohnl10}
{P{\"o}hnl}, H., \& {Paunzen}, E. 2010,
  \href{http://dx.doi.org/10.1051/0004-6361/200810855}{\aap, 514, A81}

\bibitem[{{Preibisch} \& {Feigelson}(2005)}]{Preibisch05}
{Preibisch}, T., \& {Feigelson}, E.~D. 2005,
  \href{http://dx.doi.org/10.1086/432094}{\apjs, 160, 390}

\bibitem[{{Price-Whelan} {et~al.}(2017){Price-Whelan}, {Oh}, \&
  {Spergel}}]{Price-Whelan2017}
{Price-Whelan}, A.~M., {Oh}, S., \& {Spergel}, D.~N. 2017, ArXiv e-prints,
  \href{http://arxiv.org/abs/1709.03532}{{\sffamily arXiv:1709.03532
  [astro-ph.SR]}}

\bibitem[{{Rodriguez} {et~al.}(2013){Rodriguez}, {Zuckerman}, {Kastner},
  {Bessell}, {Faherty}, \& {Murphy}}]{Rodriguez13}
{Rodriguez}, D.~R., {Zuckerman}, B., {Kastner}, J.~H., {et~al.} 2013,
  \href{http://dx.doi.org/10.1088/0004-637X/774/2/101}{\apj, 774, 101}

\bibitem[{{R{\"o}ser} {et~al.}(2016){R{\"o}ser}, {Schilbach}, \&
  {Goldman}}]{Roser16}
{R{\"o}ser}, S., {Schilbach}, E., \& {Goldman}, B. 2016,
  \href{http://dx.doi.org/10.1051/0004-6361/201629158}{\aap, 595, A22}

\bibitem[{{R{\"o}ser} {et~al.}(2017){R{\"o}ser}, {Schilbach}, {Goldman},
  {Henning}, {Moor}, \& {Derekas}}]{Roser17}
{R{\"o}ser}, S., {Schilbach}, E., {Goldman}, B., {et~al.} 2017, ArXiv e-prints,
  \href{http://arxiv.org/abs/1712.10143}{{\sffamily arXiv:1712.10143
  [astro-ph.SR]}}

\bibitem[{{Shaya} \& {Olling}(2011)}]{Shaya11}
{Shaya}, E.~J., \& {Olling}, R.~P. 2011,
  \href{http://dx.doi.org/10.1088/0067-0049/192/1/2}{\apjs, 192, 2}

\bibitem[{{Shkolnik} {et~al.}(2011){Shkolnik}, {Liu}, {Reid}, {Dupuy}, \&
  {Weinberger}}]{Shkolnik11}
{Shkolnik}, E.~L., {Liu}, M.~C., {Reid}, I.~N., {Dupuy}, T., \& {Weinberger},
  A.~J. 2011, \href{http://dx.doi.org/10.1088/0004-637X/727/1/6}{\apj, 727, 6}

\bibitem[{{Silaj} \& {Landstreet}(2014)}]{Silaj14}
{Silaj}, J., \& {Landstreet}, J.~D. 2014,
  \href{http://dx.doi.org/10.1051/0004-6361/201321468}{\aap, 566, A132}

\bibitem[{{Skiff}(2014)}]{Skiff14}
{Skiff}, B.~A. 2014, VizieR Online Data Catalog, 1

\bibitem[{{Skrutskie} {et~al.}(2006){Skrutskie}, {Cutri}, {Stiening},
  {Weinberg}, {Schneider}, {Carpenter}, {Beichman}, {Capps}, {Chester},
  {Elias}, {Huchra}, {Liebert}, {Lonsdale}, {Monet}, {Price}, {Seitzer},
  {Jarrett}, {Kirkpatrick}, {Gizis}, {Howard}, {Evans}, {Fowler}, {Fullmer},
  {Hurt}, {Light}, {Kopan}, {Marsh}, {McCallon}, {Tam}, {Van Dyk}, \&
  {Wheelock}}]{Skrutskie06}
{Skrutskie}, M.~F., {Cutri}, R.~M., {Stiening}, R., {et~al.} 2006,
  \href{http://dx.doi.org/10.1086/498708}{\aj, 131, 1163}

\bibitem[{{Taylor}(2005)}]{Taylor05}
{Taylor}, M.~B. 2005, in Astronomical Society of the Pacific Conference Series,
  Vol. 347, Astronomical Data Analysis Software and Systems XIV, ed.
  P.~{Shopbell}, M.~{Britton}, \& R.~{Ebert}, 29

\bibitem[{{Teske} {et~al.}(2015){Teske}, {Ghezzi}, {Cunha}, {Smith}, {Schuler},
  \& {Bergemann}}]{Teske15}
{Teske}, J.~K., {Ghezzi}, L., {Cunha}, K., {et~al.} 2015,
  \href{http://dx.doi.org/10.1088/2041-8205/801/1/L10}{\apjl, 801, L10}

\bibitem[{{Voges} {et~al.}(1999){Voges}, {Aschenbach}, {Boller},
  {Br{\"a}uninger}, {Briel}, {Burkert}, {Dennerl}, {Englhauser}, {Gruber},
  {Haberl}, {Hartner}, {Hasinger}, {K{\"u}rster}, {Pfeffermann}, {Pietsch},
  {Predehl}, {Rosso}, {Schmitt}, {Tr{\"u}mper}, \& {Zimmermann}}]{Voges99}
{Voges}, W., {Aschenbach}, B., {Boller}, T., {et~al.} 1999, \aap, 349, 389

\bibitem[{{Voges} {et~al.}(2000){Voges}, {Aschenbach}, {Boller}, {Brauninger},
  {Briel}, {Burkert}, {Dennerl}, {Englhauser}, {Gruber}, {Haberl}, {Hartner},
  {Hasinger}, {Pfeffermann}, {Pietsch}, {Predehl}, {Schmitt}, {Trumper}, \&
  {Zimmermann}}]{Voges00}
---. 2000, \iaucirc, 7432

\bibitem[{{Weinberg} {et~al.}(1987){Weinberg}, {Shapiro}, \&
  {Wasserman}}]{Weinberg87}
{Weinberg}, M.~D., {Shapiro}, S.~L., \& {Wasserman}, I. 1987,
  \href{http://dx.doi.org/10.1086/164883}{\apj, 312, 367}

\bibitem[{{Wilking} {et~al.}(2008{\natexlab{a}}){Wilking}, {Gagn{\'e}}, \&
  {Allen}}]{Wilking08aa}
{Wilking}, B.~A., {Gagn{\'e}}, M., \& {Allen}, L.~E. 2008{\natexlab{a}}, {Star
  Formation in the {$\rho$} Ophiuchi Molecular Cloud}, ed. B.~{Reipurth}, 351

\bibitem[{{Wilking} {et~al.}(2008{\natexlab{b}}){Wilking}, {Gagn{\'e}}, \&
  {Allen}}]{Wilking2008}
---. 2008{\natexlab{b}}, {Star Formation in the {$\rho$} Ophiuchi Molecular
  Cloud}, ed. B.~{Reipurth}, 351

\bibitem[{{Wright} {et~al.}(2010){Wright}, {Eisenhardt}, {Mainzer}, {Ressler},
  {Cutri}, {Jarrett}, {Kirkpatrick}, {Padgett}, {McMillan}, {Skrutskie},
  {Stanford}, {Cohen}, {Walker}, {Mather}, {Leisawitz}, {Gautier}, {McLean},
  {Benford}, {Lonsdale}, {Blain}, {Mendez}, {Irace}, {Duval}, {Liu}, {Royer},
  {Heinrichsen}, {Howard}, {Shannon}, {Kendall}, {Walsh}, {Larsen}, {Cardon},
  {Schick}, {Schwalm}, {Abid}, {Fabinsky}, {Naes}, \& {Tsai}}]{Wright10}
{Wright}, E.~L., {Eisenhardt}, P.~R.~M., {Mainzer}, A.~K., {et~al.} 2010,
  \href{http://dx.doi.org/10.1088/0004-6256/140/6/1868}{\aj, 140, 1868}

\bibitem[{{Zuckerman} {et~al.}(2006){Zuckerman}, {Bessell}, {Song}, \&
  {Kim}}]{Zuckerman06}
{Zuckerman}, B., {Bessell}, M.~S., {Song}, I., \& {Kim}, S. 2006,
  \href{http://dx.doi.org/10.1086/508060}{\apjl, 649, L115}

\end{thebibliography}

\end{document}